\documentclass[12pt,a4paper]{article}
\pdfoutput=1
\usepackage{jheppub2}
\usepackage{braket}
\usepackage[english]{babel}
\usepackage[utf8]{inputenc}
\usepackage{amsmath}
\usepackage{array, url}
\usepackage{fancyhdr}
\usepackage{amsfonts}
\usepackage{amsthm}
\usepackage{mathtools}
\usepackage{bm}
\usepackage{titlesec}
\usepackage{graphicx}
\graphicspath{{./figures/}}
\usepackage{color}
\usepackage{tensor}
\usepackage{mathrsfs}
\usepackage{caption}
\usepackage{subcaption}
\usepackage{etoolbox}
\usepackage{tikz}
\usetikzlibrary{positioning}
\usepackage{comment}
\usepackage{tensor}
\usepackage[title]{appendix}
\usepackage{soul,xcolor}
\usepackage{enumerate}

\numberwithin{equation}{section}

\usepackage{amsmath}

\DeclareMathOperator\tr{Tr}

\def\deltabar{{\mathchar '26\mkern -10mu\delta}}

\theoremstyle{stylename}
\usepackage{ae,aecompl}
\usepackage{xcolor}
\renewcommand*\thesection{\arabic{section}}
\definecolor{mathematica1}{rgb}{0.368417, 0.506779, 0.709798}
\definecolor{mathematica2}{rgb}{0.880722, 0.611041, 0.142051}
\definecolor{mathematica3}{rgb}{0.560181, 0.691569, 0.194885}
\definecolor{mathematica4}{rgb}{0.922526, 0.385626, 0.209179}
\definecolor{mathematica6}{rgb}{0.772079, 0.431554, 0.102387}
\definecolor{pink}{rgb}{1, 0.5, 0.5}

\newcommand{\blue}[1]{{\color{blue} #1 \color{black}}}

\newcommand{\JK}[1]{{\blue{JK: #1}}} 

\newcommand{\overbar}[1]{\mkern 1.5mu\overline{\mkern-1.5mu#1\mkern-1.5mu}\mkern 1.5mu}

\newcommand{\beq}{\begin{equation}}

\newcommand{\eeq}{\end{equation}}

\title{Cornering gravitational entropy}

\author[\dagger]{Jani~Kastikainen,}
\author[\natural]{and Andrew~Svesko}

\affiliation[\dagger]{Institute for Theoretical Physics and Astrophysics and Würzburg-Dresden Cluster of Excellence
ct.qmat, Julius-Maximilians-Universität Würzburg, Am Hubland, 97074 Würzburg, Germany}
\affiliation[\natural]{Department of Mathematics, King’s College London, Strand, London, WC2R 2LS, UK}

\emailAdd{jani.kastikainen@uni-wuerzburg.de}
\emailAdd{andrew.svesko@kcl.ac.uk}

\abstract{We present a new derivation of gravitational entropy functionals in higher-curvature theories of gravity using corner terms that are needed to ensure well-posedness of the variational principle in the presence of corners. This is accomplished by cutting open a manifold with a conical singularity into a wedge with boundaries intersecting at a corner. Notably, our observation provides a rigorous definition of the action of a
conical singularity that does not require regularization.
For Einstein gravity, we compute the R\'enyi entropy of gravitational states with either fixed-periodicity or fixed-area boundary conditions. The entropy functional for fixed-area states is equal to the corner term, whose extremization follows from the variation of the Einstein action of the wedge under transverse diffeomorphisms. 
For general Lovelock gravity the entropy functional of fixed-periodicity states is equal to the Jacobson--Myers (JM) functional, while fixed-area states
generalize to fixed-JM-functional states, having a flat spectrum. Extremization of the JM functional is shown to coincide with the variation of the Lovelock action of the wedge. For arbitrary $F$(Riemann)
gravity, under special
periodic boundary conditions, we recover the Dong--Lewkowycz entropy for fixed-periodicity states. Since the variational problem in the presence of corners is not well-posed, 
we conjecture the generalization of fixed-area
states does not exist for such theories without additional boundary conditions. Thus, our work suggests the existence of entropy functionals is tied to the existence of corner terms which make the Dirichlet variational problem well-posed.

}

\begin{document}
\setstcolor{red}

	\maketitle

\newpage
\section{Introduction and summary} \label{sec:intro}

Uncovering the microscopic origin of gravitational entropy remains an open puzzle in quantum gravity.
As such, understanding how gravitational entropy arises and which surfaces it may be attributed to is of fundamental importance. Progress along these lines has been made via the AdS/CFT correspondence, a concrete realization of the holographic principle.
In this context, gravitational entropy has a dual meaning, encapsulated by
the Ryu--Takayangi (RT) prescription for computing entanglement entropies of holographic conformal field theories (CFT) \cite{Ryu:2006bv,Ryu:2006ef},
\beq S_{\text{EE}}(A)=\underset{\mathcal{C}\,\sim\,\partial A}{\text{min}}\frac{\text{Area}\,(\mathcal{C})}{4G_{\text{N}}}\;.\label{eq:RTform}\eeq
That is, the entanglement entropy of a holographic CFT state reduced to a (spatial) subregion $A$ of the conformal boundary of anti-de Sitter (AdS) space is captured by the area of a (bulk) minimal surface $\mathcal{C}$ anchored to $\partial A$ and homologous to $A$. The RT entropy-area relation~(\ref{eq:RTform}) generalizes the Bekenstein--Hawking entropy formula specific to black holes~\cite{Bekenstein:1972tm,Bekenstein:1973ur,Hawking:1975vcx,Hawking:1976de}; in fact,
when the minimal surface is identified with a spherical black hole horizon, the entropy of the black hole equals the thermal entropy of the dual CFT \cite{Casini:2011kv}. The RT formula thus suggests the microscopics responsible for gravitational entropy are due to entangled degrees of freedom of a dual CFT.

The derivation of the RT prescription (\ref{eq:RTform}) offers further insight into how gravitational entropy arises from first principles. Initially presented by Lewkowycz and Maldacena \cite{Lewkowycz:2013nqa}, the derivation invokes the `replica trick' to compute holographic entanglement entropies.
More specifically, one considers a bulk (Euclidean) spacetime $\mathcal{M}$ with asymptotic boundary~$B$, where the dual quantum theory is expected to reside. The replica trick consists of gluing $n$-copies of~$B$ together and computing the partition function of the $n$-fold cover~$B_{n}$, which innately has a discrete global $\mathbb{Z}_{n}$ `replica symmetry' describing the cyclic permutation of the $n$-replicas.  This replica symmetry is assumed to extend to the (dominant) bulk Riemannian manifold~$(\mathcal{M}_{n},\overbar{g}_{n})$, with Euclidean metric~$\overbar{g}_{n}$ being a smooth solution to Einstein's equations. According to AdS/CFT duality, the boundary partition function $Z[B_n]$ is dual to the bulk partition function $Z[\mathcal{M}_n]$ of the replicated geometry. The key insight of \cite{Lewkowycz:2013nqa} is that the orbifold $\mathcal{M}_{n}/\mathbb{Z}_{n}$, equipped with the on-shell metric $\overbar{g}_n$, offers a better analytic continuation to non-integer $n$ than the continuation of the boundary theory. Additionally, the orbifold is regular everywhere except along a bulk codimension-2 surface $\mathcal{C}$ with a conical defect, characterized by the fixed points of $\mathbb{Z}_{n}$. 
A gravitational entropy can be ascribed to $\mathcal{C}$, where the on-shell Euclidean gravity action $I_{\mathcal{M}_{n}/\mathbb{Z}_{n}}[\overbar{g}_n]$ 
determines the form of the entropy functional (upon analytic continuation $n\to1$),
\beq S_{\text{EE}}(A)=\partial_{n}I_{\mathcal{M}_{n}/\mathbb{Z}_{n}}[\overbar{g}_n]|_{n=1}\;.\label{eq:entfuncaction}\eeq
Note the on-shell action of $(\mathcal{M}_{n}/\mathbb{Z}_{n},\overbar{g}_{n})$ does \emph{not} have any contribution from the conical defect.
Nonetheless, the gravitational entropy is ascribed to $\mathcal{C}$ as in (\ref{eq:RTform}). Additionally, which type of surfaces $\mathcal{C}$ can be assigned an entropy is determined \emph{dynamically}.
When the bulk is governed by Einstein gravity, for example, entropy is assigned to extremal surfaces, i.e., those which extremize their area, including the minimal surface in (\ref{eq:RTform}), and follows from Einstein's equations evaluated near $\mathcal{C}$.\footnote{Technically, one examines the leading divergent components of the Einstein equations arising due to the conical singularity and setting such terms to zero.}

The method of Lewkowycz and Maldacena can be seen as an extension of the Euclidean path integral derivation of black hole thermodynamics \'a la Gibbons and Hawking~\cite{Gibbons:1976ue}. Crucial to the set-up employed by Gibbons--Hawking was that the Euclidean gravity solutions have an assumed continuous $U(1)$ Killing symmetry, such that the gravitational background is in (thermal) equilibrium, as is the case for spacetimes with bifurcate Killing horizons. The technique developed in \cite{Lewkowycz:2013nqa} (see also, e.g., \cite{Fursaev:1995ef,Solodukhin:2008dh,Solodukhin:2011gn,Fursaev:2012mp,Fursaev:2013fta}), applies to Euclidean solutions without this $U(1)$ isometry, including minimal RT surfaces or event horizons of dynamical black holes far from equilibrium. In the absence of a $U(1)$ symmetry, surfaces $\mathcal{C}$ always have non-zero extrinsic curvatures. When the Euclidean solutions end up having a $U(1)$ Killing symmetry, the  Lewkowycz and Maldacena method recovers the Gibbons--Hawking result, where the Killing horizon is a minimal surface with vanishing extrinsic curvature.

In this article we examine the problem of deriving gravitational entropy using an alternative technique we dub the `corner method'. Key to this approach is to recognize the familiar conical singularity arising from the replica trick as a corner, a codimension-2 surface at the intersection of two codimension-1 boundaries, and its Euclidean action entirely encodes the gravitational entropy functional. In particular, we cut the conical singularity into a corner such that a Hayward corner term in the action is needed to have a well-posed variational problem \cite{Hayward:1993my}, a codimension-2 analog of the Gibbons--Hawking--York boundary term.\footnote{The need for a corner term was anticipated by Hawking (see the discussion above Eq. (106) in Section 8 of \cite{Hawking:1980gf}).} Advantageously, the corner method readily extends to arbitrary theories of higher-curvature gravity (provided they admit corner terms). 

 Holographic entanglement entropy for arbitrary gravity theories is formally given by (\ref{eq:entfuncaction}), however, it requires effort to determine the explicit entropy functional and the analog of area extremization. Indeed, as is the case for black hole entropy, an entropy-area relation (\ref{eq:RTform}) no longer suffices when the holographic CFT is dual to a higher-curvature theory of gravity.  One proposed prescription is known as the Camps--Dong (sometimes referred to as the Wald--Camps--Dong) formula \cite{Dong:2013qoa,Camps:2013zua}.\footnote{More precisely, Camps \cite{Camps:2013zua} only dealt with quadratic theories of gravity while Dong \cite{Dong:2013qoa} considered $F$(Riemann) gravity and coincides with Camps' result for quadratic theories.}
Explicitly, for theories of gravity whose Lagrangian density is an arbitrary diffeomorphism invariant scalar function of the bulk metric and Riemann tensor, namely $F(\text{Riemann})$ theories, the RT formula (\ref{eq:RTform}) is generalized to
\beq S_{\text{EE}}(A)=S_{\text{Wald}}(\mathcal{C})+S_{\text{anom}}(\mathcal{C})\;.\label{eq:campsdongform}\eeq
The first term is the Iyer--Wald entropy functional \cite{Wald:1993nt,Iyer:1994ys}, 
\beq S_{\text{Wald}}(\mathcal{C})=2\pi\int_{\mathcal{C}}d^{D-2}x\sqrt{\sigma}\,P^{abcd}\,\varepsilon_{ab}\varepsilon_{cd}=2\pi\int_{\mathcal{C}}d^{D-2}x\sqrt{\sigma}\,\Psi\;,\label{eq:Waldentgen} \eeq
where $\varepsilon_{ab}=2n_{[a}r_{b]}$ is the binormal on $\mathcal{C}$ satisfying $\varepsilon_{ab}\,\varepsilon^{ab}=+2$ (in Euclidean signature), and $P^{abcd}\,\varepsilon_{ab}\varepsilon_{bc}\equiv\Psi$, with $P^{abcd}\equiv\frac{\partial F}{\partial R_{abcd}}$. Initially, Wald introduced this functional to generalize the Bekenstein--Hawking entropy formula for stationary black holes in arbitrary diffeomorphism invariant theories of gravity.
The second contribution $S_{\text{anom}}$ is known as the `anomaly term',
\beq
\begin{split}
S_{\text{anom}}(\mathcal{C})&=2\pi\int_{\mathcal{C}}d^{D-2}x\sqrt{\sigma}\sum_{m}\frac{2}{q_{m}+1}\left(\frac{\partial^{2}F}{\partial R_{a_{1}b_{1}c_{1}d_{1}}\partial R_{a_{2}b_{2}c_{2}d_{2}}}\right)_{\hspace{-1mm} m}K_{f_1b_{1}d_{1}}K_{f_2b_{2}d_{2}}\\
&\times\left[(n_{a_{1}a_{2}}n_{c_{1}c_{2}}-\varepsilon_{a_{1}a_{2}}\varepsilon_{c_{1}c_{2}})\,n^{f_1f_2}+(n_{a_{1}a_{2}}\,\varepsilon_{c_{1}c_{2}}+n_{c_{1}c_{2}}\varepsilon_{a_{1}a_{2}})\,\varepsilon^{f_1f_2}\right]\;.
\end{split}
\label{eq:Sanomgen}\eeq
Here $n_{ab} = g_{ab}-\sigma_{ab}$ is the metric in the two-dimensional space orthogonal to $\mathcal{C}$ and $K_{fab}$ encodes two independent extrinsic curvatures of $\mathcal{C}$ as $ n^f K_{fab}\equiv L_{ab}$ and $r^f K_{fab}\equiv Q_{ab}$ (in Appendix \ref{app:variations} we detail our notation). The parameter $q_{m}$ (a non-negative rational number between $0$ and $1$) and its sum correspond to a weighting procedure to be performed via a scheme regularizing conical singularities of squashed cones.\footnote{The precise origin of the anomaly term arises from potential logarithmic divergences which may appear at $\mathcal{C}$ and its namesake stems from drawing an analogy with the Weyl anomaly \cite{Dong:2013qoa}.}
Two special cases of (\ref{eq:campsdongform}) include when (i) the surface $\mathcal{C}$ is a Killing horizon, or (ii) the bulk is governed by $F(R)$ gravity. In either case the anomaly term vanishes, leaving only the Wald entropy. Further, for Lovelock gravity \cite{Lovelock:1971yv}, the Camps--Dong formula (\ref{eq:campsdongform}) simplifies to the Jacobson--Myers black hole entropy functional  \cite{Jacobson:1993xs}, reproducing the correct universal contributions of the dual CFT entanglement entropy (first shown in \cite{Hung:2011xb} in the context of Gauss--Bonnet gravity and cubic Lovelock gravities in \cite{deBoer:2011wk}). Notably, the Jacobson--Myers functional is independent of extrinsic curvatures of the underlying surface.\footnote{Note that in the original derivation \cite{Jacobson:1993xs}, the Jacobson--Myers functional is independent of extrinsic curvatures, because it is evaluated on a stationary black hole horizon whose extrinsic curvatures vanish. The authors of \cite{Jacobson:1993xs} did not consider surfaces with non-zero extrinsic curvatures unlike the authors of \cite{Dong:2013qoa,Camps:2013zua}.}

The derivation and application of the Camps--Dong (\ref{eq:campsdongform}) proposal, however, poses non-trivial challenges. The first challenge we describe is known as the `splitting problem' \cite{Miao:2014nxa,Camps:2014voa,Camps:2016gfs}. As we review below, the splitting problem amounts to determining the Taylor expansion of on-shell metrics near $\mathcal{C}$. Start with an ansatz for an off-shell metric obeying regularity and replica symmetry. The Riemann curvatures evaluated at $\mathcal{C}$ exhibit a discontinuity, known as a `splitting', when $n\rightarrow 1$: the curvature in the limit $n\to1$ splits into the Riemann tensor at $n=1$ and a contribution involving an infinite number of terms owed to the ansatz for $n>1$ (e.g. Eq. (\ref{eq:splitRR'})).
The precise form of the splittings are then determined in terms of intrinsic and extrinsic data of the on-shell solution by imposing the bulk equations of motion. Hence, the splittings are theory dependent.\footnote{Historically \cite{Dong:2013qoa}, the splittings of the Riemann tensor used were not obtained from a metric satisfying the bulk equations of motion. This issue was first pointed out in \cite{Miao:2014nxa,Miao:2015iba}.}
The Taylor expansion for on-shell metrics has been solved in Einstein gravity (up to some subtleties that we will point out) and the splittings of the Riemann tensor have been determined \cite{Camps:2014voa,Miao:2014nxa,Miao:2015iba,Camps:2016gfs}. Solving the splitting problem in theories other than Einstein gravity, however, has thus far not been accomplished, because solving their equations of motion order by order is highly non-trivial, and, in addition, it is not known if a solution is guaranteed to exist. As we emphasize in this work, there is always a splitting problem (equations of motions must be solved to determine the splittings), however, in computing the entropy, the splitting problem can be ignored for certain theories.


The splittings directly enter the anomaly term (\ref{eq:Sanomgen}) of the Camps--Dong formula, where the value assigned to $q_{m}$ depends on the splitting. Notably, different splittings generally yield different entropy functionals, however, some theories are blind to said splittings.
With respect to the Camps--Dong functional, it is said there is no splitting problem for $F(R)$ or quadratic theories of gravity as the second curvature derivative of the Lagrangian in the anomaly does not produce any curvature monomials subject to splittings. Likewise, Lovelock gravity is immune to the splitting problem, since the Jacobson--Myers functional depends only on the intrinsic geometry of the surface. The splitting problem first arises for general cubic theories of gravity \cite{Camps:2016gfs} (see also \cite{Caceres:2020jrf}). Alternatively, for perturbative general higher-curvature gravities, the on-shell splittings for Einstein gravity are sufficient, allowing for an unambiguous rewriting of the Camps--Dong formula (\ref{eq:campsdongform}) without the weighted sum \cite{Bueno:2020uxs}.

Another challenge is that a potential inconsistency arises in the extremization prescription determining which surface the entropy functional is integrated over. For Einstein gravity, the surface is an extremal surface, found by extremizing the area functional, as prescribed in the RT formula (\ref{eq:RTform}). Lewkowycz and Maldacena proved the extremization of the area by removing a singularity arising from the analytic continuation $n\to1$ of Einstein's equations, 
yielding the minimal surface constraint (vanishing trace of extrinsic curvatures of $\mathcal{C}$). Thus, the minimal surface condition follows from either directly extremizing the area functional or by demanding the bulk equations of motion are divergence free. For higher-curvature theories this is no longer obviously the case \cite{Chen:2013qma,Bhattacharyya:2014yga} (see also \cite{Erdmenger:2014tba}). Already for quadratic theories the surface constraints following from extremizing the Camps--Dong formula do not generally coincide with the surface constraints found by eliminating the divergent contributions to the bulk equations of motion at $\mathcal{C}$. This has been shown explicitly for Gauss--Bonnet gravity, where the constraint coming from the leading term of the equations of motion coincides with the extremization of the JM functional, but subleading terms appear to give additional constraints \cite{Chen:2013qma,Bhattacharyya:2014yga}. Hence there is a tension in which extremization prescription is correct.\footnote{For Lovelock gravity, there is evidence the correct dual CFT entropy is given by evaluating the Jacobson--Myers functional on the surface found from extremizing the Jacobson--Myers functional \cite{Hung:2011xb,deBoer:2011wk}. Thus, one is biased to recover this surface condition, assuming AdS/CFT is valid, from the surface constraints attained from the divergent bulk equations of motion. One possibility is to impose, \emph{ad hoc}, a specific condition on the combination of the Lewkowycz--Maldacena constraints. Alternatively, the two extremization methods were found to coincide by carefully evaluating the order of limits as one approaches $\mathcal{C}$ and analytically continues $n\to 1$ \cite{Bhattacharyya:2014yga}, such that only the leading divergences to the bulk equations of motion remain. However, a physical or first principles explanation is needed to justify this delicate order of limits.}

On general grounds, Dong and Lewkowycz \cite{Dong:2017xht} argued
the surface condition found from explicitly evaluating the bulk equations of motion is equivalent to the extremization of another entropy functional
given by a boundary term evaluated at $\mathcal{C}$.
In particular, for arbitrary theories, they arrive to what we call the Dong--Lewkowycz functional
\beq S_{\text{EE}}(A)=\lim_{n\to1^{+}}S_{\text{Wald}}[g_{n,0}]\;,\label{eq:Lewkdong}\eeq
where $g_{n,0}$ is the so called ``split-metric'' (see Eq. \eqref{eq:gnsplitform} in Section \ref{sec:splitting} for the definition). 
Despite appearances, the entropy functional (\ref{eq:Lewkdong}) is claimed in \cite{Dong:2017xht} to match the Camps--Dong functional (\ref{eq:campsdongform}), but demonstrating this is non-trivial and requires solving the splitting problem in a general theory of $F$(Riemann) gravity.\footnote{To be illustrative, an explicit analysis showing equivalence was done for two-dimensional dilaton gravity with higher derivative interactions \cite{Dong:2017xht}. Thus, in our view, whether the functionals (\ref{eq:campsdongform}) and (\ref{eq:Lewkdong}) agree in general remains an open question.}

Here, using our corner method, we make progress in addressing some of these outstanding questions. In particular, we consistently formulate and derive gravitational entropy functionals and their extremization prescriptions using corner terms that are needed to ensure
well-posedness of the variational principle in the presence of corners. Our essential insight is the following identity
\begin{equation}
I_{\mathcal{M}\,\backslash\, \mathcal{C}}[g] = I_{\mathcal{W}}[g]\;.
\label{eq:mainwedge}
\end{equation}
On the left-hand side is the gravitational action of a manifold $(\mathcal{M}\,\backslash\, \mathcal{C},g)$ containing a conical singularity at a codimension-2 surface $\mathcal{C}\subset \mathcal{M}$, while the right-hand side is the action of a wedge shaped manifold $(\mathcal{W},g)$ with a corner at $\mathcal{C}$. The opening angle of the corner is determined by the angular excess/deficit of the conical singularity, and periodic boundary conditions at the edges of $\mathcal{W}$ are imposed on the metric variation. The right-hand side of \eqref{eq:mainwedge} arises by cutting open a conical singularity. Alternatively, the identity (\ref{eq:mainwedge}) can be seen as a rigorous \textit{definition} of the action of a conical singularity that does not require regularization. In particular, the corner term $I_{\mathcal{C}}[g]$ required to make the variational principle for Dirichlet boundary conditions on $(\mathcal{W},g)$ well defined can be identified with the familiar ``delta function'' contribution of the singularity,
\begin{equation}
I_{\mathcal{M}}[g] \equiv I_{\mathcal{W}}[g] + I_{\mathcal{C}}[g]\;,
\label{eq:mainwedge2}
\end{equation}
where now $\mathcal{C}$ is included on the manifold on the left-hand side. 

Our observations (\ref{eq:mainwedge}) and (\ref{eq:mainwedge2}) apply to, in principle, arbitrary theories of gravity. It is evident, however, the existence of entropy functionals is linked to whether the variational problem assuming Dirichlet boundary conditions is well-posed, i.e., the existence of (Dirichlet) corner terms. Since general theories of higher-curvature gravity do not admit a corner term with Dirichlet boundary conditions on the induced metric alone, our approach is only valid for theories that admit a well-posed Dirichlet variational problem, e.g., Lovelock theories. 
In the case of Lovelock gravity, we will show the entropy functional coming from the corner method is the Jacobson--Myers functional. Of course, it is possible to impose extra boundary conditions in addition to the Dirichlet boundary condition on the induced metric to yield a well-posed variational problem, though, at the cost of altering the corner term.\footnote{This is not so surprising: with extra boundary conditions, codimension-1 boundary terms can be shown to exist in all $F(\text{Riemann})$ theories of gravity, cf. \cite{Deruelle:2009zk,Lehner:2016vdi,Liu:2017kml,Jiang:2018sqj}. However, such boundary terms do not make the \emph{Dirichlet} variational problem well-posed. For example, $F(R)$ gravity with its boundary term \cite{Barth:1984jb,Madsen:1989rz} (see also \cite{Dyer:2008hb}) has a well-posed variational problem when both $\delta g^{ab}=0$ (Dirichlet) and $\delta R=0$ on the boundary (which amounts to also fixing certain combination of normal directed second derivatives of the metric at the boundary).} Our approach thus leads to the following observation: the corner method does not associate a unique entropy functional to all theories of gravity, i.e., the entropy depends on the boundary conditions imposed at the corner. Indeed, with special periodic boundary conditions, Eq. (\ref{eq:specialpbc}), we derive the Dong--Lewkowycz entropy (\ref{eq:Lewkdong}) for general $F$(Riemann) theories for particular states.

Our corner method also sheds light on and extends the computation of gravitational entropy of fixed-area states. In fact, thus far our discussion has been centered on entropy of Hartle--Hawking states, i.e., states prepared by a Euclidean path integral over all metrics with fixed asymptotics at infinity. Fixed-area states, by contrast, are prepared by a Euclidean path integral over metrics with the same conditions at infinity in addition to fixing the induced area of $\mathcal{C}$ in the interior \cite{Akers:2018fow,Dong:2018seb}. Correspondingly, these states may be characterized by metrics obeying different boundary conditions, (\ref{maldacena}) and (\ref{fursaev}), respectively, and yield different R\'enyi entropies, (\ref{refinedrenyiHH}) and (\ref{refinedrenyifixedarea}). Indeed,  the entropy of fixed-area states is independent of the R\'enyi index, having a flat entanglement spectrum \cite{Headrick:2010zt}. Our corner method shows the entropy functional for fixed-area states is equal to the corner term itself, while extremization of the functional follows from the variation of the action of the whole wedge under diffeomorphisms. With this perspective it is clear how to generalize fixed-area states in higher-curvature gravity, provided such theories admit a corner term. Thus, our approach implies the existence of entropy functionals is linked to the existence of corner terms which make the (Dirichlet) variational problem well posed.

Before summarizing our main results, we emphasize deriving gravitational entropy from a corner term is not new.  Historically, corner terms have been used to compute entropy of stationary black holes \cite{Banados:1993qp,Hawking:1994ii,Teitelboim:1994az,Teitelboim:1994is}. 
Further, Hayward terms were introduced in \cite{Takayanagi:2019tvn} to construct the entropy functional for fixed-area states in Einstein gravity. Gravitational R\'enyi entropy was also computed in Einstein and Jackiw--Teitelboim gravity \cite{Botta-Cantcheff:2020ywu,Arias:2021ilh} using a corner method. These works, however, do not provide a complete treatment of variational problems for the metric and the corner embedding. Aside from deriving entropy functionals for higher-curvature theories of gravity, another way our approach is distinct from previous methods is that it allows for a careful treatment of the variational principle, and allows us to directly consider the extremization prescription.

\subsection{Summary of main results}

\paragraph{Defining gravitational states and revisiting the splitting problem.} In Section \ref{sec:graventprelims}, we review the derivation of gravitational entropy of two types of states in Einstein gravity. Specifically, we define (i) Hartle--Hawking (HH) states and (ii) fixed-area states. These states are prepared by Euclidean path integrals over metrics that have $S^1\times Y$ boundary conditions at asymptotic infinity, but different conditions in the interior on the codimension-2 surface $\mathcal{C}$ where the circle $S^1$ shrinks. In particular, for Hartle--Hawking states, one requires the absence of conical singularities (regularity) on $\mathcal{C}$, while for fixed-area states one fixes the induced area of $\mathcal{C}$ \cite{Akers:2018fow,Dong:2018seb}. 

In the course of our review, we revisit and clarify the essence of the splitting problem of Hartle--Hawking metrics using the general replica symmetric metric ansatz provided in \cite{Dong:2017xht,Camps:2014voa}. In this ansatz, all metric components split into an infinite series which do not truncate, as implicitly assumed in older work investigating the splitting problem \cite{Miao:2014nxa,Miao:2015iba,Camps:2016gfs}. In Einstein gravity, we show in detail Einstein's equations require truncation of the splitting relations, reproducing \cite{Miao:2014nxa,Miao:2015iba,Camps:2016gfs}. We also clarify an additional assumption made in \cite{Miao:2014nxa,Miao:2015iba} regarding the choice of gauge. To keep this article self-contained and pedagogical, we include Appendices \ref{app:repmanifolds} and \ref{app:solvingsplittingprobs} reviewing the geometry of replicated manifolds and solving the splitting problem, respectively.

\paragraph{Entropy functionals from corner terms in Einstein gravity.} In Section \ref{sec:einsteinentropy}  we derive gravitational entropies of Hartle--Hawking and fixed-area states in Einstein gravity using the identity \eqref{eq:mainwedge}. In particular, for HH states we reproduce the RT formula from a corner localized variation of the action of the wedge, where the area extremization prescription follows from Einstein's equations as in \cite{Lewkowycz:2013nqa}. Alternatively, for fixed-area states, the RT entropy-area functional is identified with the corner term itself, having a flat-entanglement spectrum, while the extremization prescription follows from the variation of the Einstein action of the wedge under transverse diffeomorphisms of $\mathcal{C}$.
Advantageously, the corner method, for either type of gravitational state, does not require regularization of the conical singularity. This section serves as an unabridged version of \cite{Kastikainen:2023yyk}. For clarity, in Appendix \ref{app:Thetam} we compute the opening angles of the wedges obtained by cutting open manifolds endowed with metrics characterizing Hartle–
Hawking and fixed-area states at $\mathcal{C}$.


\paragraph{Corner term and stress tensor in Lovelock gravity.} A main result of our work is applying our corner method to derive gravitational entropy in higher-curvature theories. An important aspect of this is studying the (Dirichlet) variational problem on a wedge shaped manifold $(\mathcal{W},g)$ with a corner. Thus, as an interlude, in Section \ref{sec:cornhighcurv},  we consider metric variations of gravitational actions on $(\mathcal{W},g)$ in the presence of a corner (relegating details to Appendix \ref{app:actionvariation}). When the action is supplemented by boundary and corner terms, a variation with respect to the metric defines both the boundary $\widetilde{T}_{ab}$ and corner $\widehat{T}_{ab}$ stress tensors of the theory (cf. Eq. \eqref{generaltheoryvariation}). Focusing on Lovelock gravity, we recover its well-known boundary stress tensor, and derive, for the first time, the Lovelock corner stress tensor using a smoothing trick developed in \cite{Hayward:1993my,Cano:2018ckq} where the sharp corner is replaced by a smooth circular arc (with details given in Appendix \ref{app:cornerstress}).  When the action contains only a single Lovelock scalar of order $m$, i.e., pure Lovelock gravity, we find that the corner stress-tensor is
\begin{equation}
    \widehat{T}_{(m)ab} = 2m\,\Theta\,\widehat{E}_{(m-1)ab} + \mathcal{F}_{(m)ab}(L_1,Q_1)+\mathcal{F}_{(m)ab}(L_2,Q_2)\;,
\end{equation}
where $\Theta$ is the corner angle, $\widehat{E}_{(m-1)ab}$ is the induced equation of motion tensor of the corner, and the expression for $\mathcal{F}_{(m)ab}$ is defined below \eqref{lovelockcornerstress}.

We also consider metric variations of geometric quantities $(\mathcal{W},g)$ (with details left to Appendix \ref{app:variations}). In particular, we derive new formulae for the variations of extrinsic curvatures $L_{ab}$, $Q_{ab}$ of the corner when Dirichlet boundary conditions are imposed, namely,
\begin{equation}
    \delta Q_{\alpha B}^{A} = 0, \quad \delta L_{1B}^A = -\csc{\Theta}\,\delta \Theta\,L_{2 B}^A, \quad \delta L_{2 B}^A = -\csc{\Theta}\,\delta \Theta\,L_{1 B}^A\;.
    \label{eq:extvariationssummary}
\end{equation}
With these formulae we have showed, up to actions of order $m = 20$, that the corner term of Lovelock gravity derived in \cite{Cano:2018ckq} is consistent and correct by explicitly showing it cancels all terms coming from the variation of the Lovelock action on a manifold with a corner.

\paragraph{Gravitational entropy in higher-curvature gravity: Lovelock and beyond.} In Section \ref{sec:entropyfromcornerterms}, we first extend the corner method we used to compute gravitational entropy in Einstein gravity to arbitrary Lovelock gravity using the analog of the Hayward corner term. We show the Hartle--Hawking entropy functional coincides with the Jacobson--Myers functional,  even when the surface has non-zero extrinsic curvatures. Moreover, we demonstrate the variational principle for the metric in the wedge is well-posed and gives the Lovelock field equations. Solving the equations of motion to leading order, we find a condition on the embedding of $\mathcal{C}$. 
Importantly, our condition coincides with the extremization of the Jacobson--Myers functional, under certain assumptions regarding the analytic continuation of $n$.

Via our method, we further show the correct generalization of a fixed-area state in Lovelock gravity is a fixed-JM-functional state, where the Jacobson--Myers functional of the surface $\mathcal{C}$ is fixed rather than its area.  As such, we find the entropy functional of a fixed-JM-functional state is the Jacobson--Myers functional, but with a flat entanglement spectrum. The extremization prescription arises in this case from the variation of the Lovelock corner term itself and again matches with the extremization of the JM functional.



We then consider theories beyond Lovelock. Since the Dirichlet variational problem in the presence of corners is not well-posed for arbitrary $F$(Riemann) gravity, our method
implies the analog of a fixed–area state does not exist in general. Meanwhile, for Hartle--Hawking states, we can recover the Dong–Lewkowycz entropy if we employ special periodic boundary
conditions \eqref{eq:specialpbc}. Since the splitting problem remains open for arbitrary higher-curvature theories, we cannot verify whether the extremization of the Dong--Lewkowycz functional coincides with the surface constraint found from imposing the leading order equations of motion, or whether Dong–Lewkowycz coincides with the Camps--Dong formula in general. 



\vspace{3mm} 

We conclude in Section \ref{sec:discussion}, summarizing our findings and provide an outlook on some of the future directions with which to take our work.

\section{Gravitational entropy: preliminaries} \label{sec:graventprelims}


Here we review the derivation of gravitational entropy around surfaces $\mathcal{C}$ for spacetimes governed by Einstein gravity for different types of gravitational states, dubbed ``Hartle--Hawking'' and ``fixed-area'' states \cite{Akers:2018fow,Dong:2018seb}.
To clarify our approach and set up notation, first we review the now standard computation of gravitational entropy via the replica trick. 
For completeness, we then characterize the infamous splitting problem associated with computing the entropy of Hartle--Hawking states. We conclude with a summary of the derivation of gravitational entropy of Hartle--Hawking states \emph{\'a} \emph{la} Lewkowycz and Maldacena (LM)   \cite{Lewkowycz:2013nqa}, evaluating the Einstein action of an orbifold geometry. Note that, while motivated by holographic entanglement entropy in the context of AdS/CFT, the following discussion applies more broadly. 



\subsection{Gravitational entropy and states}

\noindent \textbf{Gravitational entropy.}  Consider a $D$-dimensional Riemannian manifold $(\mathcal{M},g_1)$ endowed with a Euclidean metric\footnote{The majority of the literature on this topic refers to the pair $(\mathcal{M},g)$ as $\mathcal{M}$. We find it useful to distinguish between the topological space $\mathcal{M}$ and the Riemannian manifold $(\mathcal{M},g)$.}~$g_1$.
Let $(B,\gamma)$ be the $(D-1)$-dimensional boundary of $\mathcal{M}$ with metric $\gamma$, and topology $B=S^{1}\times Y$.
Via the replica trick, 
another $(D-1)$-dimensional manifold $(B_{n},\gamma)$ is constructed by cutting and cyclically pasting together $n$-copies of $(B,\gamma)$, with $n\in\mathbb{Z}^{+}$, along $Y$.
Denoting $\tau\in(0,2\pi)$ as a (Euclidean time) coordinate parametrizing the circle $S^{1}$, the cutting-pasting procedure can be viewed as extending the range of $\tau$ to $(0,2\pi n)$. The boundary $B_{n}$ automatically has a $\mathbb{Z}_{n}$ replica symmetry since the replicas are glued cyclically. One way to understand this symmetry is that with $\tau\sim\tau +2\pi n$, there remains a symmetry in $2\pi$ shifts to $\tau$, i.e., $\tau\to \tau+2\pi s$ for $s\in\mathbb{Z}/n\mathbb{Z}$. In the AdS/CFT context, $Y$ represents a (spatial) subregion $A$ of $B$ which the CFT state is reduced to.\footnote{$Y$ is the image of $A$ under the conformal mapping that maps the modular flow of $A$ to rotations of the $S^1$.} In this context, the boundary of $A$ is often referred to as the entangling surface, and the $\mathbb{Z}_{n}$ fixed points are the fixed points of the boundary of $A$.

Now given the family $(\mathcal{M},g_1)$ of bulk geometries sharing the same boundary $(B,\gamma)$, there is a geometry $(\mathcal{M},\overbar{g}_1)$ where $\overbar{g}_{1}$ solves the bulk field equations.\footnote{Generally, there exist multiple bulk solutions, each of which contribute in a saddle-point approximation to the gravitational path integral. Here we will only focus on the dominant saddle.} The aim is to find another solution $(\mathcal{M}_{n},\overbar{g})$ to the bulk field equations whose boundary is the replicated manifold $(B_{n},\gamma)$ such that $\overbar{g}\lvert_{n = 1}\, =\overbar{g}_1$.
With respect to the bulk, the gravitational entropy $S$ has the form \cite{Lewkowycz:2013nqa}
\beq S=\partial_{n}(I_{\mathcal{M}_{n}}[\overbar{g}]-nI_{\mathcal{M}}[\overbar{g}_{1}])\big|_{n=1}=(\partial_{n}-1)I_{\mathcal{M}_{n}}[\overbar{g}]\big|_{n=1}\;,\label{eq:gengravent1}\eeq
which we can take as a definition. Here $I_{\mathcal{M}_{n}}[\overbar{g}]$ is the on-shell gravity action of the geometry $(\mathcal{M}_{n},\overbar{g})$. To motivate this formula, take $(B,\gamma)$ to be the conformal boundary of Euclidean $\text{AdS}_{D}$ and consider a holographic CFT living on $B$. The entanglement entropy of a quantum state reduced to a boundary subregion $A$ may be computed from the analytic continuation of the $n$th-R\'enyi entropy $S_n$ as\footnote{The CFT quantum state reduced to $A$ is denoted $\hat{\rho}_{A}$ and normalized such that $\text{tr}[(\hat{\rho}_{A})^{n}]=Z_{n}/(Z)^{n}$. The R\'enyi entropy is $S_{n}(A)=\frac{1}{1-n}\log\text{tr}(\hat{\rho}_{A}^{n})$, with $S_{\text{EE}}(A)=\lim_{n\to 1}S_{n}(A)$.}
\beq S_{\text{EE}}(A)=-\lim_{n\to1}\frac{1}{n-1}(\log Z_{n}-n\log Z)=-\partial_{n}(\log Z_{n}-n\log Z)\big|_{n=1}\;.\eeq
Here $Z_{n}\equiv Z[B_{n}]$ is the partition function of the CFT on the $n$-fold cover $B_{n}$. Invoking the standard AdS/CFT dictionary, where in a saddle-point approximation $Z_{n}= e^{-I_{\mathcal{M}_{n}}[\overbar{g}]}$, one easily recovers the right-hand side of (\ref{eq:gengravent1}), such that the entanglement entropy of the dual CFT is identified with the gravitational entropy.

We will find it useful to work with the refined (or modular) R\'enyi entropy, $\widetilde{S}_{n}\equiv n^{2}\partial_{n}[(n-1)S_{n}/n]$ \cite{Dong:2016fnf}.\footnote{In terms of reduced state $\hat{\rho}_{A}$, the refined entropy is defined as $\widetilde{S}_{n}\equiv-n^{2}\partial_{n}[\log\tr{(\hat{\rho}_{A}^{n})}/n]$. For a holographic CFT state this becomes $\widetilde{S}_{n}=(1-n\partial_{n})\log Z_{n}$.} Gravitationally, one has 
\begin{equation}
    \widetilde{S}_n = (n\partial_n - 1)\,I_{\mathcal{M}_n}[\overbar{g}].
\label{eq:refinedRenyi}\end{equation}
As with the standard R\'enyi entropy of a quantum state, the refined entropy coincides with the von Neumann entropy as $n\to1$. An appealing feature of $\widetilde{S}_{n}$ is that, for CFTs with gravitational duals, it obeys an area law analogous to the RT formula (\ref{eq:RTform}). Specifically, for Hartle--Hawking states the refined entropy is proportional to the area of a codimension-2 cosmic brane with an $n$-dependent tension that backreacts on the ambient geometry by creating a conical deficit \cite{Dong:2016fnf}. Below we will typically compute the refined entropy before taking the $n\to1$ limit.

\vspace{2mm}

\noindent \textbf{Gravitational states.} An input in computing entropy is the state of the system. This is also necessary to determine the on-shell solution. Relevant for us are two types of gravitational states:
\begin{enumerate}[(i)]
    \item \emph{Hartle--Hawking (HH) states}, prepared by a Euclidean gravity path integral over all metrics with fixed asymptotics at infinity.
    \item \emph{fixed-area states}, prepared by a Euclidean gravity path integral over metrics with a given fixed area $\mathcal{A}$ on a codimension-2 surface $\mathcal{C}$ in the interior and fixed asymptotics at infinity.
\end{enumerate}
More carefully, following \cite{Dong:2018seb}, consider a pure state $\lvert \psi\rangle$ in the tensor product $\mathcal{H}_{\text{CFT}}\otimes \mathcal{H}_{\text{CFT}}$, prepared via a Euclidean path integral of the CFT over half of the circle in $B = S^1\times Y$ with possible sources turned on to characterize excited states. Taking the CFT to have a holographic dual, then, in the context of AdS/CFT, there is a gravitational path integral that computes the wave function
\begin{equation}
    \langle h\lvert \psi\rangle = \int_{g\lvert_{\Sigma}\, =\, h} \mathcal{D}g\,e^{-I_{\mathcal{M}}[g]}
    \label{eq:HHpathintegral}
\end{equation}
where the integration is over all metrics with the induced metric of $\Sigma$ fixed to $h$ and with fixed asymptotics at infinity. The state $\lvert \psi\rangle$ can hence be identified with the Hartle--Hawking state.

Alternatively, a fixed-area state $\lvert \psi_\mathcal{A}\rangle$ is defined as follows. Gauge-fix a Cauchy slice $\Sigma$ (with $\partial\Sigma$ being the boundary Cauchy surface for the CFT) such that it passes through a codimension-2 surface $\mathcal{C}$ and fixes the location of $\mathcal{C}$ on $\Sigma$. Restricting the gravitational path integral \eqref{eq:HHpathintegral} to be over all metrics $g_{\mathcal{A}}$ for which $\mathcal{C}$ has a fixed area $\mathcal{A}$ defines the fixed-area state,
\begin{equation}
    \langle h\lvert \psi_{\mathcal{A}}\rangle \equiv \int_{g\lvert_{\Sigma}\, =\, h} \mathcal{D}g_{\mathcal{A}}\,e^{-I_{\mathcal{M}}[g_{\mathcal{A}}]}\;.
\label{eq:fixedareanorm}\end{equation}
We will discuss the implications of the fixed-area constraint momentarily.


Either way, the gravitational entropies of states (i) and (ii) are calculated by on-shell Euclidean gravity actions of two different metrics in a saddle-point approximation. These two metrics are obtained as solutions of the gravitational equations of motion with different boundary conditions at the codimension-2 surface $\mathcal{C}$ where the Euclidean time circle shrinks to zero size. Specifically,  let $\rho > 0$ be a radial coordinate such that $\mathcal{C}$ is located at $\rho = 0$. Assuming Euclidean time $\tau$ is periodically identified as $\tau \sim \tau + 2\pi n$, close to the shrinking point $\rho = 0$, the boundary conditions are, corresponding to states (i) and (ii),
\begin{enumerate}[(i)]	
	\item Fixed-periodicity boundary condition
	\begin{equation}
		ds^2  =d\rho^{2} +\frac{\rho^{2}}{n^2}\,d\tau^{2}+ \sigma_{AB}(\hat{x})\,d\hat{x}^Ad\hat{x}^B+\ldots.
		\label{maldacena}
	\end{equation}
	 This boundary condition was used in the proof of the RT formula by Lewkowycz and Maldacena \cite{Lewkowycz:2013nqa}. 
 \item Fixed-area boundary condition
	\begin{equation}
		ds^2  =d\rho^{2} +\rho^{2}d\tau^{2} + \sigma_{AB}(\hat{x})\,d\hat{x}^Ad\hat{x}^B +\ldots,
		\label{fursaev}
	\end{equation}
	with 
 the area $\int_{\mathcal{C}} d^{D-2}x\sqrt{\sigma} \equiv \mathcal{A}$ fixed.\footnote{The surface $\mathcal{C}$ can be anything as long as it is defined in a diffeomorphism invariant manner. An example is a surface that minimizes its area in the background metric.}  This boundary condition was used in an attempted proof of the RT formula by Fursaev \cite{Fursaev:2006ih}, however, was shown to have a flat entanglement spectrum in \cite{Headrick:2010zt}.
\end{enumerate}
In either case, the ellipses indicate subleading terms in $\rho$, and the $\hat{x}^{A}$ denote $(D-2)$-dimensional worldvolume coordinates of $\mathcal{C}$. From here on, we denote metrics obeying boundary conditions (i) as $g_{n}$, while metrics satisfying (ii) are denoted $g_{\mathcal{A}}$. Notably, the metric $g_{n}$ has no conical singularity while $g_{\mathcal{A}}$ has a conical excess.

Given a boundary condition (i) or (ii), Einstein's equations can be solved order by order in proper distance from $\mathcal{C}$, where the ellipsis in metrics (\ref{maldacena}) or (\ref{fursaev}) denote subleading terms perturbatively determined in this way. Extracting the on-shell form of the subleading terms in the metric (\ref{maldacena}) for $n>1$ is known as the \textit{splitting problem} \cite{Miao:2014nxa,Miao:2015iba,Camps:2016gfs}, which has only been solved for Einstein gravity (as we review momentarily). Fixed-area states, however, have no splitting problem because there is no explicit $n$-dependence in the metric $g_{\mathcal{A}}$.

\begin{figure}[t!]
\begin{center}
\includegraphics[scale=1.5]{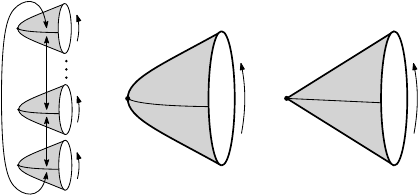}
\put(-255,127){$2\pi$}
\put(-255,66){$2\pi$}
\put(-255,22){$2\pi$}
\put(1,92){$2\pi$}
\put(-131,92){$2\pi n$}
\caption{Gravitational states. \textbf{Left:} Replicated $n$-manifold $(\mathcal{M}_n,\overbar{g}_{\mathcal{A}})$ preparing fixed-area states. \textbf{Center:} Manifold $(\mathcal{M}_n,\overbar{g}_{n})$ preparing Hartle--Hawking states. \textbf{Right:} The quotient geometry $(\mathcal{M}_{n}/\,\mathbb{Z}_{n},\overbar{g}_{n})$ for computation of HH entropy.}\vspace{-7mm}
\label{fig:gravstates}
\end{center}
\end{figure}
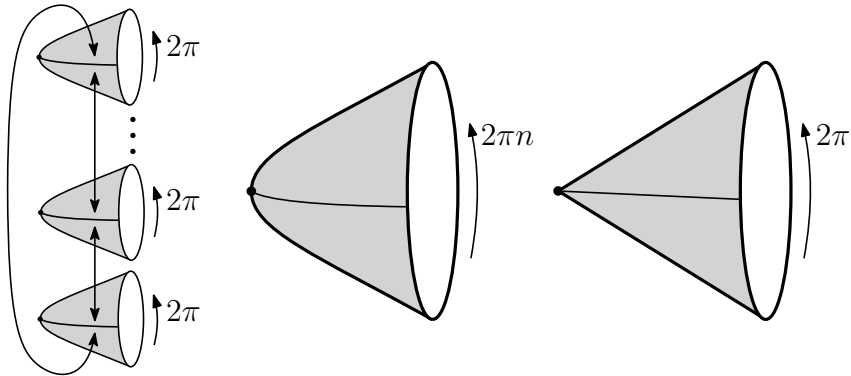

In the first condition (i), the proper circumference of the Euclidean time circle has been fixed to be $2\pi$ so that the manifold near $\mathcal{C}$ is regular and looks locally like $\mathbb{R}^2\times \mathcal{C}$ (without an angular deficit). This produces the solution $\overbar{g}_n$ from which the on-shell value of the induced metric $\sigma_{AB}$ of $\mathcal{C}$ is determined to be $\overbar{\sigma}_{n}$. In the second condition (ii), the proper area of $\mathcal{C}$ is fixed and the bulk metric near $\mathcal{C}$ in the transverse directions is locally a replicated manifold (with multiple sheets of $ \mathbb{R}^{2} $ cyclically glued together) with a conical excess at $\mathcal{C}$. In fact, the fixed-area solution is locally equal to the $n = 1$ Hartle--Hawking solution $\overbar{g}_{\mathcal{A}} = \overbar{g}_1$. For an illustration, refer to Figure \ref{fig:gravstates}.


A physical way of distinguishing the two boundary conditions (i) and (ii) is that they describe different ensembles, corresponding to a pair of conjugate variables in Einstein gravity: the area of $\mathcal{C}$ and the angular deficit (or excess) around $\mathcal{C}$ \cite{Carlip:1993sa}. Consequently, the entropy associated with either state is computed differently. To wit,
 first consider Hartle--Hawking states (i) and denote the solution satisfying the boundary condition \eqref{maldacena} by $\overbar{g}_n$. 
 Key to the construction in \cite{Lewkowycz:2013nqa} is to assume that the on-shell metric $\overbar{g}_n$ preserves replica symmetry, i.e.,  the replica symmetry is not broken spontaneously.\footnote{To leading order in $\rho \rightarrow 0$ the metric \eqref{maldacena} respects full continuous $U(1)$-symmetry, and discrete $\mathbb{Z}_n$ replica symmetry at asymptotic infinity, where $\rho \rightarrow \infty$, but these could be broken at finite $\rho$.} To perform the analytic continuation of $n$ necessary to evaluate the derivative in entropy (\ref{eq:gengravent1}), one typically considers the orbifold space $\mathcal{M}_{n}/\,\mathbb{Z}_{n}$ where the Euclidean time coordinate $\tau$ is periodically identified as $\tau \sim \tau + 2\pi$. Equivalently then, the orbifold is $\mathcal{M}_{n}/\,\mathbb{Z}_{n} = \mathcal{M}\,\backslash\, \mathcal{C} $ with periodicity $\tau \sim \tau + 2\pi$ and with $\mathcal{C}$ removed. It follows then\footnote{This can be shown via the $\tau$-integral in the bulk action, $\int_0^{2\pi n}d\tau = n\int_0^{2\pi}d\hat{\tau}$, with $\hat{\tau} = \tau \slash n$.}\textsuperscript{,}\footnote{A pedantic comment on notation. It is common practice to write $I_{\mathcal{M}_n}[\overbar{g}_n]$ on the left-hand side of (\ref{replicasymmetry}). We prefer to instead write $I_{\mathcal{M}_{n}\backslash\,\mathcal{C}}$ since we treat $(\mathcal{M}_{n},\overbar{g}_{n})$ topologically as a punctured disc such that while integrals range over $\rho\in(0,\infty)$, the $\rho=0$ surface is not included. This is consistent with the set-up of \cite{Dong:2013qoa}.}
\begin{equation}
	I_{\mathcal{M}_n\backslash\, \mathcal{C}}[\overbar{g}_n] = n\,I_{\mathcal{M}\,\backslash\, \mathcal{C}}[\overbar{g}_n].
	\label{replicasymmetry}
\end{equation}
The refined Rényi entropy \eqref{eq:refinedRenyi} of a Hartle--Hawking state can thus be written as
\begin{equation}
	\widetilde{S}_n = n^{2}\partial_{n}\,I_{\mathcal{M}\,\backslash\, \mathcal{C}}[\overbar{g}_n],\quad (\text{HH state}).
	\label{refinedrenyiHH}
\end{equation}
Since the integration region on $\mathcal{M}\,\backslash\, \mathcal{C}$ is independent of $n$, the $n$-derivative acts only on the components of the metric $\overbar{g}_n$, and can thus be viewed as an on-shell variation of the action induced by changing $n$ infinitesimally. This requires an analytic continuation of the on-shell actions to non-integer $n$.

Alternatively, now consider the gravitational entropy of a fixed-area state (ii). Denote the solution satisfying the fixed-area boundary condition \eqref{fursaev} by $\overbar{g}_\mathcal{A}$ which we emphasize is independent of $n$ since $\overbar{g}_{\mathcal{A}}=\overbar{g}_{1}$. The refined Rényi entropy is then
\begin{equation}
	\widetilde{S}_n = (n\partial_n-1)\,I_{\mathcal{M}_n}[\overbar{g}_\mathcal{A}],\quad (\text{fixed-area state}),
	\label{refinedrenyifixedarea}
\end{equation}
where the $n$-derivative acts only on the integration region $\mathcal{M}_n$. The manifold $(\mathcal{M}_n,\overbar{g}_{\mathcal{A}})$ has a conical singularity at $\mathcal{C}$ with an angular excess $\Delta \tau = 2\pi\,(n-1)$ since Euclidean time has periodicity $\tau \sim \tau + 2\pi n$. 

An important distinction between the R\'enyi entropies of the Hartle--Hawking and fixed-area states is that the former is $n$-dependent while the latter is $n$-independent. This is a consequence of the fact that the saddle point solution $\overbar{g}_{\mathcal{A}}$ to the fixed-area path integral (\ref{eq:fixedareanorm}) need only satisfy the bulk equations of motion away from $\mathcal{C}$, and is allowed to develop a conical defect (or excess) at $\mathcal{C}$ \cite{Dong:2018seb}.
Then, when computing the (boundary) R\'enyi entropies $S_{n}$, one glues cyclically together $n$-copies of the path integral representing $|\psi_{\mathcal{A}}\rangle\langle \psi_{\mathcal{A}} \lvert$, where each copy obeys the same fixed-area constraint. The dominant saddle  to the $n$-fold path integral is locally identical to $\overbar{g}_{\mathcal{A}}$, but has a conical defect on $\mathcal{C}$ with an $n$-dependent opening angle. As we review below (see also \cite{Dong:2018seb}), the result is that the refined R\'enyi entropy (\ref{refinedrenyifixedarea}) is equal to $\widetilde{S}_{n}=4\pi\,\mathcal{A}$, independent of $n$. Consequently, the R\'enyi entropy $S_{n}$ is also independent of $n$:
\beq S_{n}=\frac{n}{n-1}\int_{1}^{n}dm\,\frac{\widetilde{S}_{m}}{m^{2}}=4\pi\,\mathcal{A}\;.\eeq
Hence, the entanglement spectrum of a fixed area state $|\psi_{\mathcal{A}}\rangle$ is said to be `flat' in that 
the reduced density matrix characterizing the state can be
approximated as proportional to the identity operator, a normalized projector (see Appendix A of \cite{Akers:2018fow} for a proof).\footnote{Analogously, one can imagine starting with a thermal state in the canonical ensemble, whose R\'enyi entropy is known to have non-trivial $n$-dependence. The fixed-area state is akin to restricting the thermal state to a small energy window, thereby moving to the microcanonical ensemble. The microcanonical ensemble has a flat entanglement spectrum and thus an (approximately) $n$-independent R\'enyi entropy.} 
The gravitational entropy of fixed-area states has been explicitly demonstrated for Einstein gravity \cite{Dong:2018seb}, and AdS Jackiw--Teitelboim (JT) gravity \cite{Arias:2021ilh}. Using corner terms, we will upgrade this derivation in the case of Lovelock gravity. Notably, in a first attempt at proving the Ryu--Takayanagi formula, Fursaev computed an $n$-independent R\'enyi entropy \cite{Fursaev:2006ih}, though his derivation was not accepted as a complete proof of the RT formula as it did not match with the CFT calculation of R\'enyi entropy in $\text{AdS}_3\slash \text{CFT}_2$ \cite{Headrick:2010zt}.

\subsection{The splitting problem for Hartle--Hawking states in Einstein gravity}\label{sec:splitting}

\textbf{Note to reader:} here we summarize the splitting problem and carry out its resolution in the context of Einstein gravity, highlighting aspects which we believe have not been emphasized in prior literature. Solving this problem is necessary to compute the entropy of Hartle--Hawking states, particularly in higher-curvature gravity. However, since this discussion becomes technical, a first-time reader may skip ahead to Section \ref{sec:einsteinentropy} and still be able to follow the core narrative of the rest of the article.

For either Hartle--Hawking or fixed-area states, the aim is to solve bulk equations of motion for metrics (\ref{maldacena}) or (\ref{fursaev}), respectively, in an expansion of powers of $\rho$ about $\mathcal{C}$. Determining the coefficients of the Taylor expansion for on-shell metrics $\overbar{g}_n$ for metrics $g_{n}$ obeying Hartle--Hawking boundary conditions is subtle, leading to the `splitting problem' \cite{Camps:2014voa,Miao:2014nxa,Miao:2015iba,Camps:2016gfs}.\footnote{There is no splitting problem for fixed-area states, because in the fixed-area saddle $(\mathcal{M}_n,\overbar{g}_{\mathcal{A}})$, the on-shell metric $\overbar{g}_{\mathcal{A}} = \overbar{g}_1$ is the Hartle--Hawking metric evaluated at $n=1$. Hence Riemann tensors of $\mathcal{C}$ are unambiguously determined by $\overbar{g}_1$ for all $n$ and there is no discontinuity.}  Often in the literature it is said there is no splitting problem in Einstein gravity but there is for general cubic theories and beyond. However, the essence of the splitting problem is that expressing curvature invariants of the metric $g_{n}$ with $n>1$ in terms of curvature invariants of $g_{n=1}$ is non-trivial: the $n=1$ invariants are generally not the $n\to1$ limit of the $n>1$ invariants, there is a discontinuity (see, e.g., Eq. (\ref{eq:splitRR'})). Knowledge of the discontinuity in the Riemann curvature (known as ``splitting'') is necessary for the eventual computation of the gravitational entropy. Stated this way, it is clear the splitting question arises for Hartle--Hawking states in any theory, but is only a problem for theories whose splittings cannot be determined due to the complicated nature of solving the gravitational equations of motion near $\mathcal{C}$.



To be more precise, let us summarize the splitting problem of Hartle--Hawking states as it appears in Einstein gravity (for further details, see Appendix \ref{app:solvingsplittingprobs}). 
Since Hartle--Hawking states satisfy the boundary condition \eqref{maldacena} with a factor of $n^{-2}$ multiplying $d\tau^2$, we will work with a general Hartle--Hawking metric $g_n$ of the form
\begin{equation}
ds^{2} = d\rho^{2} +\frac{\rho^2}{n^2}\,d\tau^2 -\frac{4T^{(n)}}{n^2}\rho^4d\tau^2 - \frac{4V_{A}^{(n)}}{n}\,\rho^2d\tau d\hat{x}^A+h_{AB}^{(n)}\,d\hat{x}^Ad\hat{x}^B\;,
\label{generalgn}
\end{equation}
where $h_{AB}^{(n)} \equiv \sigma_{AB} + H_{AB}^{(n)}$ and  $T^{(n)},V_{A}^{(n)},H_{AB}^{(n)}$ are arbitrary functions of $z,\bar{z},\hat{x}^A$, while $\sigma_{AB}$ is independent of $z,\bar{z}$ (it is the induced metric of $\mathcal{C}$). Introducing complex coordinates $(z,\bar{z}) = (\rho e^{i\tau\slash n},\rho e^{-i\tau\slash n})$ for convenience, the metric \eqref{generalgn} takes the form
\begin{equation}
    ds^{2} = dzd\bar{z}+T^{(n)}\,(\bar{z}dz - z d\bar{z})^2+ 2iV_{A}^{(n)}\,(\bar{z}dz - z d\bar{z})\, d\hat{x}^A + h_{AB}^{(n)}\,d\hat{x}^Ad\hat{x}^B.
    \label{generalgncomplex}
\end{equation}
The metric functions have series expansions in $z,\bar{z}$, whose precise form for on-shell metrics will be determined by requiring Einstein's equations be satisfied. 
To simplify matters, we will only consider expansions compatible with replica symmetry $\tau \rightarrow \tau + 2\pi k$, i.e., we restrict to metrics $g_n$ that respect replica symmetry and look for solutions $\overbar{g}_n$ of Einstein's equations that do not spontaneously break replica symmetry. The most general expansions compatible with replica symmetry are \cite{Dong:2017xht}
\begin{align}
    T^{(n)} &= \mathcal{T}_0+\mathcal{T}_1\, (z\bar{z}) + \ldots\nonumber\\
    V_{A}^{(n)} &= \mathcal{U}_A + \mathcal{V}_{(z)A}\,z^n + \mathcal{V}_{(\bar{z})A}\,\bar{z}^n + \ldots\label{gnexpansiontext} \\
    H_{AB}^{(n)} &= 2\mathcal{K}_{(z)AB}\,z^n + 2\mathcal{K}_{(\bar{z})AB}\,\bar{z}^n+\mathcal{Q}_{(zz)AB}\, z^{2n} + \mathcal{Q}_{(\bar{z}\bar{z})AB}\,\bar{z}^{2n} + 2\mathcal{Q}_{(z\bar{z})AB}\,(z\bar{z}) + \ldots\nonumber
\end{align}
Each coefficient $\mathcal{T}_{i}$, $\mathcal{U}_{A}$, etc., has its own expansion in powers of $(z\bar{z})^{n-1}$ with coefficients that are independent of $z,\bar{z}$. For example,
\begin{align}
   \mathcal{T}_0 = \sum_{k=0}^{\infty}\mathcal{T}_{0,k}\,(z\bar{z})^{k(n-1)},  
    \label{Qnexpansion}
\end{align}
where $\mathcal{T}_{0,k}$ are $z,\bar{z}$ independent. The fact the coefficients are expanded in powers of $(z\bar{z})^{n-1}$ with an $n-1$ exponent follows from the constraint that at $n = 1$ the coefficients must be independent of $z,\bar{z}$.\footnote{Note the exponent can be $k\alpha_n$ for some real function $\alpha_{n}$ of $n$ such that $\alpha_1 = 0$. The choice $\alpha_n = n-1$ leads to a consistent solution of Einstein's equations.} Generalization of the expansion \eqref{Qnexpansion} that includes replica non-symmetric terms can be found in \cite{Camps:2014voa}. At leading order, the metric $g_n$ \eqref{generalgn} is thus explicitly
\beq ds^{2}=d\rho^{2}+\frac{\rho^{2}}{n^{2}}\,d\tau^{2}+\bigl(\sigma_{AB}+2\mathcal{K}_{0(z)AB}\,\rho^{n}e^{i\tau}+2\mathcal{K}_{0(\bar{z})AB}\,\rho^{n}e^{-i\tau}+\ldots\bigr)\,d\hat{x}^{A}d\hat{x}^{B}+\ldots\;,\label{eq:linegn0}\eeq
where $\mathcal{K}_{0(p)AB}$ is the leading term in the expansion of $\mathcal{K}_{(p)AB}$. Borrowing notation from \cite{Dong:2017xht}, we refer to the leading term in (\ref{eq:linegn0}) as the metric $g_{n,0}$ while the subleading terms can be organized in an infinite expansion in powers of $(z\bar{z})$,
\begin{equation}
    g_n = g_{n,0} + g_{n,1}\,(z\bar{z})^{n-1} + g_{n,2}\,(z\bar{z})^{2(n-1)} + \ldots\;.
    \label{eq:gnsplitform}
\end{equation}
Importantly, the metric \eqref{generalgncomplex} is defined only for integer $n\geq 1$, because otherwise there is a jump in the extrinsic curvature between $\tau = 0$ and $\tau = 2\pi n$ surfaces.

When $n = 1$, the expansion of $g_1$ simplifies. We will denote the expansions of the component functions of $g_1$ as
\begin{align}
    T^{(1)}&\equiv T  =  T_0+T_1\,(z\bar{z}) + \ldots\nonumber\\
    V_A^{(1)} &\equiv V_{A} = U_A + V_{(z)A}\,z + V_{(\bar{z})A}\,\bar{z} + \ldots \label{Vexpnis1}\\
    H_{AB}^{(1)} &\equiv H_{AB} = 2K_{(z)AB}\,z + 2K_{(\bar{z})AB}\,\bar{z}+Q_{(zz)AB}\, z^2 + Q_{(\bar{z}\bar{z})AB}\,\bar{z}^2 + 2Q_{(z\bar{z})AB}\,(z\bar{z}) + \ldots
    \nonumber
\end{align}
where all coefficients are independent of $z$ and $\bar{z}$ since $n = 1$. In particular, the $n = 1$ coefficients are the $n\rightarrow 1$ limit of the $n>1$ coefficients \eqref{gnexpansiontext} and take the form of an infinite series, e.g., 
\begin{equation}
   T_0 = \sum_{k=0}^{\infty}\mathcal{T}_{0,k}\lvert_{n=1}\;.
   \label{T0text}
\end{equation}
This is also evident from \eqref{eq:gnsplitform} which at $n = 1$ becomes the infinite series
\begin{equation}
    g_{1}=g_{1,0}+g_{1,1}+g_{1,2}+\ldots
\end{equation}
The fact the $n=1$ expansion coefficients are infinite sums of subleading coefficients of the $n>1$ metric is the essence of the splitting problem.


The coefficients in the expansions of $g_n$ and $g_1$ can be related to geometric data at $\mathcal{C}$. Denote the extrinsic curvature of $\mathcal{C}$ in $g_n$ by $K'_{(p)AB}$ for $p=z,\bar{z}$, and the Riemann tensor components of $g_n$ at $\mathcal{C}$ by $R'_{abcd}$. Due to the absence of linear terms $z,\bar{z}$ terms in the expansion of $H^{(n)}_{AB}$, the extrinsic curvature vanishes, $K'_{(p)AB}\sim\partial_{p}H^{(n)}_{AB} = 0$ (at $\rho=0$). All non-zero components of the Riemann tensor of $g_n$ at $\mathcal{C}$ are given in Appendix \ref{app:solvingsplittingprobs}. The two components that will turn out to be non-trivial from the point of view of the splitting problem are given by
\begin{equation}
    R'_{z\bar{z}z\bar{z}}\lvert_{\mathcal{C}}\, = -3\mathcal{T}_{0,0}\ ,\quad R_{zA\bar{z}B}'\lvert_{\mathcal{C}}\,=\frac{i}{2}\,(\partial_A\mathcal{U}_{0B}-\partial_B\mathcal{U}_{0A})-2\mathcal{Q}_{0(z\bar{z})AB}
    \label{gntensors}
\end{equation}
when $n > 1$.  Meanwhile, the coefficients $K_{(z)AB}$ and $K_{(\bar{z})AB}$ in \eqref{Vexpnis1} are equal to extrinsic curvatures of $\mathcal{C}$ in the metric $g_1$ (explaining our notation). In Appendix \ref{app:solvingsplittingprobs} we also determine the non-zero Riemann tensor $R_{abcd}$ components of metric $g_1$. In particular,
\begin{equation}
    R_{z\bar{z}z\bar{z}}\lvert_{\mathcal{C}}\, = -3T_0\ ,\quad R_{zA\bar{z}B}\lvert_{\mathcal{C}}\,= \frac{i}{2}\,(\partial_AU_B-\partial_BU_A)+K_{(z)A}^{C}K_{(\bar{z})CB}+K_{(\bar{z})A}^{C}K_{(z)CB}-2Q_{(z\bar{z})AB}\ .
    \label{g1riemannT0}
\end{equation}
Comparing curvatures (\ref{gntensors}) and (\ref{g1riemannT0}), it is clear the geometric quantities are discontinuous at $n = 1$: for example the extrinsic curvature vanishes for $n>1$, but is non-zero at $n=1$.

It is possible to write the $n = 1$ Riemann tensors $R_{abcd}$ with the $n\rightarrow 1$ limit of the $n>1$ components $R'_{abcd}$. For example, combining \eqref{g1riemannT0} with \eqref{gntensors}  using (\ref{T0text}) gives
\begin{equation}
    R_{z\bar{z}z\bar{z}}'\big\lvert_{n=1}\, = R_{z\bar{z}z\bar{z}} +\sum_{k=1}^{\infty}3\mathcal{T}_{0,k}\big\lvert_{n=1},
\label{eq:splitRR'}\end{equation}
(where the evaluation on $\mathcal{C}$ is implicit). Similar relations for other components of the Riemann tensor can be found in Appendix \ref{app:solvingsplittingprobs}. We see that the $n\rightarrow 1$ limit of the $n>1$ Riemann tensor splits into two parts: the $n = 1$ Riemann tensor and a part involving an infinite number of $n>1$ coefficients. Again, there is a clear discontinuity between the Riemann curvature of the $g_{n=1}$ and $g_{n>1}$ metrics in the analytic continuation $n\to1$. The precise form of the coefficients $\mathcal{T}_{0,k}$ in terms of intrinsic and extrinsic geometric data of $\mathcal{C}$ are then determined by imposing Einstein's equations.\footnote{Historically, fixing coefficients $\mathcal{T}_{0,k}$ was accomplished via a prescription of `minimal regulation' \cite{Camps:2013zua,Dong:2013qoa}: (i) take coordinates adapted to a generic codimension-2 surface by shooting geodesics orthogonal to it and then, in the resulting metric expanded near this surface (ii) replace any holomorphic factors $z$ and $dz$ not paired with their anti-holomorphic counterparts by  $z^{n}$ and $d(z^{n})$. While the resulting metric is replica symmetric and regular for integers $n\geq1$, this is not, however, the  only prescription which would have produced an equally valid metric (see, e.g., \cite{Camps:2016gfs}). The ambiguity in a prescription propagates as ambiguities in fixing $\mathcal{T}_{0,k}$. Further, the minimal regulation prescription is not consistent with Einstein's equations.}




\paragraph{Alternative form of the HH metric.} The Hartle--Hawking metric $g_n$ \eqref{generalgncomplex} in $(z,\bar{z})$ coordinates has been used in \cite{Camps:2014voa,Camps:2016gfs}, but it is also common to write it in an alternative coordinate system \cite{Lewkowycz:2013nqa,Dong:2013qoa,Miao:2014nxa,Miao:2015iba}. To this end, for completeness, let us introduce a new set of complex coordinates $(\zeta,\bar{\zeta})$ via
\begin{equation}
    z=n\zeta^{1/n},\quad \bar{z} = n\bar{\zeta}^{1/n}\;.
\end{equation}
In these coordinates, the Hartle--Hawking metric $g_n$ \eqref{generalgncomplex} takes the form
\beq 
\begin{split}
 ds^{2}&=e^{2A_\epsilon}\,d\zeta d\bar{\zeta}+e^{4A_\epsilon}\,T^{(n)}\,(\bar{\zeta}d\zeta-\zeta d\bar{\zeta})^{2}+2ie^{2A_\epsilon}\,V_{A}^{(n)}\,(\bar{\zeta}d\zeta-\zeta d\bar{\zeta})\,d\hat{x}^{A}+h^{(n)}_{AB}\,d\hat{x}^{A}d\hat{x}^{B}\;,
\end{split}
\label{eq:metreghigh}\eeq
where the conformal factor 
\begin{equation}
    A_\epsilon = -\frac{\epsilon}{2}\log(\zeta\bar{\zeta}),\quad \epsilon\equiv1-\frac{1}{n}\;.
\end{equation}
In $(\zeta,\bar{\zeta})$ coordinates, the expansions \eqref{gnexpansiontext} become\footnote{We keep the factor of $e^{2A_\epsilon}$ multiplying $\mathcal{Q}_{(\zeta\bar{\zeta})AB}$ explicit while in \cite{Dong:2013qoa} it is ``secretly'' absorbed into $\mathcal{Q}_{(\zeta\bar{\zeta})AB}$.}
\begin{align}
    T^{(n)} &= \mathcal{T}_0+ e^{2A_\epsilon}\,\mathcal{T}_1\,(\zeta\bar{\zeta}) + \ldots\nonumber\\
    V_{A}^{(n)} &= \mathcal{U}_A + \mathcal{V}_{(\zeta)A}\,\zeta + \mathcal{V}_{(\bar{\zeta})A}\,\bar{\zeta} + \ldots\label{gnacexpansiontext} \\
    H_{AB}^{(n)} &= 2\mathcal{K}_{(\zeta)AB}\,\zeta + 2\mathcal{K}_{(\bar{\zeta})AB}\,\bar{\zeta}+\mathcal{Q}_{(\zeta\zeta)AB}\, \zeta^2 + \mathcal{Q}_{(\bar{\zeta}\bar{\zeta})AB}\,\bar{\zeta}^{2} + 2e^{2A_\epsilon}\mathcal{Q}_{(\zeta\bar{\zeta})AB}\,(\zeta\bar{\zeta}) + \ldots\nonumber
\end{align}
where the coefficients have expansions in powers of $e^{-2kA_{\epsilon}}$, for example,
\begin{equation}
    \mathcal{T}_k = \sum_{l=0}^{\infty}\mathcal{T}_{k,l}\,e^{-2lA_\epsilon}\,
    \label{eq:altTkexp}.
\end{equation}
These forms of the expansion follow from
\begin{equation}
    (z\bar{z}) = n^2 e^{2A_\epsilon}(\zeta\bar{\zeta})\,\quad (z\bar{z})^{k(n-1)} = n^{2k(n-1)} e^{-2kA_{\epsilon}}\,,
    \label{eq:zzbarzetazetabar}
\end{equation}
and we have redefined the coefficients to absorb explicit factors of $n$ coming from the coordinate change. We can see that at leading order in an expansion around $\rho = 0$,
\begin{equation}
    T^{(n)} = \mathcal{T}_{0,0} + e^{-2A_\epsilon}\,\mathcal{T}_{0,1}+\ldots,\quad e^{2A_\epsilon}\mathcal{Q}_{(\zeta\bar{\zeta})AB} = e^{2A_\epsilon}\mathcal{Q}_{0(\zeta\bar{\zeta})AB}+\mathcal{Q}_{1(\zeta\bar{\zeta})AB} + \ldots\;,
    \label{eq:leadingTandQac}
\end{equation}
which match with the expansions used in \cite{Miao:2014nxa,Miao:2015iba}.\footnote{Note that in \cite{Miao:2014nxa,Miao:2015iba}, the numbering of the coefficients with zero and one are flipped.} 

\paragraph{Solving the splitting problem.} Vacuum Einstein's equations, $R_{ab}  = 0$, which, when solved for a Hartle--Hawking metric $g_n$, gives the solution $\overbar{g}_n$. For general Hartle--Hawking metrics, Einstein's equations can only be solved perturbatively in an expansion around $\mathcal{C}$, which we carry out explicitly in Appendix \ref{app:solvingsplittingprobs}. Without loss of generality, we solve Einstein's equations under the assumption
\begin{equation}
    \mathcal{U}_{0A} = 0,
\end{equation}
which was also assumed in \cite{Miao:2014nxa,Miao:2015iba,Camps:2016gfs}.\footnote{In \cite{Miao:2014nxa,Miao:2015iba} this assumption is made to recover the Wald entropy from the Camps--Dong proposal when applied to stationary black hole backgrounds.} This is allowed, because $\mathcal{U}_{0A}$ can be removed by a diffeomorphism under which it transforms as a gauge field as explained in Appendix \ref{app:solvingsplittingprobs}.

Solving Einstein's equations perturbatively in an expansion around $\rho = 0$ does not fix the on-shell value of its induced metric $\sigma$ as a function of $n$, but gives a solution for all choices of $\sigma$. However if one has access to the full non-perturbative solution $\overbar{g}_n$ the induced metric is determined as a function of $n$ to a particular value $\overbar{\sigma}_n$. We illustrate this explicitly in Appendix \ref{subapp:Einsteinsplitting} using the Schwarzschild black hole solution $\overbar{g}_n$ which is known for all $\rho$. Henceforth we will set $\sigma$ to its on-shell value $\overbar{\sigma}_n$ when writing down the solutions for subleading coefficients.

The Ricci tensor components $R_{ab}'$ of the metric $g_n$ \eqref{generalgn} have been computed in Appendix \ref{app:solvingsplittingprobs}. Firstly, off-shell, the Ricci tensor has the component
\begin{equation}
    R_{zz}' = -n(n-1)\,\sigma^{AB}\,\mathcal{K}_{0(z)AB}\,z^{n-2} + \ldots
    \label{eq:Rzzleading}
\end{equation}
so that the Einstein equation component $R_{zz}' = 0$ imposes the relation
\begin{equation}
\overbar{\sigma}_n^{AB}\,\overbar{\mathcal{K}}_{0(p)AB} = 0, \quad p = z,\overbar{z}\;,
    \label{einsteinleading}
\end{equation}
where $\overbar{\mathcal{K}}_{0(p)AB} $ indicates it is the coefficient in the expansion of the on-shell metric $\overbar{g}_n$. The subleading terms in the Einstein's equations, moreover, determine higher order coefficients in terms of $n$ (assuming $\sigma = \overbar{\sigma}_{n}$) and the traceless part of $\overbar{\mathcal{K}}_{0(p)AB}$. The on-shell leading coefficients are given by
\begin{equation}
\overbar{\mathcal{T}}_{0,0} = \frac{1}{24}\,\overbar{\widehat{R}}',\quad \overbar{\mathcal{Q}}_{0(z\bar{z})AB} = \frac{1}{4}\,\overline{\widehat{R}}_{AB}'\ , 
\label{eq:leadingcoeffs}\end{equation}
where $\overline{\widehat{R}}_{ABCD}'$ is the Riemann tensor of the induced metric $\overbar{\sigma}_n$, while at next to leading order 
\begin{gather}
    \overbar{\mathcal{T}}_{0,1} = \frac{n^{2}}{6}\,\overbar{\mathcal{K}}_{0(z)}^{AB}\overbar{\mathcal{K}}_{0(\bar{z})AB},\quad\overbar{\mathcal{Q}}_{1(z\bar{z})AB} = \overbar{\mathcal{K}}_{0(z)A}^{C}\overbar{\mathcal{K}}_{0(\bar{z})BC}+\overbar{\mathcal{K}}_{0(\bar{z})A}^{C}\overbar{\mathcal{K}}_{0(z)BC}\ ,\\
    \overbar{\sigma}_n^{AB}\,\overbar{\mathcal{Q}}_{0(pp)AB} = \frac{n}{2n-1}\,\overbar{\mathcal{K}}_{0(p)}^{AB}\overbar{\mathcal{K}}_{0(p)AB},\quad p = z,\bar{z}\;,
    \label{eq:nexttoleading}
\end{gather}
where we have used \eqref{einsteinleading} and all indices are understood to be contracted with $\overbar{\sigma}_{nAB}$. 

These solutions are not enough to fix the splitting because they involve an infinite series of higher-order coefficients as in \eqref{eq:splitRR'}, however, Einstein's equations seem to require that the following higher-order coefficients vanish on-shell:
\begin{align}
    \overbar{\mathcal{U}}_{kA} = \overline{\mathcal{K}}_{k(p)AB}= \overline{\mathcal{Q}}_{k(pp)AB}= 0,\quad k\geq 1\ ,\quad \overbar{\mathcal{T}}_{0,k} = \overbar{\mathcal{Q}}_{k(z\bar{z})AB}= 0,\quad k\geq 2\ .
    \label{eq:vanishinghigher}
\end{align}
This has been proven in \cite{Camps:2014voa} except for the $\mathcal{Q}$ coefficients which were not included in their metric ansatz. We have also checked it numerically for low values of $k$. Regardless, Einstein's equations can certainly be consistently solved assuming it is the case (as is done implicitly in \cite{Camps:2016gfs}). First, \eqref{eq:vanishinghigher} implies
\begin{equation}
    \overbar{\mathcal{K}}_{0(p)AB}\big\lvert_{n=1}\, = \overbar{K}_{(p)AB}
    \label{eq:curlyKK}
\end{equation}
equals the extrinsic curvature of $\mathcal{C}$ in the on-shell metric $\overbar{g}_1$. The $n = 1$ limit of the equation \eqref{einsteinleading} is thus given by
\begin{equation}   
\overbar{\sigma}_1^{AB}\,\overbar{K}_{(p)AB} = 0, \quad p = z,\overbar{z},
\label{minimalareacondition}
\end{equation}
which is the area minimization condition for $\mathcal{C}$ in the metric $\overbar{g}_1$.\footnote{This means that for $n>1$ the extrinsic curvatures of $\mathcal{C}$ in the metric $\overbar{g}_n$ vanish identically, but at $n=1$, only the their traces vanish.} In addition, the relation \eqref{eq:curlyKK} implies that \eqref{eq:nexttoleading} at $n = 1$ become
\begin{equation}
\overbar{\mathcal{Q}}_{1(z\bar{z})AB}\big\lvert_{n=1}\, = \overbar{K}_{(z)A}^{C}\overbar{K}_{(\bar{z})BC}+\overbar{K}_{(\bar{z})A}^{C}\overbar{K}_{(z)BC},\quad \overbar{\mathcal{T}}_{0,1}\big\lvert_{n=1}\, = \frac{1}{6}\,\overbar{K}_{(z)}^{AB}\overbar{K}_{(\bar{z})AB}
    \label{eq:campsequations}
\end{equation}
where the indices are understood to be contracted with $\overbar{\sigma}_{1AB}$. These are the same relations as originally found in \cite{Camps:2016gfs}, using $(z,\overbar{z})$ coordinates, and in \cite{Miao:2014nxa,Miao:2015iba}, using $(\zeta,\overbar{\zeta})$ coordinates.

Second, \eqref{eq:vanishinghigher} implies the difference of the components $R'_{abcd}\big\lvert_{n=1}$ and $R_{abcd}$ truncates. Setting $n = 1$ in \eqref{gntensors} and substituting to \eqref{g1riemannT0} gives
\begin{align}
    \overbar{R}_{z\bar{z}z\bar{z}}'\big\lvert_{n=1}\, &= \overbar{R}_{z\bar{z}z\bar{z}} +3\overbar{\mathcal{T}}_{0,1}\big\lvert_{n=1}\;,\nonumber\\
\overbar{R}_{zA\bar{z}B}'\big\lvert_{n=1}\,&=\overbar{R}_{zA\bar{z}B}-\overbar{K}_{(z)a}^{C}\overbar{K}_{(\bar{z})CB}-\overbar{K}_{(\bar{z})A}^{C}\overbar{K}_{(z)CB}+2\overline{\mathcal{Q}}_{1(z\bar{z})AB}\big\lvert_{n=1}\;.
    \label{eq:truncatedsplittings}
\end{align}
Substituting \eqref{eq:campsequations} to \eqref{eq:truncatedsplittings} gives the Einstein gravity splitting relations
\begin{align}
    \overbar{R}'_{z\bar{z}z\bar{z}}\big\lvert_{n=1}\, &= \overbar{R}_{z\bar{z}z\bar{z}}+ \frac{1}{2}\,\overbar{K}_{(z)}^{AB}\,\overbar{K}_{(\bar{z})AB}\;,\nonumber\\
    \overbar{R}_{zA\bar{z}B}'\big\lvert_{n=1}\, &= \overbar{R}_{zA\bar{z}B} +\overbar{K}_{(z)A}^{C}\,\overbar{K}_{(\bar{z})BC}+\overbar{K}_{(\bar{z})A}^{C}\,\overbar{K}_{(z)BC}\;.
    \label{eq:Einsteinsplittings}
\end{align}
The splitting relations for rest of the Riemann tensor components are trivial in the sense that they are completely fixed by the vanishing of the higher-order coefficients \eqref{eq:vanishinghigher}. Their explicit expressions are given in \eqref{eq:trivialsplittings}. The relations \eqref{eq:Einsteinsplittings} are non-trivial, because they depend on $\overbar{\mathcal{T}}_{0,1}\lvert_{n=1}$ and $\overbar{\mathcal{Q}}_{1(z\bar{z})AB}\lvert_{n=1}$ which are required by Einstein's equations to be functions of extrinsic curvatures.

Now the derivation of the gravitational entropy functional is done by evaluating gravitational actions of the on-shell metric $\overbar{g}_n$ so that the refined Rényi $\widetilde{S}_n$ functional is constructed from $\overbar{g}_{nab}\lvert_{\mathcal{C}}$ and $\overbar{R}'_{abcd}\lvert_{\mathcal{C}}$.\footnote{Recall that no extrinsic curvatures of $\mathcal{C}$ appear as they vanish in the metric $\overbar{g}_n$ for $n>1$.} 
Using the splittings, the entropy $ S = \lim_{n\rightarrow 1}\widetilde{S}_n$ can be written in terms of $\overbar{R}_{abcd}\lvert_{\mathcal{C}}$ and extrinsic curvatures $\overbar{K}_{(p)AB}$. The splittings are not relevant for the computation of entropy in Einstein gravity because no Riemann tensors appear in $\widetilde{S}_n$, but they are needed in higher-curvature theories of gravity (see Section \ref{subsec:FRiemannentropies}).\footnote{Note that in a given higher curvature theory of gravity one should use the splittings relations obtained by solving equations of motion of that theory.}

\subsection{Two-roads to entropy of Hartle--Hawking states}\label{subsec:boundarytermmethod}


To contrast with our derivation of entropy functionals presented in the following section, let us recap the computation of gravitational entropy of Hartle--Hawking states and a bulk governed by Einstein gravity, as first accomplished by Lewkowycz--Maldacena \cite{Lewkowycz:2013nqa}. We describe two methods which differ in how the $n$-derivative of the on-shell action is evaluated. In the first method, the derivative is evaluated directly on-shell after cutting a hole around the conical singularity. In the second method, the derivative is evaluated off-shell after adding and subtracting a manifold whose conical singularity (and curvature singularity) has been regularized. In Einstein gravity, the result is independent of regularization of the off-shell geometry and the two methods produce the same answer for the R\'enyi entropy.


\vspace{2mm}



\noindent \textbf{Boundary term method.} First, consider the entropy \eqref{refinedrenyiHH} given in terms of the action of the orbifold $(\mathcal{M}\,\backslash\,\mathcal{C},\overbar{g}_n)$, endowed with the Hartle--Hawking solution $\overbar{g}_{n}$ \eqref{generalgn}, where now $\tau\sim\tau+2\pi$, giving rise to a conical singularity at $\rho=0$. Now recall the refined R\'enyi entropy (\ref{refinedrenyiHH}), where as we described, the $n$-derivative acts only on the components of the metric. Thence we need to compute the variation of the action of $(\mathcal{M}\,\backslash\,\mathcal{C},\overbar{g}_n)$ with respect to the metric, and as shown in \cite{Lewkowycz:2013nqa,Dong:2016fnf}, this can be done by cutting a small hole of size $\epsilon\rightarrow 0$ around $\mathcal{C}$ introducing a boundary at $\rho = \epsilon$. The variation of the action is then given by\footnote{There is no boundary term at large $\rho$ assuming boundary conditions such that varying the action yields only the equations of motion with no additional boundary terms.}
\begin{equation}
    \delta_gI_{\mathcal{M}\,\backslash\,\mathcal{C}}[g_{n}]=-\int_{\rho >0}d^Dx\sqrt{g_n}\,G_{ab}'\,\delta g^{ab}+\lim_{\epsilon\rightarrow 0}\int_{\rho=\epsilon} d^{D-2}\hat{x}d\tau\sqrt{\gamma_{n}}\,\hat{n}^{a}\,(\nabla^{b}\delta g_{ab}-g_{n}^{bc}\,\nabla_{a}\delta g_{bc})
\end{equation}
where $G_{ab}'$ is the Einstein tensor of $g_n$, $\gamma_{n}$ is the induced metric of the boundary at $\rho = \epsilon$ with outward pointing unit normal $\hat{n}^{a}=\delta^{a}_{\rho}$, and $\delta g_{ab} = - g_{ac}g_{bd}\,\delta g^{cd}$. Setting $g_n$ on-shell gives
\begin{equation}
    \delta_gI_{\mathcal{M}\,\backslash\,\mathcal{C}}[\overbar{g}_{n}]=\lim_{\epsilon\rightarrow 0}\int_0^{2\pi} d\tau\int_{\rho=\epsilon} d^{D-2}\hat{x}\sqrt{\overbar{\gamma}_{n}}\,\hat{n}^{a}\,(\nabla^{b}\delta g_{ab}-g_{n}^{bc}\,\nabla_{a}\delta g_{bc})
\end{equation}
To compute the entropy, we need to consider the variation $\delta g_{ab} = \partial_{n}g_{nab} $ due to a variation $\delta n$. To ensure this variation is small near $\mathcal{C}$, one must work with $n>1$, taking the limit $n\to1$ at the end of the calculation.\footnote{To see this, note the $AB$-components of the metric $g_{n}$ have terms like $H^{(n)}_{AB}\sim \mathcal{K}_{(z)AB}\rho^{n}=\mathcal{K}_{0(z)AB}\rho^{n}+\mathcal{K}_{1(z)AB}\rho^{2(n-1)}\rho^{n}+\ldots$. Then, $\partial_{n}H^{(n)}_{AB}\sim\mathcal{K}_{0(z)AB}\rho^{n}\log(\rho)+3\mathcal{K}_{1(z)AB}\rho^{2(n-1)}\rho^{n}\log(\rho)$, which will not be small as $n\to1$ for fixed, small $\rho$.} Keeping in mind $n>1$, it is sufficient to work with the leading order metric $g_{n,0}$ in \eqref{eq:gnsplitform}, because rest of the terms go to zero. This means that the $\sqrt{\overbar{\gamma}_n} = \frac{\rho}{n}\sqrt{\overbar{\sigma}_{n}} $ up to corrections where $\overbar{\sigma}_{n}$ is the induced metric of $\mathcal{C}$ in $\overbar{g}_{n}$.\footnote{We have chosen not to `split' the induced metric $\sigma_{n}$, however, one could do so, e.g., \cite{Dong:2017xht} (such a splitting could be incorporated into the expansion of $H_{AB}^{(n)}$). In such an event $\overbar{\sigma}_{n}$ is replaced with $\overbar{\sigma}_{n,0}$, the induced metric in $\overbar{g}_{n,0}$, such that $\overbar{\sigma}_{1,0}$ is distinct from $\overbar{\sigma}_{1}$.}
Then, for $n>1$, it follows that the second contribution in the boundary term vanishes at $\rho=0$, while the first contribution is $\hat{n}^{a}\nabla^{b}\partial_{n}g_{nab}=\frac{2}{n\rho}$. Therefore, on-shell, 
\beq \partial_{n}I_{\mathcal{M}\,\backslash\,\mathcal{C}}[\overbar{g}_{n}]=\lim_{\epsilon\rightarrow 0}\int_0^{2\pi}d\tau\int d^{D-2}\hat{x}\sqrt{\overbar{\sigma}_{n}}\,\biggl(\frac{\epsilon}{n}\biggr)\left(\frac{2}{\epsilon n}\right)=\frac{4\pi}{n^{2}}\int_{\mathcal{C}} d^{D-2}\hat{x}\sqrt{\overbar{\sigma}_{n}}\;.\eeq
Applying the entropy formula (\ref{refinedrenyiHH}) gives
\begin{equation}
\widetilde{S}_n = 4\pi\int_{\mathcal{C}}d^{D-2}x\sqrt{\overbar{\sigma}_{n}}\;.
\end{equation}
Taking the $n\to1$ limit yields the entropy functional 
\beq S=4\pi\,\text{Area}\,(\mathcal{C})\;.\label{eq:areafuncg10}\eeq
By the above analysis, in taking the $n\to1$ limit, the entropy $S$ is the Bekenstein--Hawking entropy of the ``split metric'' $\overbar{g}_{1,0}$. This method has been generalized to higher-curvature theories of gravity in \cite{Dong:2017xht} where the result is the Wald entropy of $\overbar{g}_{1,0}$ (see Section \ref{sec:entropyfromcornerterms}). 

Note, moreover, the $zz$-component of the Einstein equation has a potential divergence \eqref{eq:Rzzleading} as $\rho\to0$. A similar relation holds for the conjugate component $R_{\bar{z}\bar{z}}$. The analytic continuation to $n\rightarrow 1$ gives \eqref{minimalareacondition} since the stress-tensor from the matter sector (when present) is not expected to be divergent.
In other words, the codimension-2 surface is a minimal surface $K_{(z)}=K_{(\bar{z})}=0$, such that gravitational entropy (\ref{eq:areafuncg10}) recovers the Ryu--Takayanagi formula.

\vspace{2mm}


\noindent \textbf{Regularized squashed cone method.} Recall the original definition of the entropy \eqref{eq:gengravent1} in terms of the smooth Hartle--Hawking solution $(\mathcal{M}_n,\overbar{g}_n)$ and the solution $(\mathcal{M}_n,\overbar{g}_1)$ which has a conical excess at $\mathcal{C}$. To evaluate the difference $I_{\mathcal{M}_{n}}[\overbar{g}_{n}]-nI_{\mathcal{M}}[\overbar{g}_{1}]$, one may add and subtract an off-shell regularized squashed cone $(\mathcal{M}_n,g_{n,a}^{\text{reg}})$ (with a regularization parameter $a$) where $g_{n,a}^{\text{reg}}$ equals $\overbar{g}_1$ for $\rho > a$, but deviates from it for $\rho < a$ such that $(\mathcal{M}_n,g_{n,a}^{\text{reg}})$ has no conical excess at $\rho = 0$. The regularized metric $g_{n,a}^{\text{reg}}$ is constructed from $\overbar{g}_1$ following \cite{Fursaev:2013fta,Dong:2013qoa} and by definition $\lim_{a\rightarrow 0}g_{n,a}^{\text{reg}} \equiv \overbar{g}_1$(see Appendix \ref{app:repmanifolds} for a review).\footnote{Regularized by introducing a regularization parameter $a$ into the conformal factor $A\to-\frac{\epsilon}{2}\log(\zeta\bar{\zeta}+a^{2})$.} Thus, the difference is recast as $(I_{\mathcal{M}_{n}}[\overbar{g}_{n}]-I_{\mathcal{M}_{n}}[g_{n,a}^{\text{reg}}])+(I_{\mathcal{M}_{n}}[g_{n,a}^{\text{reg}}]-nI_{\mathcal{M}}[\overbar{g}_{1}])$. Assuming $g_{n,a}^{\text{reg}}$ differs from the on-shell metric $\overbar{g}_{n}$ by an $\mathcal{O}(n-1)$ amount (not worrying about the discontinuity of the extrinsic curvatures for non-integer $n$), the first parenthesis vanishes up to this order and the entropy becomes
\begin{equation}
    S = \partial_n(I_{\mathcal{M}_{n}}[g_{n,a}^{\text{reg}}]-nI_{\mathcal{M}}[\overbar{g}_{1}])\lvert_{n=1}\,.
    \label{eq:midstepoffshellemethod}
\end{equation}
If the above assumptions on $g_{n,a}^{\text{reg}}$ can be satisfied for arbitrary small $a$, we can take the $a\rightarrow 0$ limit of and use integral identity for Ricci scalars of regularized squashed cones \cite{Fursaev:2013fta}
\beq \lim_{a\rightarrow 0}I_{\mathcal{M}_{n}}[g_{n,a}^{\text{reg}}]=nI_{\mathcal{M}}[\overbar{g}_{1}]+4\pi\,(n-1)\int_{\mathcal{C}} d^{D-2}x\sqrt{\overbar{\sigma}_1}\;,\eeq
where there is a plus sign in the second term due to the minus sign in the definition of the Euclidean action. The entropy \eqref{eq:midstepoffshellemethod} coincides with \eqref{eq:areafuncg10}.

\section{Entropy from corner terms in Einstein gravity}\label{sec:einsteinentropy}

Here we analyze gravitational entropy of both Hartle--Hawking and fixed-area states using a Hayward (corner) term. The Hayward term is necessary to have a well-posed variational principle for spacetimes with codimension-2 corners \cite{Hayward:1993my}. The idea holographic entanglement entropy arising from a corner was previously considered for Einstein gravity \cite{Takayanagi:2019tvn} and Jackiw--Teteilboim gravity \cite{Botta-Cantcheff:2020ywu,Arias:2021ilh}.
As we will explain, our derivation is different from these approaches. Namely, we consider replicated geometries with conical excess, which can be thought of as a spacetime with a corner, a feature recognized by Dowker \cite{Dowker:1994bj}. Further, we provide a careful treatment of the variational principle and work directly with the action of a wedge shaped manifold. Consequently, our approach readily extends to higher-curvature theories of gravity, the subject of subsequent sections. This section serves as an expanded version of our previous work \cite{Kastikainen:2023yyk}.



\subsection{Entropy of Hartle--Hawking states}\label{subsec:einsteinHH}

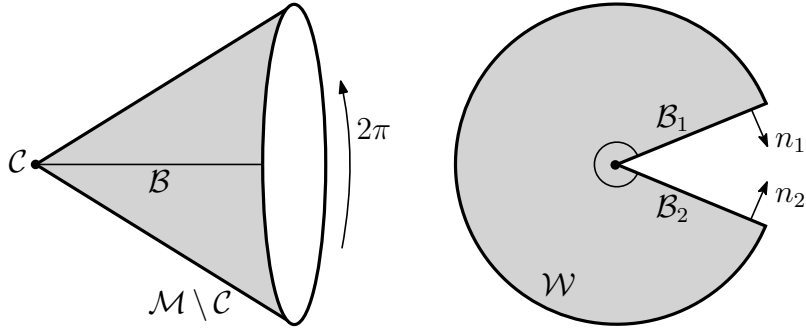
\begin{figure}[t!]
\begin{center}
\includegraphics[scale=1.5]{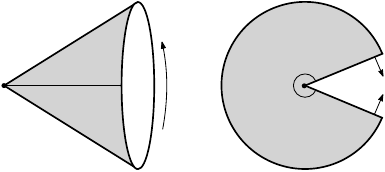}
\put(-300,62){$\mathcal{C}$}
\put(-245,54){$\mathcal{B}$}
\put(-245,6){$\mathcal{M}\,\backslash\,\mathcal{C}$}
\put(-163,75){$2\pi$}
\put(-90,13){$\mathcal{W}$}
\put(-45,80){$\mathcal{B}_{1}$}
\put(-45,44){$\mathcal{B}_{2}$}
\put(2,50){$n_{2}$}
\put(2,72){$n_{1}$}
\caption{Cutting the manifold $(\mathcal{M}\backslash\mathcal{C},g_{n})$ into a wedge $(\mathcal{W},g_{n})$.}\vspace{-7mm}
\label{fig:wedgecone}
\end{center}
\end{figure}

We will start by considering a general off-shell metric $g_n$ that satisfies the Hartle--Hawking boundary condition \eqref{maldacena} on the space $\mathcal{M}\,\backslash \,\mathcal{C}$ with the periodicity condition $\tau \sim \tau + 2\pi$. The key idea is to notice that cutting $\mathcal{M}\,\backslash \,\mathcal{C}$ open along a codimension-1 surface $\mathcal{B}$, such that $\partial \mathcal{B} = \mathcal{C}$, produces a wedge shaped space $\mathcal{W}$ which has two boundaries $\mathcal{B}_{\alpha}$ (with $\alpha = 1,2$) meeting at a corner $\mathcal{C} = \mathcal{B}_{1}\cap \mathcal{B}_{2}$.\footnote{We treat $\mathcal{W}$ as an open set which does not include its boundaries or the corner.} See Figure \ref{fig:wedgecone} for an illustration. Importantly, this cutting has no effect on the value of the gravitational action because it amounts to removing a sliver of measure zero from the integration region $\mathcal{M}$. Therefore, 
\begin{equation}
I_{\mathcal{M}\,\backslash\,\mathcal{C}}[g_n] = I_{\mathcal{W}}[g_n],
\label{cutHH}
\end{equation}
which states that the gravitational action of the manifold $(\mathcal{M}\,\backslash \,\mathcal{C},g_n)$ is equal to the action of a manifold $(\mathcal{W},g_n)$ with a corner whose opening angle is determined by $ n $. 


Consider now the Euclidean Einstein--Hilbert action of the wedge $(\mathcal{W},g_n)$ supplemented by Gibbons--Hawking--York boundary terms (but not a Hayward corner term)
\begin{equation}
I_{\mathcal{W}}[g_{n}] =-\int_{\mathcal{W}}d^{D}x\sqrt{g_{n}}\,R - \sum_{\alpha=1}^{2}\int_{\mathcal{B}_{\alpha}}d^{D-1}x\sqrt{h_{\alpha }}\,2K_\alpha\;.
\label{fullEinsteinaction}
\end{equation}
Here $h_{\alpha ij}$ is the induced metric on $\mathcal{B}_{\alpha}$, $K_{\alpha ij} = h_{\alpha i}^k h_{\alpha j}^{l} \nabla_k n_{\alpha l}$ is its extrinsic curvature and $n_{\alpha}^a$ is the outward-pointing unit normal vector of $\mathcal{B}_{\alpha}$, obeying $n_{\alpha}\cdot n_{\alpha}=1$. The boundaries $\mathcal{B}_1$ and $\mathcal{B}_2$ are located at $\tau = 0$ and $\tau = 2\pi$ respectively. Despite $\mathcal{W}$ not including boundaries $\mathcal{B}_{\alpha}$, we are nonetheless allowed to include boundary terms to the action because the induced metrics and extrinsic curvatures of the edges $\tau = 0,2\pi$ of the wedge are related by\footnote{The relative minus sign in the extrinsic curvatures is due to the fact that the normal vectors point in opposite directions.}
\begin{equation}
    h_{1ab}\lvert_{\mathcal{B}_1}\, = h_{2ab}\lvert_{\mathcal{B}_2},\quad K_{1ab}\lvert_{\mathcal{B}_1}\, = -K_{2ab}\lvert_{\mathcal{B}_2}
    \label{periodicBCnonpert}
\end{equation}
due to replica symmetry $\tau \rightarrow \tau + 2\pi k$ of $g_n$ for integer $n\geq 1$. Thus, the boundary terms in fact cancel one another in \eqref{fullEinsteinaction} provided $n$ is an \emph{integer} (we have merely added zero). For non-integer $n$, however, replica symmetry is broken leading to a discontinuity in the derivative of the metric when $\tau = 0$ and $\tau = 2\pi$ surfaces are identified. As such, we work at integer $n$ and only analytically continue values of on-shell actions at the very end of the computation. 

\paragraph{Variational principle for the metric.} The variation of \eqref{fullEinsteinaction} with respect to the metric without imposing any boundary conditions is given by (see Appendix \ref{app:actionvariation})
\begin{equation}
\delta_{g_{n}} I_{\mathcal{W}} = -\int_{\mathcal{W}}\hspace{-2mm}d^{D}x\sqrt{g_{n}}\,G_{nab}\,\delta g_{n}^{ab}-\sum_{\alpha=1}^{2}\int_{\mathcal{B}_{\alpha}}\hspace{-2mm}d^{D-1}x\sqrt{h_{\alpha }}\,\widetilde{T}_{\alpha ab}\,\delta h^{ab}_{\alpha}+2\int_{\mathcal{C}}d^{D-2}x\sqrt{\sigma}\,\delta_{g_{n}} \Theta_n,
\label{fulleinsteinvariation1}
\end{equation}
where $G_{nab}$ is the Einstein tensor, $\tilde{T}_{\alpha ab}$ is the boundary stress tensor of Einstein gravity,
\begin{equation}
\widetilde{T}_{\alpha ab} = K_{\alpha ab} - K_{\alpha } h_{\alpha ab}
\label{Einsteinstresstensors}
\end{equation}
and the corner angle $\Theta_{n}$ is given by
\begin{equation}
    \cos{\Theta_n} = g_{n ab}\,n_1^an_2^b.
\end{equation}
Since the embedding of the first boundary is $\tau = 0$ and the second is $\tau = 2\pi$, we get explicitly (see Appendix \ref{app:Thetam})
\begin{equation}
\Theta_n = \pi\,(1-2n^{-1}).
\label{Thetan}
\end{equation}

We can now use our identity \eqref{cutHH} to obtain expressions for the action of the manifold $(\mathcal{M}\backslash\mathcal{C},g_{n})$ and its variation. On the boundaries $\mathcal{B}_{1,2}$ of the wedge we impose periodic boundary conditions,
\begin{equation}
\delta h_{1}^{ab}|_{\mathcal{B}_{1}}= \delta h_{2}^{ab}|_{\mathcal{B}_{2}},
\label{periodicBCpert}
\end{equation}
which ensure that the metric variation is continuous across the cut. Combining with \eqref{periodicBCnonpert} we see the boundary variations in
\eqref{fulleinsteinvariation1} cancel each other when $n\geq1$ is an integer to give
\begin{equation}
\delta_gI_{\mathcal{M}\,\backslash\,\mathcal{C}}[g_n] =-\int_{\mathcal{M}}d^{D}x\sqrt{g_n}\,G_{n ab}\,\delta g^{ab}  +2\int_{\mathcal{C}}d^{D-2}x\sqrt{\sigma}\,\delta_g \Theta_n \;.
\label{Mvariation}
\end{equation}
We want to extremize the action \eqref{replicaactioncorner} over metrics that satisfy the Hartle--Hawking boundary condition (i) \eqref{maldacena} at $\mathcal{C}$. Hence the metric variation is such that $\delta n = 0$, or 
\begin{equation}
    \delta_g \Theta_n = 0\ .
\end{equation}
For such variations the second term in \eqref{Mvariation} vanishes so that the variational principle imposes Einstein's equations $ G_{n ab} = 0 $ on the metric $g_n$ everywhere outside of $\mathcal{C}$.

As shown in Section \ref{sec:splitting}, the Ricci tensor near $\mathcal{C}$ contains the divergent term \eqref{eq:Rzzleading} which leads to the condition \eqref{minimalareacondition} constraining $\mathcal{C}$ to be a minimal area surface in $\overbar{g}_1$. Hence, the variational principle for the Hartle--Hawking metric also fixes the embedding of $\mathcal{C}$. 



\paragraph{The entropy functional.} The variational principle for the metric above fixes the on-shell metric $\overbar{g}_n$ and the on-shell embedding $\overbar{E}_n$ of $\mathcal{C}$ in $\mathcal{M}$. Together they determine the on-shell induced metric $\overbar{\sigma}_n$ of $\mathcal{C}$. Since the metric is on-shell and satisfies the periodic boundary conditions \eqref{periodicBCpert}  
we get from \eqref{Mvariation},
\begin{equation}
\delta_g I_{\mathcal{M}\,\backslash\,\mathcal{C}}[\overbar{g}_n] = 2\int_{\mathcal{C}}d^{D-2} x\sqrt{\overbar{\sigma}_n}\, \delta_g \overbar{\Theta}_{n},
\label{generalHHvariation}
\end{equation}
which is valid for integer $n\geq 1$ to ensure cancellation of boundary terms. However, to compute the entropy \eqref{refinedrenyiHH}, we consider metric variations corresponding to variations $\delta n$. This requires analytic continuation of $n$ to non-integer values. We will simply extend \eqref{generalHHvariation} to non-integer $n$ so that the entropy \eqref{refinedrenyiHH} can be written as
\begin{equation}
	\widetilde{S}_n  = 2n^{2}\int_{\mathcal{C}}d^{D-2}x\sqrt{\overbar{\sigma}_n}\,\partial_n\overbar{\Theta}_{n}\;.
	\label{entropyfromcorner}
\end{equation}
Using corner angle \eqref{Thetan}, the refined R\'enyi entropy explicitly becomes
\begin{equation}
\widetilde{S}_n = 4\pi\int_{\mathcal{C}}d^{D-2}x\sqrt{\overbar{\sigma}_n}, 
\label{einsteinentropy}
\end{equation}
i.e., the area functional of the corner $\mathcal{C}$ in the solution $\overbar{g}_n$. Taking the limit $n\to1$, we recover the RT formula (\ref{eq:RTform}), with the minimization prescription determined by Einstein's equations, in line with \cite{Lewkowycz:2013nqa}. 

\subsection{Entropy of fixed-area states}\label{sec:einsteinFA}

Let us now compute the entropy functional of fixed-area states where the area of $\mathcal{C}$ is fixed.
To this end, consider a general off-shell metric $g_{\mathcal{A}}$ that satisfies the boundary condition (ii) \eqref{fursaev} with a $\sigma$ whose area is fixed to a constant,
\begin{equation}
    \int_{\mathcal{C}}d^{D-2}x\sqrt{\sigma}\equiv \mathcal{A}\ .
\end{equation}
As in the previous section, we cut the replicated manifold $(\mathcal{M}_n ,g_{\mathcal{A}})$ open which produces a manifold $(\mathcal{W}_n,g_{\mathcal{A}})$ with boundaries $\mathcal{B}_{\alpha}$ (with $\alpha = 1,2$) meeting at a corner $\mathcal{C} = \mathcal{B}_{1}\cap \mathcal{B}_{2}$ (the notation $\mathcal{W}_{n}$ will become clear momentarily, but is indicative of the fact boundary $\mathcal{B}_{2}$ depends on $n$). The Einstein--Hilbert action satisfies
\begin{equation}
I_{\mathcal{M}_n}[g_{\mathcal{A}}] = I_{\mathcal{W}_n}[g_{\mathcal{A}}]+I_{\mathcal{C}}[g_{\mathcal{A}}]\;,
\label{cutreplica}
\end{equation}
where the action $I_{\mathcal{W}_{n}}$ on the right-hand side again is given by \eqref{fullEinsteinaction}. In contrast with the Hartle--Hawking states, for fixed-area boundary conditions we must include a Hayward corner term to have a well-posed variational problem since $(\mathcal{W}_{n},g_{\mathcal{A}})$ has a corner inside its interior.

Further, unlike Hartle--Hawking states, fixed-area states require us to consider a variational problem for the embedding of $\mathcal{C}$ separately.
We thus determine the gravitational entropy of fixed-area states in three steps: (1) variational principle for the metric, (2) variational principle for the embedding of $\mathcal{C}$ and (3) demonstrate the entropy functional as the on-shell action of these solutions. 
	
\paragraph{Variational principle for the metric.}

The variational principle for fixed-area metrics $ g_{\mathcal{A}} $ (area-preserving metric variations) on the replicated space $(\mathcal{M}_n,g_{\mathcal{A}})$ is not well defined because at $\mathcal{C}$ there is an angular excess. This is evident after the cutting procedure \eqref{cutreplica}: the variation of the Einstein--Hilbert action with respect to the metric includes a term localized at the corner $ \mathcal{C}$ which has to be cancelled to make the fixed-area variational problem well defined. This is why we supplemented the wedge action with a Hayward term \cite{Hayward:1993my} in (\ref{cutreplica}) (see also Section \ref{sec:cornhighcurv}). The Hayward term provides the necessary energy density to support the conical excess present on  $(\mathcal{M}_n\,\backslash\,\mathcal{C},g_{\mathcal{A}})$. In particular, we add to $I_{\mathcal{W}_{n}}[g_{\mathcal{A}}]$ the following corner contribution
\begin{equation}
I_{\mathcal{C}}[g_{\mathcal{A}}]=-2\int_{\mathcal{C}}d^{D-2}x\sqrt{\sigma}\,(\Theta_{1\slash n} + \pi).
\label{fullEinsteinactioncorner}
\end{equation}
 The corner angle, defined by
\begin{equation}
\cos{\Theta_{1\slash n}} = g_{\mathcal{A}ab}\,n_{1}^{a} n_{2}^{b}\ ,
\label{cornerangle}
\end{equation}
 is $n$-dependent because the embedding of the boundary $\mathcal{B}_2$ depends on $n$. Namely, the two boundaries are located at $\tau = 0$ and $\tau = 2\pi n$, which explicitly gives (see Appendix \ref{app:Thetam})
\begin{equation}
\Theta_{1\slash n} = \pi\,(1-2n) \,.
\label{Thetaoneovern}
\end{equation}
Observe this amounts to setting $n\rightarrow 1\slash n$ in \eqref{Thetan}. We see at $n=1$ the angle does not vanish, $\Theta_1 = -\pi$. This is why we have included a ``counterterm'' proportional to $ 2\pi \mathcal{A}$ in (\ref{fullEinsteinactioncorner}), with a coefficient such that the combination $\Theta_{1\slash n}+\pi = 2\pi\,(1-n)$ vanishes at $n = 1$, and hence $I_{\mathcal{C}}[g_{\mathcal{A}}]=0$ at $n=1$ (when there is no corner). This condition at $n = 1$ uniquely fixes the coefficient of the area counterterm to $2\pi$.

Via equation \eqref{cutreplica}, we can obtain the action of the replicated manifold $(\mathcal{M}_n,g_{\mathcal{A}})$ from the action of a wedge \eqref{fullEinsteinactioncorner}. Due to replica symmetry condition (\ref{periodicBCnonpert}), the codimension-1 boundary terms in 
\eqref{fullEinsteinactioncorner} cancel one another to give
\begin{equation}
I_{\mathcal{M}_n}[g_{\mathcal{A}}] =-n\int_{\mathcal{M}\backslash\mathcal{C}}d^{D}x\sqrt{g_{\mathcal{A}}}\,R -4\pi\,(1-n)\int_{\mathcal{C}}d^{D-2}x\sqrt{\sigma}\ ,
\label{replicaactioncorner}
\end{equation}
where we have used \eqref{Thetaoneovern},  and replica symmetry to pull out the factor of $n$ in the $\tau$-integration in the bulk integral.
We observe this is the formula for the distributional contribution of a squashed conical excess to the Ricci scalar originally derived in \cite{Fursaev:2013fta}. Note that the formula in \cite{Fursaev:2013fta} is derived using a regularization method which includes additional regularization dependent terms that enter at higher orders in $ n-1 \ll 1 $ (except in two dimensions or when the extrinsic curvatures vanish).
Notably, our corner method does not yield such terms.

We now seek to extremize the action \eqref{replicaactioncorner} over metrics that have a fixed area at $\mathcal{C}$. This is achieved by considering variations for which the change in the induced metric of $\mathcal{C}$ is traceless, $\sigma_{ab}\,\delta \sigma^{ab} = 0$, as this ensures $\delta \sqrt{\sigma} = 0$ (fixed area). The variation of \eqref{replicaactioncorner} can be obtained from the variation of the action on a wedge \eqref{fullEinsteinactioncorner} plus corner term, which, without imposing any boundary conditions, is given by (see Appendix \ref{app:actionvariation})
\begin{equation}
\hspace{-2mm} \delta_g I_{\mathcal{M}_n} = -\int_{\mathcal{W}_n}\hspace{-3mm}d^{D}x\sqrt{g}\,G_{ab}\,\delta g^{ab}-\sum_{\alpha=1}^{2}\int_{\mathcal{B}_{\alpha}}\hspace{-2mm}d^{D-1}x\sqrt{h_{\alpha }}\,\widetilde{T}_{\alpha ab}\,\delta h^{ab}_{\alpha}-\int_{\mathcal{C}}d^{D-2}x\sqrt{\sigma}\,\widehat{T}_{ab}\,\delta \sigma^{ab},
\label{fulleinsteinvariation}
\end{equation}
where $G_{ab}$ is the Einstein tensor of $g_{\mathcal{A}}$, the boundary stress tensor is given by \eqref{Einsteinstresstensors} and the corner stress tensor of Einstein gravity is (including the contribution of the counterterm)
\begin{equation}
\widehat{T}_{ab} = -(\Theta_{1\slash n}+\pi)\,\sigma_{ab}\ .
\label{Einsteincornerstresstensor}
\end{equation}
The boundary terms in \eqref{fulleinsteinvariation} again cancel by imposing periodic boundary conditions (\ref{periodicBCpert}), leaving
\begin{equation}
\delta_gI_{\mathcal{M}_n}[g_{\mathcal{A}}] =-\int_{\mathcal{M}_n}d^{D}x\sqrt{g_{\mathcal{A}}}\,G_{ab}\,\delta g^{ab}  -4\pi\,(1-n)\int_{\mathcal{C}}d^{D-2}x\sqrt{\sigma}\,\sigma_{ab}\,\delta \sigma^{ab}\ .
\label{replicaactioncornervariation}
\end{equation}
where we have used \eqref{Thetaoneovern}. For area preserving variations the second term in \eqref{replicaactioncornervariation} vanishes so that the variational principle imposes Einstein's equations $ G_{ab} = 0 $ on $g_{\mathcal{A}}$ everywhere on the replica manifold. Hence, the Hayward term, which was historically introduced to make the variational problem assuming Dirichlet boundary conditions well defined, also makes the fixed-area variational problem well defined.




\paragraph{Variational principle for the embedding.} 

Unlike for Hartle--Hawking states, Einstein's equations for $g_{\mathcal{A}}$ do not give constraints on the embedding of $\mathcal{C}$: the term \eqref{eq:Rzzleading} in the Ricci tensor $R_{zz}$ responsible for the minimization constraints vanishes for $g_{\mathcal{A}}$ since $n = 1$. Hence the variational problem for the embedding in the on-shell metric $\overbar{g}_{\mathcal{A}}$ has to be considered separately. 



Recall $\mathcal{M}_n$ consists of $n$-copies of $\mathcal{M}$ glued together cyclically around $\mathcal{C}\subset \mathcal{M}$. We denote the embedding of $ \mathcal{C} $ in $\mathcal{M}$ as $x^a = E^a(\hat{x})$, where $\hat{x}^A$ are worldvolume coordinates of $\mathcal{C}$. Define a set of tangent vectors $e^a_A = \partial_A E^a$ that can be used to pull-back tensors to $\mathcal{C}$ (see Appendix \ref{app:hypersurfaceconventions} for more details). Instead of varying the embedding directly, we will keep the embedding fixed $\delta e^a_A  = 0 $ and vary the background metric in the wedge $\mathcal{W}$ by an infinitesimal diffeomorphism
\begin{equation}
 \delta_\xi g^{ab} = \nabla^{a}\xi^{b}+\nabla^{b}\xi^{a},\quad \xi^{a} =  \xi^{n}_{\alpha}\,n^{a}_{\alpha} + \xi^{r}_{\alpha}\,r^{a}_{\alpha},
 \label{transversediffeo}
\end{equation}
where we have taken the vector field $\xi$ to be normal to $ \mathcal{C} $, and $r_{\alpha}$ are vectors tangent to $\mathcal{B}_{\alpha}$ obeying $r_{\alpha}\cdot r_{\alpha}=1$ and $r_{\alpha}\cdot n_{\alpha} = 0$. We will assume that the vector field $\xi$ and its derivative are continuous across the cut so that the diffeomorphism descends to a transverse variation of the embedding of the conical excess $\mathcal{C}$ in $\mathcal{M}_n$.

Under the diffeomorphism \eqref{transversediffeo}, the induced metric of $\mathcal{C}$ changes as \cite{Bhattacharyya:2014yga}
\begin{align}
\sigma_{c}^{a}\sigma_{d}^{b}\,\delta_{\xi}\sigma^{cd} &= 2\,\xi^n_\alpha\, L^{ab}_\alpha+ 2\,\xi^r_\alpha\, Q^{ab}_\alpha
\label{inducedmetricvariation}
\end{align}
where the extrinsic curvatures of $\mathcal{C}$ in the metric $\overbar{g}_{\mathcal{A}}$ are defined as\footnote{Note that the extrinsic curvatures of $\mathcal{C}$ are non-vanishing in fixed-area metrics $g_{\mathcal{A}}$.}
\begin{equation}
	L_{\alpha ab} = \sigma_{a}^{c}\,\sigma_{b}^{d}\,\nabla_{c}n_{\alpha d}, \quad Q_{\alpha ab} = \sigma_{a}^{c}\,\sigma_{b}^{d}\,\nabla_{c}r_{\alpha d},\quad \alpha = 1,2.
 \label{cornerextrinsiccurvatures}
\end{equation}
Applying the diffeomorphism \eqref{transversediffeo} to the general variation \eqref{replicaactioncornervariation} when the metric is on-shell $\overbar{g}_{\mathcal{A}}$ gives
\begin{equation}
\delta_\xi I_{\mathcal{M}_n}[\overbar{g}_{\mathcal{A}}] = -8\pi\,(1-n)\int_{\mathcal{C}}d^{D-2}x\sqrt{\sigma}\,(\xi^n_1\,\sigma_{ab}\,L_{1}^{ab}+\xi^r_1\,\sigma_{ab}\,Q^{ab}_1).
\end{equation}
Requiring the variation to vanish gives the conditions
\begin{equation}
\sigma^{ab}L_{1ab} =\sigma^{ab}Q_{1ab} = 0,
\end{equation}
which are equivalent to the conditions obtained by extremizing the area functional in the background $\overbar{g}_{\mathcal{A}}$. Thus, we have derived the area extremization prescription for fixed-area states from the variation of the Einstein action of the wedge under transverse diffeomorphisms of $\mathcal{C}$. Equivalently, the extremization coincides with the vanishing of the Hamiltonian charge generating $\xi$ \cite{Balasubramanian:2023dpj}.


\paragraph{The entropy functional.} The two variational principles above determine the on-shell fixed-area metric $\overbar{g}_{\mathcal{A}}$ and the on-shell embedding $\overbar{E}^a$ of $\mathcal{C}$ in $\mathcal{M}$. Using \eqref{replicaactioncorner} we get the on-shell action (recall (\ref{replicaactioncorner}))
\begin{equation}
I_{\mathcal{M}_n}[\overbar{g}_{\mathcal{A}}] = n\,I_{\mathcal{M}\backslash\, \mathcal{C}}[\overbar{g}_{\mathcal{A}}] -4\pi\,(1-n)\,\mathcal{A}.
\label{replicaactioneinstein2}
\end{equation}
where in the first term we have used that $\overbar{g}_{\mathcal{A}}$ respects replica symmetry (which is always guaranteed and not an assumption unlike in the case of Hartle--Hawking states). Substituting (\ref{replicaactioneinstein2}) into the refined R\'enyi entropy \eqref{refinedrenyifixedarea}, all terms linear in $ n $ vanish giving 
\begin{equation}
\widetilde{S}_n = 4\pi\,\mathcal{A},
\label{fixedareaentropy}
\end{equation}
which is independent of $n$, thereby having a flat entanglement spectrum.

\vspace{3mm}

Before we move on to discuss the situation of higher-curvature theories of gravity, let us briefly take stock in what we have shown thus far. By using Hayward corner terms, we were able to develop
a first principles prescription to compute gravitational
R\'enyi entropy of both Hartle–Hawking and fixed–area states
in Einstein gravity. While our discussion thus far has been at the level of the bulk, our method provides an alternative derivation of the RT formula. Specificaly, upon invoking AdS/CFT duality, where we take the
bulk to be asymptotically AdS, our computations directly translate to (refined) R\'enyi entropies
of holographic CFTs living on the AdS conformal boundary. In the limit $n\to1$, we recover the
Ryu–Takayanagi relation (\ref{eq:RTform}) for Hartle–Hawking states,
and the analog for fixed–area states, in line with Fursaev's original derivation \cite{Fursaev:2006ih}.

We point that previous work has also used the Hayward term to
construct entropy functionals of fixed–area states \cite{Takayanagi:2019tvn} and R\'enyi entropies in Einstein and Jackiw–Teitelboim gravity \cite{Botta-Cantcheff:2020ywu,Arias:2021ilh}. However, a complete consideration of variational problems for the metric and the corner embedding
was lacking in these works. In particular, our formulation, for which we realize the Euclidean action of the wedge encodes information about the action on manifolds $(\mathcal{M}\,\backslash\,\mathcal{C},\bar{g}_{n})$ or $(\mathcal{M}_{n},\overbar{g}_{\mathcal{A}})$ (Eqs. (\ref{cutHH}) and (\ref{cutreplica}), respectively),
allows for a careful and systematic treatment of the variational principle. As we detail below, this becomes crucial when considering the extension to higher-curvature theories of gravity.


\section{Corner term and corner stress tensor in Lovelock gravity} \label{sec:cornhighcurv}

Here we study Lovelock gravity on manifolds with corners. We review the derivation of the Dirichlet corner term in general Lovelock gravity presented in \cite{Cano:2018ckq} and apply the same methodology to derive the Lovelock corner stress tensor which so far has not been done in the literature. We confirm the corner term of \cite{Cano:2018ckq} is correct by explicitly showing its variation cancels all corner localized terms coming from the variation of the Lovelock action. The technology introduced in this section will be applied in Section \ref{sec:entropyfromcornerterms} where we derive entropy functionals in higher-curvature gravity. 


\subsection{Corners in higher-curvature gravity}

Begin by considering a general diffeomorphism invariant $F(\text{Riemann}) $ theory of gravity on a wedge $(\mathcal{W},g)$ (as defined in Section \ref{sec:einsteinentropy}) with Euclidean action
\begin{equation}
I_{\mathcal{W}}[g] = -\int_{\mathcal{W}} \hspace{-1mm} d^{D} x \sqrt{g}\,F(g^{ab},R_{abcd}).
\label{Friemann}
\end{equation}
$F(\text{Riemann})$ constitutes a wide class of theories including general relativity, Lovelock gravity \cite{Lovelock:1971yv}, as well as $F(R)$ and more general higher-curvature theories (see, e.g., \cite{Bueno:2016ypa} for a review). The variation of the action under Dirichlet boundary conditions is given by (see Appendix \ref{app:actionvariation})
\begin{equation}
    \delta I_{\mathcal{W}}[g] =  -\int_{\mathcal{W}} \hspace{-1mm} d^{D} x \sqrt{g}\,E_{ab}\,\delta g^{ab}+\sum_{\alpha =1}^{2}\int_{\mathcal{B}_{\alpha}} \hspace{-2mm} d^{D-1} x \sqrt{h_\alpha}\,4\Psi_{\alpha ab}\,\delta K^{ab}_\alpha+\int_{\mathcal{C}}d^{D-2} x\sqrt{\sigma}\, \Psi\,\delta\Theta,
    \label{cornervartext}
\end{equation}
where we have defined the tensors
\begin{equation}
	\Psi_{ab} \equiv P_{acbd}\,n^{c}n^{d},\quad P_{abcd}  \equiv \frac{\partial F}{\partial R^{abcd}},
	\label{Psiabtext}
\end{equation}
and introduced the Wald entropy density \cite{Wald:1993nt,Iyer:1994ys}
\begin{equation}
    \Psi \equiv 4\Psi_{\alpha ab}\,r^a_{\alpha}r^b_{\alpha} = 4P_{abcd}\,n_{\alpha}^{[a} r^{b]}_\alpha n_{\alpha}^{[c} r^{d]}_\alpha.
\label{eq:Waldentdensitytext}\end{equation}
For the variational problem on the wedge to be well defined under Dirichlet boundary conditions $\delta h_{\alpha}^{ab}\lvert_{\mathcal{B}_\alpha}\, = \delta \sigma^{ab}\lvert_{\mathcal{C}}=0$, the boundary and corner variations in \eqref{cornervartext} must be cancelled by supplementing the action (\ref{Friemann}) with boundary and corner terms. The supplemented action takes the form
\begin{equation}
I_{\mathcal{W}}[g]+I_{\mathcal{C}}[g] = -\int_{\mathcal{W}}d^{D}x\sqrt{g}\,F -\sum_{\alpha=1}^{2}\int_{\mathcal{B}_{\alpha}}d^{D-1}x\sqrt{h_{\alpha }}\,B_\alpha -\int_{\mathcal{C}}d^{D-2}x\sqrt{\sigma}\,C,
\label{startingaction}
\end{equation}
where we implicitly include the boundary term as a contribution to $I_{\mathcal{W}}[g]$ (as in the previous section). The boundary term $B_\alpha$ and the corner term $C$ must obey 
\begin{equation}
    \delta_g B_\alpha = 4\Psi_{\alpha ab}\,\delta_g K^{ab}_\alpha,\quad \delta_g C = \Psi\,\delta_g\Theta,
    \label{boundarycornertermrequirement}
\end{equation}
in order to cancel the remaining boundary and corner terms in variation (\ref{cornervartext}) under Dirichlet boundary conditions. Hence, the Dirichlet variation of \eqref{startingaction} is given by
\begin{equation}
    \delta_g (I_{\mathcal{W}}[g]+I_{\mathcal{C}}[g]) =  -\int_{\mathcal{W}} \hspace{-1mm} d^{D} x \sqrt{g}\,E_{ab}\,\delta g^{ab},\quad \text{Dirichlet BC.}
\end{equation}
Finally, boundary $\widetilde{T}_{\alpha ab}$ and corner $\widehat{T}_{ab}$ stress tensors of the theory are defined by the variation of \eqref{startingaction} without any boundary conditions imposed,
\begin{equation}
\begin{split}
\delta_g (I_{\mathcal{W}}[g]+I_{\mathcal{C}}[g]) &= -\int_{\mathcal{W}}d^{D}x\sqrt{g}\,E_{ab}\,\delta g^{ab}-\sum_{\alpha=1}^{2}\int_{\mathcal{B}_{\alpha}}d^{D-1}x\sqrt{h_{\alpha }}\,\widetilde{T}_{\alpha ab}\,\delta h^{ab}_{\alpha}\\
&-\int_{\mathcal{C}}d^{D-2}x\sqrt{\sigma}\,\widehat{T}_{ab}\,\delta \sigma^{ab}.
\label{generaltheoryvariation}
\end{split}
\end{equation}
In arbitrary $F(\text{Riemann})$ theory of gravity, a variational formula of this type is \emph{not} guaranteed to exist meaning that boundary and corner stress tensors are not well defined. In fact, it only exists if the theory admits boundary and corner terms for Dirichlet boundary conditions.\footnote{If the Dirichlet problem is to be well defined, the general variation has to be of the form \eqref{generaltheoryvariation} for all boundary and corner contributions to vanish upon imposing the Dirichlet condition. Induced covariant derivatives of $\delta h^{ab}_\alpha$ or $\delta \sigma^{ab}$ might also appear, but they can be integrated by parts.} Lovelock gravity is an example of a theory which admits boundary and corner terms. This should not come as a surprise for arbitrary $F$(Riemann) gravity: the analog of Gibbons--Hawking--York boundary terms are not known to exist assuming Dirichlet
boundary conditions alone \cite{Madsen:1989rz} (see also, \cite{Dyer:2008hb,Deruelle:2009zk,Guarnizo:2010xr,Teimouri:2016ulk,Bueno:2016dol,Liu:2017kml}), e.g.. for $F(R)$ gravity, a boundary (and corner) term can be written down by 
for which the supplemented action is stationary by fixing induced metric variations and $\delta R=0$ at the boundary.

\subsection{Derivations of the corner term and stress tensor in Lovelock gravity}\label{subsec:cornerderivations}


The above variational formulae in principle determine the corner term and the corner stress tensor in a given theory, if they exist. However, the computation of the variation \eqref{generaltheoryvariation} is complicated and requires to keep track of total derivative terms on the two boundaries $\mathcal{B}_\alpha$ (see \cite{Chakraborty:2017zep} for calculations in Lovelock gravity). Luckily, they can be derived using a simpler method provided the boundary term $B$ and boundary stress tensor $\widetilde{T}_{ab}$ of the theory are known. In this approach, the corner is smoothed out into a single boundary by introducing a regulator $\epsilon$ where the sharp corner corresponds to $\epsilon = 0$ \cite{Hayward:1993my,Cano:2018ckq}. Then the corner term and the corner stress tensor are obtained from $B$ and $\widetilde{T}_{ab}$, respectively, in the $\epsilon\rightarrow 0$ limit.

\begin{figure}[t!]
\begin{center}
\includegraphics[scale=0.9]{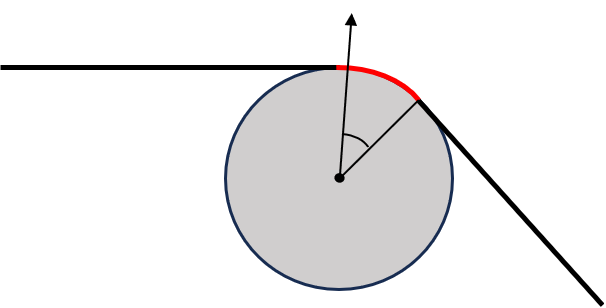}
\put(-210,118){$\mathcal{B}_{2,\epsilon}$}
\put(-29,40){$\mathcal{B}_{1,\epsilon}$}
\put(-100,112){$\mathcal{C}_{\epsilon}$}
\put(-113,85){$\theta$}
\put(-100,70){$u$}
\put(-118,138){$n$}
\caption{Smoothing out a corner. Replace corner $\mathcal{C}$ with a circular arc $\mathcal{C}_{\epsilon}$ of size $\epsilon$. The smoothed wedge $(\mathcal{W}_{\epsilon},g)$, endowed with metric (\ref{eq:smoothwedgemet}) in terms of polar coordinates $(u,\theta)$, with 1-form $n=Ndu$ normal to $\mathcal{C}_{\epsilon}$. }\vspace{-7mm}
\label{fig:smoothcorn}
\end{center}
\end{figure}
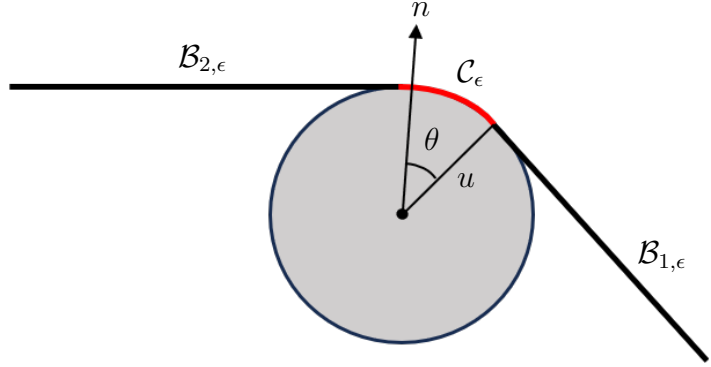


To this end, introduce a one-parameter family of smoothed out wedges $\mathcal{W}_\epsilon$ (such that $\mathcal{W}_0 \equiv \mathcal{W}$) which has a single smooth boundary $\partial \mathcal{W}_\epsilon$ obtained by replacing the corner with a circular cap $\mathcal{C}_\epsilon$ (see Figure \ref{fig:smoothcorn}). The boundary is thus divided into three parts
\begin{equation}
    \partial \mathcal{W}_\epsilon = \mathcal{B}_{1,\epsilon}\cup \mathcal{B}_{2,\epsilon}\cup \mathcal{C}_\epsilon
\end{equation}
such that
\begin{equation}
    \lim_{\epsilon\rightarrow 0}\mathcal{B}_{\alpha,\epsilon} = \mathcal{B}_{\alpha},\quad \lim_{\epsilon\rightarrow 0}\mathcal{C}_\epsilon = \mathcal{C}.
\end{equation}
For general theories, the gravitational action characterizing the smooth wedge with a single boundary is
\begin{equation}
    I_\epsilon = -\int_{\mathcal{W}_{\epsilon}}d^{D}x\sqrt{g}\,F - \int_{\partial \mathcal{W}_\epsilon}d^{D-1}x\sqrt{Z}\,B:,
    \label{Iepsilon}
\end{equation}
where $Z$ is the induced metric of $\partial \mathcal{W}_\epsilon$. 

Varying with respect to the metric yields
\begin{equation}
    \delta_g I_\epsilon = -\int_{\mathcal{W}_{\epsilon}}d^{D}x\sqrt{g}\,E_{ab}\,\delta g^{ab} - \int_{\partial \mathcal{W}_\epsilon}d^{D-1}x\sqrt{Z}\,\widetilde{T}_{ab}\,\delta Z^{ab}\;.
    \label{deltaIepsilon}
\end{equation}
Assuming the variation $\delta_g$ and the sharp limit commute, the $\epsilon\rightarrow 0$ limit of \eqref{Iepsilon} and \eqref{deltaIepsilon} gives \eqref{startingaction} and \eqref{generaltheoryvariation}, respectively. Hence, we can identify the corner term and the corner stress tensor as
\begin{align}
    \int_{\mathcal{C}}d^{D-2}x\sqrt{\sigma}\,C &= \lim_{\epsilon\rightarrow 0}\int_{\mathcal{C}_\epsilon}d^{D-1}x\sqrt{Z}\,B\\
    \int_{\mathcal{C}}d^{D-2}x\sqrt{\sigma}\,\widehat{T}_{ab}\,\delta \sigma^{ab} &= \lim_{\epsilon\rightarrow 0}\int_{\mathcal{C}_\epsilon}d^{D-1}x\sqrt{Z}\,\widetilde{T}_{ab}\,\delta Z^{ab}.
    \label{cornerlimits}
\end{align}
To compute these limits, we introduce a set of polar coordinates $(u,\theta)$ so that $\mathcal{C}_\epsilon$ is the circular arc $0 < \theta< \Theta$ at $ u = \epsilon$. In the limit $\epsilon \rightarrow 0$, the origin $u = 0$ is near the corner $\mathcal{C}$ so that the metric $g$ can be written in the form \cite{Hayward:1993my,Cano:2018ckq}
\begin{equation}
    ds^2 = N^2du^2 + Z_{ij}\,dx^idx^j = N^2du^2 +M^2d\theta^2 + \sigma_{AB}\,d\hat{x}^Ad\hat{x}^B + \ldots,\quad \epsilon\rightarrow 0,
\label{eq:smoothwedgemet}\end{equation}
where $\sigma_{AB}$ is the induced metric of $\mathcal{C}$. In these coordinates $\sqrt{Z} = M\sqrt{\sigma}$ so that \eqref{cornerlimits} can be written more explicitly as
\begin{align}
    C &= \lim_{\epsilon\rightarrow 0}\int_0^{\Theta}d\theta\,M\,B\\
   \widehat{T}_{ab}\,\delta \sigma^{ab} &= \lim_{\epsilon\rightarrow 0}\int_0^{\Theta}d\theta\,M\,\widetilde{T}_{ab}\,\delta Z^{ab}.
   \label{eq:CTtildelimits}
\end{align}

Let us now apply this method to derive $C$ and $\widehat{T}_{ab}$ in Lovelock gravity, leaving computational details for Appendix \ref{app:cornerstress}. The derivation of the corner term in this way was originally done in  \cite{Hayward:1993my,Cano:2018ckq}, however, the corner stress tensor has not been derived before in the literature.

Begin with the Euclidean action of pure Lovelock gravity of order $m$ \cite{Lovelock:1971yv},
\begin{equation}
I_{\mathcal{M}} = -\int_{\mathcal{M}}d^D x \sqrt{g}\, \mathcal{R}_{(m)},
\label{purelovelock}
\end{equation}
where $\mathcal{R}_{(m)} $ is the Lovelock scalar
\begin{equation}
\mathcal{R}_{(m)} \equiv \frac{1}{2^m}\delta^{a_1b_1\ldots a_mb_m}_{c_1d_1\ldots c_md_m}R^{c_1d_1}_{a_1b_1}\cdots R^{c_md_m}_{a_mb_m},
\label{eq:lovelockscalar}
\end{equation}
with the generalized Kronecker-delta symbol, defined as
\begin{equation}
\delta^{a_1\ldots a_n}_{b_1\ldots b_n} \equiv n!\,\delta^{a_1}_{[b_1}\cdots \delta^{a_n}_{b_n]}.
\label{eq:genkron}
\end{equation}
For $m=0$ we set $\mathcal{R}_{(0)} = 1$, while $m=1$ gives the Einstein-Hilbert term and $m=2$ results in the Gauss--Bonnet term, $\mathcal{R}_{(2)}=(R^{2}-4R_{ab}^{2}+R_{abcd}^{2})$. General Lovelock gravity has an action comprised of a linear combination of pure Lovelock gravity actions with generic coefficients. Famously, Lovelock gravity is the unique theory of pure gravity whose equations of motion are second-order in the metric and the equation of motion tensor of pure Lovelock gravity is explicitly
\begin{equation}
     E^a_{(m)b} = -\frac{1}{2^{m+1}}\delta^{aa_1b_1\ldots a_mb_m}_{b c_1d_1\ldots c_md_m}R^{c_1d_1}_{a_1b_1}\cdots R^{c_md_m}_{a_mb_m}.
     \label{lovelockEOM}
\end{equation}

The GHY-like boundary term and stress tensor are given by \cite{Myers:1987yn,Teitelboim:1987zz,Miskovic:2007mg},
\begin{align}
B_{(m)} &=2m\int_{0}^{1} ds\, \delta^{ii_1j_1\ldots i_{m-1}j_{m-1}}_{jk_1l_1\ldots k_{m-1}l_{m-1}}K^{j}_{i}\prod_{p=1}^{m-1}\left(\frac{1}{2}\widetilde{R}^{k_pl_p}_{i_pj_p} - s^{2}K_{i_p}^{k_p}K_{j_p}^{l_p}\right)\label{eq:LovelockGH}\\
\widetilde{T}_{(m)j}^{i} &= -m\int_{0}^{1} ds\, \delta^{iki_1j_1\ldots i_{m-1}j_{m-1}}_{jlk_1l_1\ldots k_{m-1}l_{m-1}}K^{l}_{k}\prod_{p=1}^{m-1}\left(\frac{1}{2}\widetilde{R}^{k_pl_p}_{i_pj_p} - s^{2}K_{i_p}^{k_p}K_{j_p}^{l_p}\right)
\label{lovelockboundarystress}
\end{align}
We use these boundary quantities to compute corner term and stress tensor via (\ref{eq:CTtildelimits}), the details of which are relegated to Appendix \ref{app:cornerstress}. For Einstein gravity, the derivation reproduces the Hayward term and the corner stress tensor \eqref{Einsteincornerstresstensor} (without the area counterterm), obtained by varying the action directly. For general Lovelock gravity, the derivation of the corner term using the smoothing method is \cite{Cano:2018ckq} (see also Appendix \ref{app:lovelockcorner})
\begin{equation}
    C_{(m)} = 2m\Theta \,\widehat{\mathcal{R}}_{(m-1)}+ \mathcal{F}_{(m)}(L_1,Q_1) + \mathcal{F}_{(m)}(L_2,Q_2),
    \label{Cmtext}
\end{equation}
where
\begin{equation}
    \mathcal{F}_{(m)}(L,Q) = \sum_{l=2}^{m}\frac{m!(l-1)!}{2^{m-l-1}(m-l)!}R^{m-l}\,\text{Im}\,\left[\sum_{j=0}^{l-2}\frac{(L-iQ)^{j}(L+iQ)^{2l-2-j}}{j!\,(2l-2-j)!\,(l-j-1)}\right].
    \label{canoF}
\end{equation}
To keep expressions compact, here we have not written the indices explicitly and it is understood that all free indices are to be contracted with the generalized Kronecker delta symbol \eqref{eq:genkron}. Due to the contraction, all extrinsic curvature factors behave as commuting numbers. Further,  corner stress tensor of order $m$ is (see Appendix \ref{app:lovelockcorner} for details)
\begin{equation}
    \widehat{T}_{(m)ab} = 2m\,\Theta\,\widehat{E}_{(m-1)ab} + \mathcal{F}_{(m)ab}(L_1,Q_1)+\mathcal{F}_{(m)ab}(L_2,Q_2),
    \label{lovelockcornerstress}
\end{equation}
where $\widehat{E}_{(m)ab}$ is the equation of motion tensor \eqref{lovelockEOM} of pure Lovelock gravity constructed from $\sigma$ and the tensor $\mathcal{F}_{(m)B}^A(L,Q)$ has the same form as \eqref{canoF}, but it is contracted with $-\frac{1}{2}\,\delta^{AA_1B_1\ldots A_{m-1}B_{m-1}}_{BC_1D_1\ldots C_{m-1}D_{m-1}}$ with two extra free indices $A$ and $B$.

To be illustrative, consider pure Gauss--Bonnet gravity, $m=2$. Then, 
\begin{equation}
    \mathcal{F}_{(2)}(L,Q) = 4\,\delta^{AB}_{CD} L^C_A Q^D_B,
    \label{GBcurlyF}
\end{equation}
such that the corner term is \cite{Cano:2018ckq}
\begin{equation}
    C_{(2)} = 4\,\left(\Theta\,\widehat{R} + L_1Q_1+ L_2Q_2 -L_{1ab}Q_{1}^{ab}  -L_{2ab}Q_{2}^{ab}\right).
\end{equation}
Moreover, in Appendix \ref{app:GBcorner} we show that the corner stress tensor in pure Gauss--Bonnet gravity is
\begin{equation}
\widehat{T}_{(2)ab} = 4\,\Theta\,\widehat{G}_{ab} + \mathcal{F}_{(2)ab}(L_1,Q_1)+\mathcal{F}_{(2)ab}(L_2,Q_2),
\label{GBcornerstresstext}
\end{equation}
where $ \widehat{G}_{ab} $ is the Einstein tensor of the induced metric $ \sigma_{ab} $ of the corner and we have defined the tensor
\begin{equation}
    \mathcal{F}^A_{(2)B}(L,Q) \equiv -4\delta^{AA_1B_1}_{BC_1D_1}L_{A_1}^{C_1}Q_{B_1}^{D_1}.
\end{equation}

\subsection{Explicit variation of the Lovelock corner term}\label{sec:cornertermvariation}

Above we used the smoothing method to derive $C$ and $\widehat{T}_{ab}$ in pure Lovelock gravity. Originally, however, they are defined by the variational expressions \eqref{boundarycornertermrequirement} and \eqref{generaltheoryvariation}. It remains to be seen whether the variation $\delta_g$ and the sharp limit $\epsilon\rightarrow 0$ commute with each other. In other words, we would like to check that the Lovelock corner term \eqref{Cmtext} satisfies condition \eqref{boundarycornertermrequirement}, and, moreover, that the corner stress tensor \eqref{lovelockcornerstress} arises in the variation \eqref{generaltheoryvariation}. Outside of Einstein gravity, these consistency checks have not been proven before. Here we explicitly show that the corner term indeed satisfies \eqref{boundarycornertermrequirement} as required. Proving that the corner stress tensor arises from the variation is more complicated and  left for future work.

Our aim is to verify $\delta_{g}C=\Psi\delta\Theta$ in general Lovelock gravity. Begin with the right hand side. The Wald entropy density \eqref{eq:Waldentdensitytext} of pure Lovelock gravity (at order $m$) \eqref{purelovelock} is
\begin{equation}
\Psi_{(m)} = \frac{4m}{2^{m}}\,\delta^{A_1B_1\ldots A_{m-1}B_{m-1}}_{C_1D_1\ldots C_{m-1}D_{m-1}}R^{C_1D_1}_{A_1B_1}\cdots R^{C_{m-1}D_{m-1}}_{A_{m-1}B_{m-1}} = 2m\,\mathfrak{R}_{(m-1)},
\label{lovelockpsi}
\end{equation}
where $A,B,\ldots = 1,\ldots D-2$ label tangential directions along the corner $\mathcal{C}$ and $\mathfrak{R}_{(m)}$ is the Lovelock scalar constructed from the corner projected Riemann tensor
\begin{equation}
    \mathfrak{R}_{abcd} \equiv \sigma_a^{a_1}\,\sigma_b^{b_1}\,\sigma_c^{c_1}\,\sigma_d^{d_1}\,R_{a_1b_1c_1d_1},
    \label{eq:cornerprojectedriemann}
\end{equation}
which satisfies $\mathfrak{R}_{ABCD} = R_{ABCD}$. Using the codimension-2 Gauss--Codazzi equation
\begin{equation}
	R^{CD}_{AB} = \widehat{R}^{CD}_{AB}-2L^{C}_{1[A}L^{D}_{1B]}- 2Q^{C}_{1[A}Q^{D}_{1B]},
	\label{gausscodazzi}
\end{equation} 
we may decompose the $m$th Wald entropy density as
\begin{equation}
\Psi_{(m)} = 2m\,\widehat{\mathcal{R}}_{(m-1)} + \Phi_{(m)},
\label{eq:Waldentdens2}
\end{equation}
where $\widehat{\mathcal{R}}_{(m)}$ is the intrinsic Lovelock scalar of the corner and all extrinsic curvature dependence is included in
\begin{equation}
\Phi_{(m)}(L_1,Q_1) = \frac{4m}{2^{m}}\sum_{k=1}^{m-1}\binom{m-1}{k}\widehat{R}^{m-k-1}(-2L^{2}_1 - 2Q^{2}_1)^{k},
\label{lovelockphi}
\end{equation}
where all indices are understood to be contracted with a generalized Kronecker delta symbol $\delta^{A_1B_1\ldots A_{m-1}B_{m-1}}_{C_1D_1\ldots C_{m-1}D_{m-1}}$. Hence, the defining equation \eqref{boundarycornertermrequirement} of the corner term can be written as
\begin{equation}
    \delta C_{(m)} = 2m\,\widehat{\mathcal{R}}_{(m-1)}\,\delta\Theta + \Phi_{(m)}\,\delta\Theta\;.
\label{eq:varCmLove1}\end{equation}

Now recall the Lovelock corner $C_{(m)}$ \eqref{Cmtext}. Using Dirichlet boundary conditions, we find the variation (\ref{eq:varCmLove1}) requires
\begin{equation}
    \delta\mathcal{F}_{(m)}(L_1,Q_1) + \delta\mathcal{F}_{(m)}(L_2,Q_2)  =\Phi_{(m)}(L_1,Q_1)\,\delta\Theta.
    \label{correctcorner}
\end{equation}
To check this relation is satisfied, we use (see Appendix \ref{app:dirichletvariations})
\begin{equation}
    \delta Q_{\alpha B}^{A} = 0, \quad \delta L_{1B}^A = -\csc{\Theta}\,\delta \Theta\,L_{2 B}^A, \quad \delta L_{2 B}^A = -\csc{\Theta}\,\delta \Theta\,L_{1 B}^A,
    \label{eq:extvariations}
\end{equation}
which are valid under Dirichlet boundary conditions. To be illustrative,  first consider pure Gauss--Bonnet gravity $m = 2$, relation  \eqref{correctcorner} becomes
\begin{equation}
    \delta \mathcal{F}_{(2)}(L_1,Q_1)+\delta \mathcal{F}_{(2)}(L_2,Q_2) = 4\,Q_1\,\delta L_1+ 4\,Q_{2}\,\delta L_{2} = -4\csc{\Theta}\,\delta\Theta\,(L_2Q_1 + L_1 Q_2)\;,
    \label{eq:GBLHS}
\end{equation}
where all indices are understood to be contracted with $\delta^{AB}_{CD}$. Using the rotation relation (see Appendix \ref{app:hypersurfaceconventions})
\begin{align}
L_{2AB} &= L_{1AB}\cos{\Theta}+Q_{1AB}\sin{\Theta}\\
Q_{2AB} &= L_{1AB}\sin{\Theta}-Q_{1AB}\cos{\Theta},
\label{rotextrLQtext}
\end{align}
we obtain
\begin{equation}
    L_2Q_1 + L_1 Q_2 = \sin{\Theta}\,(L_1^2 + Q_1^2).
\end{equation}
Hence \eqref{eq:GBLHS} can be written as
\begin{equation}
    \delta \mathcal{F}_{(2)}(L_1,Q_1)+\delta \mathcal{F}_{(2)}(L_2,Q_2) = -4\,(L_1^2 + Q_1^2)\,\delta\Theta.
\end{equation}
Comparing with \eqref{lovelockphi}, which for Gauss--Bonnet gravity is
\begin{equation}
    \Phi_{(2)}(L_1,Q_1) = -4\,(L_1^2+Q_1^2),
\end{equation}
we see that \eqref{correctcorner} is satisfied. It becomes cumbersome to analytically verify \eqref{correctcorner} for general $m$, however, we have checked the relation exactly holds up to $m = 20$ using Mathematica. Hence, we are confident the corner term \eqref{Cmtext} derived using the smoothing method is correct and cancels all corner variations of the action under Dirichlet boundary conditions.


\section{Entropy in Lovelock gravity and beyond} \label{sec:entropyfromcornerterms}

As noted in the introduction, the Camps--Dong entropy functional (\ref{eq:campsdongform}) is a proposal on how to compute the entanglement entropy for holographic CFTs dual to general $F$(Riemann) theories. In Section \ref{sec:einsteinentropy} we showed how the Hayward term can be used to derive gravitational entropy functionals of Einstein gravity for Hartle--Hawking and fixed-area states. Here we extend the derivation to Lovelock gravity using the corner terms and corner stress-tensor described in Section \ref{sec:cornhighcurv}. We also comment on entropy functionals for general $F$(Riemann) theories of gravity. 



\subsection{Entropy of Hartle--Hawking states}\label{sec:HHlovelock}

We follow the same logic as in the case of Einstein gravity in Section \ref{subsec:einsteinHH}. Let us consider a Hartle--Hawking metric $g_n$ satisfying the boundary condition \eqref{maldacena} and consider its Lovelock gravity action inside a wedge
\begin{equation}
I_{\mathcal{W}}[g_n] =-\int_{\mathcal{W}}d^{D}x\sqrt{g_n}\,\mathcal{R}_{(m)}'- \sum_{\alpha=1}^{2}\int_{\mathcal{B}_{\alpha}}d^{D-1}x\sqrt{h_{\alpha }}\,B_{(m)\alpha}\;,
\label{highercurvaturevariation}
\end{equation}
where $\mathcal{R}_{(m)}'$ denotes the Lovelock scalar \eqref{eq:lovelockscalar} of $g_n$ and $B_{(m)\alpha}$ are the Lovelock boundary terms \eqref{eq:LovelockGH} supported on the two edges of the wedge. As in Einstein gravity, we are allowed to include boundary terms at the edges of the wedge since they cancel due to replica symmetry of $g_n$ for integer $n$.

\paragraph{Variational principle for the metric.} The variation of the Lovelock action \eqref{highercurvaturevariation} with boundary terms included at the two edges (computed in Appendix \ref{subapp:lovelockactionvars}) gives
\begin{align}
    &\delta_gI_{\mathcal{W}}[g_n] =\label{lovelockvariation}\\
    &-\int_{\mathcal{W}} \hspace{-1mm} d^{D} x \sqrt{g}\,E_{(m)ab}'\,\delta g^{ab}-\sum_{\alpha=1}^{2}\int_{\mathcal{B}_{\alpha}}d^{D-1}x\sqrt{h_{\alpha }}\,\widetilde{T}_{(m)\alpha ab}\,\delta h^{ab}_{\alpha}+\int_{\mathcal{C}}d^{D-2} x\sqrt{\sigma}\, \Psi_{(m)}'\,\delta_g\Theta_n,\nonumber 
\end{align}
where the angle $\Theta_n$ is given by \eqref{Thetan}, $E_{(m)ab}'$ is the equation motion tensor \eqref{lovelockEOM} of $g_n$, and $\widetilde{T}_{(m)\alpha ab}$ is the boundary stress tensor \eqref{lovelockboundarystress}. By identifying the edges of the wedge, the variation \eqref{lovelockvariation} reduces to the variation of the action evaluated on $(\mathcal{M}\,\backslash\,\mathcal{C},g_n)$. Since the boundary terms in \eqref{lovelockvariation} vanish by periodic boundary conditions \eqref{periodicBCpert}, we get
\begin{equation}
    \delta_g I_{\mathcal{M}\,\backslash\,\mathcal{C}}[g_n] = -\int_{\mathcal{M}} \hspace{-1mm} d^{D} x \sqrt{g_{n}}\,E_{(m)ab}'\,\delta g^{ab}+\int_{\mathcal{C}}d^{D-2} x\sqrt{\sigma}\, 2m\,\mathfrak{R}_{(m-1)}'\,\delta_g\Theta_n,
    \label{lovelockvariationM}
\end{equation}
where $\mathfrak{R}_{(m-1)}'$ is the Lovelock scalar constructed from the projected Riemann tensor $\mathfrak{R}'_{abcd}$ of $g_n$ and we used \eqref{lovelockpsi}. Under variations that keep the angle fixed $\delta_g\Theta_n = 0$, the variational principle imposes equations of motion $E_{(m)ab}' = 0$ of pure Lovelock gravity on $g_n$ and the solution is denoted by $\overbar{g}_n$.




\paragraph{The entropy functional.} Setting the metric on-shell in the variation \eqref{lovelockvariationM} gives
\begin{equation}
    \delta_g I_{\mathcal{M}\,\backslash\,\mathcal{C}}[\overbar{g}_n] =2m\int_{\mathcal{C}}d^{D-2} x\sqrt{\overbar{\sigma}_n}\, \overbar{\mathfrak{R}}_{(m-1)}'\,\delta_g\Theta_n,
\end{equation}
where $\overbar{\mathfrak{R}}_{(m)}'$ denotes the projected Lovelock scalar of $\overbar{g}_n$. Applying this formula to the variation $\delta_g = \partial_n$, the refined Rényi entropy \eqref{refinedrenyiHH} of Lovelock gravity is
\begin{equation}
    \widetilde{S}_n = 4\pi m\int_{\mathcal{C}}d^{D-2} x\sqrt{\overbar{\sigma}_n}\,\overbar{\mathfrak{R}}_{(m-1)}'\ .
    \label{firstentropy}
\end{equation}
This formula further simplifies because extrinsic curvatures of $\mathcal{C}$ in the metric $\overbar{g}_n$ vanish for any integer $n > 1$. By the Gauss--Codazzi equation \eqref{gausscodazzi}, we find the Jacobson--Myers functional of $\overbar{\sigma}_n$,
\begin{equation}
    \widetilde{S}_n = 4\pi m\int_{\mathcal{C}}d^{D-2} x\sqrt{\overbar{\sigma}_{n}}\,\overbar{\widehat{\mathcal{R}}}_{(m-1)}'\;,
    \label{JMfunctional}
\end{equation}
where $\overbar{\widehat{\mathcal{R}}}_{(m)}'$ is the Lovelock scalar of $\overbar{\sigma}_{n}$. The entropy $S = \lim_{n\rightarrow 1}\widetilde{S}_n$ is the analytic continuation of this expression to $n = 1$. Since \eqref{JMfunctional} is valid for integers $n>1$, the natural analytic continuation is to simply set $n = 1$. Hence we find
\begin{equation}
    S = 4\pi m\int_{\mathcal{C}}d^{D-2} x\sqrt{\overbar{\sigma}_{1}}\,\overbar{\widehat{\mathcal{R}}}_{(m-1)}\;,
    \label{JMfunctionaln1}
\end{equation}
which is the Jacobson--Myers functional of the induced metric $\overbar{\sigma}_1$ of $\mathcal{C}$ in $\overbar{g}_1$.


\paragraph{Extremization prescription.} As with the entropy functional,  the embedding of $\mathcal{C}$ in $\overbar{g}_{1}$ is expected to change for other theories of gravity. For Einstein gravity, the minimal surface condition is found by imposing Einstein's equations at leading order in an expansion near $\rho=0$ \cite{Lewkowycz:2013nqa}. Let us show how the leading term in the expansion of $E_{(m)\bar{z}}^z = 0$ determines the embedding of $\mathcal{C}$ in $\overbar{g}_1$. In Appendix \ref{app:lovelockextremization}, we show that the solution $\overbar{g}_n$ satisfies the condition
\begin{equation}
    \overbar{\mathfrak{E}}_{(m-1)ab}'\,\overbar{\mathcal{K}}_{0(p)}^{ab} = 0,\quad p =z,\overbar{z},\quad n>1\ ,
    \label{lovelockextremization}
\end{equation}
where $\overline{\mathfrak{E}}_{(m)ab}'$ is the equation of motion tensor \eqref{lovelockEOM} constructed from the corner projected Riemann tensor $\overline{\mathfrak{R}}_{abcd}'$ of $\overbar{g}_n$. Again for integer $n> 1$, the extrinsic curvatures of $\mathcal{C}$ vanish and \eqref{lovelockextremization} reduces to
\begin{equation}
    \overbar{\widehat{E}}_{(m-1)ab}\,\overbar{\mathcal{K}}_{0(p)}^{ab} = 0,\quad p =z,\overbar{z},\quad n>1\ ,
    \label{eq:lovelockextr}
\end{equation}
where $\overbar{\widehat{E}}_{(m)ab}$ is the equation of motion tensor of $\overbar{\sigma}_n$. 

The condition \eqref{eq:lovelockextr} was derived for integer $n> 1$ and the natural analytic continuation is to simply  set $n = 1$, however, this requires solving the splitting problem for the higher-order coefficients in pure Lovelock gravity that relates $\overbar{\mathcal{K}}_{0(p)ab} \lvert_{n=1}$ to the components of the metric $\overbar{g}_1$. This problem has not been solved for general Lovelock theories, though one might expect the answer is the same as in Einstein gravity \eqref{eq:curlyKK}, where $\overbar{\mathcal{K}}_{0(p)ab}\lvert_{n=1}$ simply equals the extrinsic curvature of $\mathcal{C}$ in $\overbar{g}_1$ as shown in \cite{Camps:2014voa}. In \cite{Camps:2014voa} it was also proven to hold in Gauss--Bonnet gravity when the Gauss--Bonnet coupling is treated perturbatively. In previous works \cite{Chen:2013qma,Bhattacharyya:2014yga,Erdmenger:2014tba} that study the extremization problem in Lovelock gravity, the equality was assumed to be true from the beginning. In the lack of a general proof, we will assume \eqref{eq:curlyKK} also holds for solutions of Lovelock gravity. Then equation \eqref{eq:lovelockextr} at $n = 1$ reduces to
\begin{equation}
    \overbar{\widehat{E}}_{(m-1)ab}\,\overbar{K}_{(p)}^{ab} = 0,\quad p =z,\overbar{z},\quad n = 1\ ,
\end{equation}
which matches the condition obtained by extremizing the Jacobson--Myers functional \eqref{JMfunctionaln1} with respect to the embedding of $\mathcal{C}$ in the on-shell $n = 1$ background $\overbar{g}_1$.\footnote{One can see this by computing the variation of \eqref{JMfunctionaln1} under a transverse diffeomorphism \eqref{inducedmetricvariation}.} 


\subsection{Entropy of fixed-JM-functional states}



The analog of an Einstein gravity fixed-area state in Lovelock gravity is a fixed-JM-functional state. Such a state is a solution $\overbar{g}_{\mathcal{S}}$ of equations of motion of Lovelock gravity for which the JM functional of $\mathcal{C}$ is fixed to a given value $\mathcal{S}$. We will now study the gravitational entropy of these states as we did for Einstein gravity in Section \ref{sec:einsteinFA}.

\paragraph{Variational principle for the metric.} Let $g_{\mathcal{S}}$ be an off-shell metric that satisfies the boundary condition (ii) \eqref{fursaev} with an induced metric $\sigma$ of $\mathcal{C}$ whose JM functional is fixed as
\begin{equation}
    4\pi m\int_{\mathcal{C}}d^{D-2} x\sqrt{\sigma}\,\widehat{\mathcal{R}}_{(m-1)}\equiv \mathcal{S}\ ,
\end{equation}
where $\widehat{\mathcal{R}}_{(m-1)}$ is the Lovelock scalar of $\sigma$. We consider the replica manifold $(\mathcal{M}_n,g_{\mathcal{S}})$, and as done in Section \ref{sec:einsteinFA}, we cut it open along a codimension-2 surface into a wedge shaped manifold $(\mathcal{W}_n,g_{\mathcal{S}})$. Then, 
\beq I_{\mathcal{M}_{n}}[g_{\mathcal{S}}]=I_{\mathcal{W}_{n}}[g_{\mathcal{S}}]+I_{\mathcal{C}}[g_{\mathcal{S}}]\;,\label{eq:wedgeactlovegen}\eeq
where 
the  action of $(\mathcal{W}_n,g_{\mathcal{S}})$ supplemented with boundary and corner terms is
\beq
\begin{split}
I_{\mathcal{W}_n}[g_{\mathcal{S}}] +I_{\mathcal{C}}[g_{\mathcal{S}}]&= -\int_{\mathcal{W}_n}d^{D}x\sqrt{g_{\mathcal{S}}}\,\mathcal{R}_{(m)}- 2m\int_{\mathcal{C}}d^{D-2}x\sqrt{\sigma}\,(\Theta_n + \pi)\,\widehat{\mathcal{R}}_{(m-1)}\\
&-\sum_{\alpha = 1}^{2}\biggl[\int_{\mathcal{B}_{\alpha}}d^{D-1}x\sqrt{h_{\alpha }}\,B_{(m)\alpha} +\int_{\partial \mathcal{B}_\alpha}\hspace{-2mm} d^{D-2}x\sqrt{\sigma}\,\mathcal{F}_{(m)}(L_\alpha,Q_\alpha)\biggr],
\label{lovelockwedgeaction}
\end{split}
\eeq
where $ \Theta_n = \pi\,(1-2n) $ as in \eqref{Thetan} and we have written the $\mathcal{F}_{(m)}$ terms of the corner term \eqref{Cmtext} as integrals over $\partial \mathcal{B}_\alpha$ instead of over $\mathcal{C}$.\footnote{Because $\mathcal{F}_{(m)}$ terms can be taken to be supported on $\partial\mathcal{B}_\alpha$, the corner term is additive under gluing of multiple wedges to each other \cite{Cano:2018ckq}.} As in Einstein gravity, we have also included a corner counterterm proportional to the JM functional whose coefficient is fixed to be $\pi$.

Using the general formula \eqref{generaltheoryvariation}, the variation of \eqref{lovelockwedgeaction} without imposing any boundary conditions is
\beq
\begin{split}
&\delta_g (I_{\mathcal{W}_n}[g] +I_{\mathcal{C}}[g])\\
&= -\int_{\mathcal{W}_n}d^{D}x\sqrt{g}\,E_{(m)ab}\,\delta g^{ab}- 2m\int_{\mathcal{C}}d^{D-2}x\sqrt{\sigma}\,(\Theta_n + \pi)\,\widehat{E}_{(m-1)ab}\,\delta \sigma^{ab}\\
&-\sum_{\alpha = 1}^{2}\biggl[\int_{\mathcal{B}_{\alpha}}d^{D-1}x\sqrt{h_{\alpha }}\,\widetilde{T}_{(m)\alpha ab}\,\delta h^{ab}_\alpha +\int_{\partial \mathcal{B}_\alpha}\hspace{-2mm} d^{D-2}x\sqrt{\sigma}\,\mathcal{F}_{(m)ab}(L_\alpha,Q_\alpha)\,\delta h^{ab}_\alpha\biggr],
\label{lovelockwedgevariation}
\end{split}
\eeq
where we have used that the corner stress tensor of Lovelock gravity is explicitly \eqref{lovelockcornerstress}. As discussed in Section \ref{sec:cornertermvariation}, we have not proven this variational formula directly, but we assume that the variation commutes with the sharp limit of the smoothing method so that we can use the corner stress tensor \eqref{lovelockcornerstress}.

For the angle $\Theta_n = \pi\,(1-2n) $, the rotation relation \eqref{rotextrLQtext} becomes
\begin{align}
L_{2AB} &= Q_{1AB}\sin{(2\pi n)}-L_{1AB}\cos{(2\pi n)}\\
Q_{2AB} &= Q_{1AB}\cos{(2\pi n)}+L_{1AB}\sin{(2\pi n)}
\label{rotextrLQtext2}
\end{align}
so that for integer $n$,
\begin{equation}
L_{2AB} = -L_{1AB},\quad Q_{1AB} = Q_{2AB}.
\end{equation}
Since $\mathcal{F}_{(m)}(-L,Q) = -\mathcal{F}_{(m)}(L,Q) $, we see that the boundary terms supported on $\mathcal{B}_\alpha$ and $\partial \mathcal{B}_\alpha $ in \eqref{lovelockwedgeaction} and \eqref{lovelockwedgevariation} cancel under the replica symmetry conditions \eqref{periodicBCnonpert} and periodic boundary conditions. Therefore, via (\ref{eq:wedgeactlovegen}), the action of $(\mathcal{M}_n ,g_{\mathcal{S}})$ becomes
\begin{equation}
I_{\mathcal{M}_n}[g_{\mathcal{S}}] = -\int_{\mathcal{M}_n}d^{D}x\sqrt{g_{\mathcal{S}}}\,\mathcal{R}_{(m)}-4\pi m\,(1-n)\int_{\mathcal{C}}d^{D-2}x\sqrt{\sigma}\,\widehat{\mathcal{R}}_{(m-1)}.
\label{lovelockreplicaaction}
\end{equation}
and its variation
\begin{equation}
\delta_g I_{\mathcal{M}_n}[g_{\mathcal{S}}] =-\int_{\mathcal{M}_n}d^{D}x\sqrt{g_{\mathcal{S}}}\,E_{(m)ab}\,\delta g^{ab}-4\pi m\,(1-n)\int_{\mathcal{C}}d^{D-2}x\sqrt{\sigma}\,\widehat{E}_{(m-1)ab}\,\delta \sigma^{ab}.
\label{lovelockreplicavariation}
\end{equation}

Consider then the fixed-JM-functional variational problem corresponding to metric variations that satisfy
\begin{equation}   
\delta\bigl(\sqrt{\sigma}\,\widehat{\mathcal{R}}_{(m-1)}\bigr)\big\lvert_{\mathcal{C}}\, = 0\ \quad \Longleftrightarrow \quad\widehat{E}_{(m-1)ab}\,\delta \sigma^{ab}\big\lvert_{\mathcal{C}}\, = 0\ .
\end{equation}
Under such variations, we see from \eqref{lovelockreplicavariation} that the term localized on $\mathcal{C}$ vanishes and the variational principle $ \delta_g I_{\mathcal{M}_n}[g_{\mathcal{S}}] = 0 $ imposes the equations of motion of Lovelock gravity on $g_{\mathcal{S}}$. Hence, the fixed-JM-functional variational problem is well defined on the replica manifold.


\paragraph{The entropy functional.} Having solved the fixed-JM-functional variational problem for $\overbar{g}_{\mathcal{S}}$, the action \eqref{lovelockreplicaaction} evaluated on $\overbar{g}_{\mathcal{S}}$ gives the refined Rényi entropy via
\begin{equation}
    \widetilde{S}_n = (n\partial_n-1)\,I_{\mathcal{M}_n}[\overbar{g}_\mathcal{S}],\quad (\text{fixed-JM-functional state}),
\end{equation}
assuming $I_{\mathcal{M}_n}[\overbar{g}_\mathcal{S}]$ is analytically continued to non-integer $n$. Notice that the extrinsic curvature terms localized at $\partial \mathcal{B}_\alpha $ do not cancel in the action \eqref{lovelockwedgeaction} if $ n $ is not an integer: for non-integer $n$ the action contains additional terms proportional to $\sin{(\pi n)}$ and it is well known that such terms constitute an ambiguity of the analytic continuation. This ambiguity is fixed by imposing growth conditions in the complex plane $ n \rightarrow \pm i \infty $ that forbids the appearance of $\sin{(\pi n)}$ terms, an application of Carlson's theorem for uniqueness of analytic continuation. Hence, the part of \eqref{lovelockwedgeaction} obeying these growth conditions is expected to be the correct (unique) analytic continuation of the action to non-integer $n$, given by \eqref{lovelockreplicaaction} with $n\in \mathbb{R}$. Thus,
\begin{equation}
\widetilde{S}_n = 4\pi m\,\int_{\mathcal{C}}d^{D-2} x \sqrt{\sigma}\,\widehat{\mathcal{R}}_{(m-1)}(\overbar{g}_{\mathcal{S}}) \equiv \mathcal{S}.
\label{donglovelock}
\end{equation}
This precisely coincides with the Camps--Dong functional of pure Lovelock gravity \cite{Dong:2013qoa}, i.e., the Jacobson--Myers entropy. Note, moreover, $\tilde{S}_{n}$ is independent of $n$, indicating fixed-JM-functional states have a flat spectrum.

\paragraph{Variational principle for the embedding.} The extremization prescription that determines the location of the codimension-2 surface $ \mathcal{C} $ works out, \emph{mutatis mutandis}, as in Einstein gravity (Section \ref{sec:einsteinFA}). For on-shell variations under a diffeomorphism $\xi$ \eqref{transversediffeo} transverse to $ \mathcal{C} $, the formula \eqref{lovelockreplicavariation} yields
\begin{equation}
\delta_\xi I_{\mathcal{M}_n}[g_{\mathcal{S}}] =-4\pi m\,(1-n)\int_{\mathcal{C}}d^{D-2}x\sqrt{\sigma}\,(\xi^{n}_{1}\,\widehat{E}_{(m-1)ab}\,L_1^{ab} + \xi^{r}_{1}\,\widehat{E}_{(m-1)ab}\,Q_1^{ab}).
\end{equation}
Requiring this variation to vanish gives the conditions
\begin{equation}
\widehat{E}_{(m-1)ab}\,L_1^{ab} = \widehat{E}_{(m-1)ab}\,Q_1^{ab} = 0,
\label{regularextremization}
\end{equation}
which match the conditions obtained from the variation of the JM functional itself.

\subsection{Generalizations to higher-curvature theories of gravity}\label{subsec:FRiemannentropies}

In the previous two sections, we analyzed gravitational entropy of Hartle--Hawking and fixed-JM-functional states in Lovelock gravity. We will now attempt to generalize the analysis to $F$(Riemann) theories of gravity.

\paragraph{Generalizations of fixed-area states.} The existence of states analogous to fixed-area and fixed-JM-functional states in general higher-curvature gravity requires the existence of a corner term. A general $F$(Riemann) theory of gravity does not admit a corner term and a well defined variational principle in the presence of corners, assuming Dirichlet boundary conditions alone. Hence, the action of a fixed-functional state cannot be defined in general. 

\paragraph{Hartle--Hawking states.} For Einstein and Lovelock gravities, we used the action of the wedge manifold $(\mathcal{W},g_n)$ to \textit{define} the action of the $\mathbb{Z}_n$-quotient manifold $(\mathcal{M}\,\backslash \,\mathcal{C},g_n)$ that computes Hartle--Hawking entropy. Let us attempt to do the same for arbitrary $F$(Riemann) gravity and consider the action \eqref{Friemann} of $(\mathcal{W},g_n)$ without including boundary terms. The variation of the action with respect to the metric without imposing any boundary conditions has been computed in Appendix \ref{app:actionvariation} and the result is given in \eqref{eq:actionvarnoBCs}. 
Imposing also periodic boundary conditions \eqref{periodicBCpert} on the metric variation implies vanishing of certain corner localized terms and what remains is (for details see Appendix \ref{app:actionvariation}, namely, \eqref{eq:actionvarnoBCs})
\begin{equation}
    \delta_g I_{\mathcal{W}}[g_n] =  -\int_{\mathcal{W}} \hspace{-1mm} d^{D} x \sqrt{g_n}\,E_{ab}'\,\delta g^{ab}-\sum_{\alpha =1}^{2}\int_{\mathcal{B}_{\alpha}} \hspace{-2mm} d^{D-1} x \sqrt{h_\alpha}\,n_{\alpha a}\,\deltabar W^{a}_{\alpha}+\int_{\mathcal{C}}d^{D-2} x\sqrt{\sigma}\, \Psi'\,\delta_g\Theta_n
    \label{cornervarPBCtext}
\end{equation}
where $E_{ab}'$ is the equation of motion tensor, $\Psi'$ is the Wald entropy density of the metric $g_n$, and boundary variation \cite{Jiang:2018sqj}
\begin{align}
	n_{\alpha a}\,\deltabar W^{a}_{\alpha} =-4\Psi_{\alpha ab}\,\delta K^{ab}_{\alpha}+(2n^{a}_{\alpha}\nabla^{d}P_{\alpha acbd}+6\Psi_{\alpha ab}\,K^{a}_{\alpha c})\,\delta h^{bc}_{\alpha} - 2n^{d}_{\alpha}P_{\alpha acbd} D^{a}_{\alpha}\delta h^{bc}_{\alpha},
 \label{eq:boundvariation}
\end{align}
with $\Psi_{ab}$ defined in \eqref{Psiabtext}.


Recall that for Einstein and Lovelock gravities we included boundary terms to the action of the wedge. We did this such that the boundary terms of the variation of the action are proportional to the boundary stress tensor $\widetilde{T}_{ab}\,\delta h^{ab}$, which cancel (for integer $n$) because $\delta g^{ab}$ satisfies the periodic boundary condition \eqref{periodicBCpert}. In general $F$(Riemann) gravity, however, we cannot do this since boundary terms for Dirichlet boundary conditions do not necessarily exist. When the theory admits a Dirichlet boundary term $B$, as in (\ref{startingaction}), such that the total boundary variation is
\begin{equation}
    \delta(\sqrt{h_\alpha}\,B_\alpha) + \sqrt{h_\alpha}\,n_{\alpha a}\,\deltabar W^{a}_{\alpha} =\sqrt{h_\alpha}\,\Bigl[\widetilde{T}_{\alpha ab}\,\delta h^{ab}_\alpha + (\text{total derivative})\Bigr],
    \label{eq:boundarytermincludedvar}
\end{equation}
then generic periodic boundary conditions which remove the total boundary variation (\ref{eq:boundarytermincludedvar}) are
\begin{gather}
    h_{1ab}\lvert_{\mathcal{B}_1}\, = h_{2ab}\lvert_{\mathcal{B}_2},\quad \widetilde{T}_{1ab}\lvert_{\mathcal{B}_1}\, = -\widetilde{T}_{2ab}\lvert_{\mathcal{B}_2}\nonumber\\
    \delta h_{1}^{ab}\lvert_{\mathcal{B}_1}\, = \delta h_{2}^{ab}\lvert_{\mathcal{B}_2}\;.
    \label{eq:generalperiodic}
\end{gather}
For Einstein and Lovelock gravities, continuity of the normal derivative of the metric $K_{1ab}\lvert_{\mathcal{B}_1}\, = -K_{2ab}\lvert_{\mathcal{B}_2}$ implies continuity of the boundary stress tensor, in line with (\ref{eq:generalperiodic}), however, this may not be true in more general theories where continuity of higher-order derivatives is also required.\footnote{This is related to the definition of junction conditions in higher-curvature gravities \cite{Balcerzak:2007da}.} If the theory does not admit Dirichlet boundary terms it is not clear what type of periodicity conditions ensure cancellation of $n_{\alpha a}\,\deltabar W^{a}_{\alpha}$ in \eqref{cornervarPBCtext}.




It is amusing to ignore this problem and proceed by assuming the boundary variations in \eqref{cornervarPBCtext} can be cancelled by some periodic boundary conditions without necessarily including boundary terms to the action. Namely, with
\beq n_{1a}\,\deltabar W^{a}_{1}=-n_{2a}\,\deltabar W^{a}_{2a}\;.\label{eq:specialpbc}\eeq
and setting the metric $g_n$ on-shell yields
\begin{equation}
    \delta_gI_{\mathcal{M}\,\backslash\,\mathcal{C}}[\overbar{g}_n] = \delta_g I_{\mathcal{W}}[\overbar{g}_n] =\int_{\mathcal{C}}d^{D-2} x\sqrt{\overbar{\sigma}_n}\, \overbar{\Psi}'\,\delta_g\overbar{\Theta}_n\;,
    \label{eq:onshellvarwald}
\end{equation}
where $\overbar{\Psi}'$ is the Wald entropy density associated with on-shell metric $\overbar{g}_n$. Note that $\overbar{\Psi}'$ is finite at $\mathcal{C}$ provided $n\geq 2$: the Riemann tensor components of a Hartle--Hawking metric $g_n$ can be divergent only within the interval $1 < n < 2$. Hence, \eqref{eq:onshellvarwald} is valid for all integers $n\geq 2$ and we get the refined R\'enyi entropy functional
\begin{equation}
    \widetilde{S}_n = 2\pi\int_{\mathcal{C}}d^{D-2} x\sqrt{\overbar{\sigma}_n}\,\overbar{\Psi}',\quad n\geq 2.
    \label{firstentropygeneralgravity}
\end{equation}
This is simply the Wald entropy of $\mathcal{C}$ in the metric $\overbar{g}_n$.\footnote{This expression simplifies further since some components of the Riemann tensor of $\overbar{g}_n$ and extrinsic curvatures of $\mathcal{C}$ in $\overbar{g}_n$ vanish for $n\geq 2$. We list the non-zero components in \eqref{eq:gntensorsapp}, such that \eqref{firstentropygeneralgravity} will be given terms of $\mathcal{U}_{0A}$, $\overline{\mathcal{T}}_{0,0}$, $\overbar{\mathcal{Q}}_{0(pq)AB}$ and $\overbar{\widehat{R}}_{ABCD}'$.}

It remains to perform the analytic continuation of \eqref{firstentropygeneralgravity} to $n = 1$ to recover entropy $S$. The natural analytic continuation is to simply set $n = 1$ which gives
\begin{equation}
    S = \int_{\mathcal{C}}d^{D-2} x\sqrt{\overbar{\sigma}_1}\,\overbar{\Psi}'\big\lvert_{n = 1}.
    \label{eq:donglewkowycz}
\end{equation}
This is precisely the entropy functional (\ref{eq:Lewkdong}) of Dong and Lewkowycz \cite{Dong:2017xht}, originally derived using the boundary term method we reviewed in Section \ref{subsec:boundarytermmethod} for Einstein gravity. 

As we discussed in the introduction, it remains to be seen whether the Dong--Lewkowycz functional coincides with the Camps--Dong functional (except in Lovelock gravity where they are both equal to the Jacobson--Myers functional).
Recall the Camps--Dong formula is a sum of the Wald entropy functional of $\mathcal{C}$ in the metric $\overbar{g}_1$ and an anomaly term that depends on the extrinsic curvatures of $\mathcal{C}$ in $\overbar{g}_1$. Writing the Dong--Lewkowycz functional \eqref{eq:donglewkowycz} in terms of the Riemann tensor of $\overbar{g}_1$ requires solving the splitting problem in $F$(Riemann) theories of gravity which is not guaranteed to have a solution. For the splitting problem to be well posed, the equations of motion of the theory should determine all coefficients of the Hartle--Hawking metric $g_n$ in the expansion near $\mathcal{C}$. Since in general the equations of motion are not even second order, it is not clear if this will be true. If we include an Einstein term to the action and treat higher-curvature corrections to Einstein gravity perturbatively we can assume that $\overbar{g}_n$ is a solution of only Einstein's equations. Einstein gravity splittings were applied in \cite{Bueno:2020uxs} to rewrite the Camps--Dong formula in a simple form and it is an interesting future problem to do this for the Dong--Lewkowycz functional \eqref{eq:donglewkowycz} to check if they match.


Lastly, let us consider the extremization prescription for the entropy functionals in $F$(Riemann) gravity. 
In general, one expects the coefficient of the leading divergence in the equations of motion to give a condition for the embedding of $\mathcal{C}$ in $\overbar{g}_1$ when analytically continued to $n = 1$. There are two issues with this approach. First, due to their higher-order nature, it is not clear if the equations of motion always give a well defined condition on the embedding. Second, the analytic continuation of the conditions to $n = 1$ requires solving the splitting problem in general (and in particular for the $\mathcal{K}_{(p)AB}$ as at the end of Section \ref{sec:HHlovelock}). The conditions coming from the equations of motion of quadratic $F$(Riemann) gravities and Lovelock gravity were considered in \cite{Chen:2013qma,Bhattacharyya:2014yga,Erdmenger:2014tba,Camps:2014voa} and we plan to return to this issue in future work.

\section{Discussion} \label{sec:discussion}

In this article we derived gravitational entropy functionals for higher-curvature theories of gravity using Dirichlet corner terms that are required to ensure the well-posedness of the variational principle in the presence of corners. Our key insight was to recognize that, at the level of the action, a manifold with a conical singularity is equivalent to a wedge shaped manifold with boundaries intersecting at a codimension-2 corner. This is because when we cut open the conical singularity along a codimension-1 surface into a wedge, only a sliver of measure zero is removed from the integration region of the conically singular manifold, having no effect on the gravitational action. Our approach also allows for a careful treatment of the variational problems for the metric and the embedding of the conical defect $\mathcal{C}$ which we have used to determine the extremization prescription of the entropy.

We focused on computing entropy of two different types of gravitational states, Hartle--Hawking states and fixed-area states (or their higher-curvature analog). In the case of Lovelock gravity, the entropy of Hartle--Hawking states was shown to equal the Jacobson--Myers functional, consistent with previous work \cite{Hung:2011xb,deBoer:2011wk}, and coincides with the Camps--Dong holographic entanglement entropy functional. Via our method the variational principle of the wedge action is well-posed,
where Lovelock’s field equations provide a surface constraint for $\mathcal{C}$ that matches the extremization of the JM
functional. Moreover, we showed fixed-area states naturally generalize to fixed-JM states, where the Jacobson--Myers functional of $\mathcal{C}$ is held fixed, i.e., states prepared by a Euclidean gravity path integral over metrics with a fixed JM functional at $\mathcal{C}$ in the interior. As for Einstein gravity, the refined R\'enyi entropy is independent of $n$, having a flat entanglement spectrum. Further, the extremization prescription follows from the variation of the Lovelock
action of the wedge and is equivalent to extremization
of the Jacobson--Myers functional.

For arbitrary theories of  $F$(Riemann) gravity,
corner terms are not known to exist assuming Dirichlet
boundary conditions alone (corner terms can be uncovered if one relaxes strictly working with Dirichlet boundary conditions). Thus, the Dirichlet variational problem in
the presence of corners is not well-posed.  Consequently, our method
suggests fixed-area state analogs do not exist in arbitrary higher-curvature theories. For Hartle--Hawking states, however, under special periodic boundary
conditions, we derived the Dong–Lewkowycz entropy functional, i.e., the Iyer--Wald entropy evaluated in $\overbar{g}_{n}$. To check whether this functional corresponds to the Camps--Dong formula in general requires one to solve the splitting problem in $F$(Riemann) gravity, which remains open. Combined, our work suggests the existence of entropy functionals is inextricably linked to the existence of boundary/corner terms which make the Dirichlet variational problem well-posed.




\vspace{2mm}

There are multiple possible future directions to take this work and apply the corner method we developed here, some of which we list below. 

\vspace{2mm}



\vspace{2mm}

\paragraph{Pontryagin gravity type theories.} In this work we focused on $F(\text{Riemann})$ theories of gravity whose actions do not involve the dual of the Riemann tensor, which would require inclusion of the Levi--Civita tensor to the Lagrangian. The simplest example is Pontryagin gravity in four-dimensions \cite{Jackiw:2003pm,Alexander:2009tp}, but there exists   an infinite family of such theories in higher dimensions \cite{Zanelli:1999fs,Zanelli:2012px}. Pontryagin gravity is known to have a Dirichlet boundary term \cite{Grumiller:2008ie} which suggests that it should also have a corner term obtained by taking a sharp limit of the boundary term. The detailed analysis will be presented elsewhere.



\vspace{2mm}

\paragraph{Dynamical black hole entropy.} A relevant example of a spacetime without a $U(1)$ symmetry is a dynamical black hole, i.e., a black hole with a non-stationary horizon. The corner method employed here, in principle, should be able to supply a definition of dynamical black hole entropy. Notably, the Camps--Dong formula for holographic entanglement entropy obeys the second law \cite{Chatterjee:2011wj,Kolekar:2012tq,Sarkar:2013swa,Bhattacharjee:2015yaa,Wall:2015raa}, at least for linear non-stationary perturbations to stationary horizons, and is thus a candidate for dynamical entropy.\footnote{Wall's construction \cite{Wall:2015raa} does not imply whether the Camps--Dong entropy \emph{increases}, just that it does not change over time (unless one matter obeying the null energy condition). Further, Wall's formula does not agree with the Jacobson--Kang--Myers entropy \cite{Jacobson:1995uq} (which was shown to obey the second law) for arbitrary $f(R)$ gravity; they only agree for $R^{2}$ theories in the linear approximation. For effective higher-curvature theories, Walls method was extended to higher order in perturbations by adding terms (at least) quadratic in quantities with positive boost weight \cite{Hollands:2022fkn,Davies:2022xdq,Davies:2023qaa}.} It would be interesting to extend our analysis here to see whether dynamical black hole entropy follows from a corner term, and whether it coincides with the Camps--Dong formula in its regime of applicability.

\vspace{2mm}

\paragraph{Covariant entropy} An important extension of the RT prescription is its covariant version \cite{Hubeny:2007xt,Wall:2012uf,Dong:2016hjy} accounting for time-dependent CFT states. We expect the corner method is capable of being appropriately modified to account for this extensions. We would need to consider Lorentzian spacetimes with boundaries that contain both timelike and spacelike corners. Such corner terms have already been shown to exist in Lovelock gravity \cite{Cano:2018ckq}, and it is a matter of seeing how the covariant extremization prescription arises from the appropriate wedge action. It would also be worth seeing if the corner method developed here gives additional insight in understanding gravitational entropy at the level of an algebra of observables \cite{Colafranceschi:2023urj}.

\vspace{2mm}

\paragraph{Quantum corrections.} The RT formula has a quantum-corrected generalization \cite{Faulkner:2013ana,Engelhardt:2014gca,Akers:2019lzs}. When accounting for quantum corrections, our approach suggests to formulate the computation of quantum effective actions in a wedge and analyse possible corner contributions (effective actions on squashed cone manifolds were considered in \cite{Dowker:1994bj}). This includes computing heat kernel coefficients in wedges and computing the logarithmic contribution to entanglement entropy of conformal and non-conformal field theories alike. It would be worth comparing this outlined approach to the derivation given in \cite{Dong:2017xht}, where the extremization prescription follows from the variational principle of the action and the entropy functional also lacks an anomaly term. 

\vspace{2mm}

\paragraph{Islands and entropy of Hawking radiation.} The island formula -- a particular application of the quantum-corrected RT formula extremizing the generalized entropy -- suggests the von Neumann entropy of Hawking radiation follows a unitary Page curve \cite{Penington:2019npb,Almheiri:2019psf,Almheiri:2019yqk,Almheiri:2019hni}. A derivation of the island formula using the replica trick has so far only been explicitly carried out for two-dimensional Jackiw--Teitelboim (JT) gravity \cite{Almheiri:2019qdq,Penington:2019kki,Goto:2020wnk,Colin-Ellerin:2021jev}, where the transition in the Page curve is understood as a competition between two different saddles dominating the path integral. It would be interesting to extend the corner method to compute the entropy of Hawking radiation to elucidate competition of saddles from another perspective. 
It would also be interesting to compare to an alternative derivation of the island formula \cite{Pedraza:2021ssc,Svesko:2022txo}, which 
follows from minimizing the gravitational microcanonical action of an entanglement wedge.

\vspace{2mm}

\paragraph{Entropy, corners, and edge modes.} We have derived gravitational entropy using corner terms but we have not addressed its microphysical origins. Via AdS/CFT, the gravitational entropy has a dual meaning as the entanglement entropy of a holographic CFT. A more gravitational perspective is that the entropy has a statistical interpretation in terms of entanglement of `edge modes' (see, e.g., \cite{Kabat:1995eq,Donnelly:2011hn,Donnelly:2012st,Donnelly:2016auv,Speranza:2017gxd}). Edge modes arise due to the nonfactorization of the Hilbert space since gauge constraints relate entangling degrees of freedom across tensor factors of the Hilbert space.
Hence, there is a connection between the existence of corners, edge modes and entropy, and it was argued in \cite{Takayanagi:2019tvn}, in the context of Einstein gravity, that the Hayward corner term strengthens the notion gravitational entropy emerges due to edge modes
Our work extends this view to higher-curvature gravities which admit a corner term. It would be interesting to make the edge mode interpretation more precise. 

\vspace{2mm}

\paragraph{Distributional geometry and other applications.} One of our key observations is that the action of a
conical singularity does not require regularization if one instead works with the action of a wedge. Notably, via the wedge action (possibly supplemented with a corner term), we derived the distributional nature of the Einstein-Hilbert term without producing higher-order regularization terms. In light of this observation, it may be of interest to revisit the regularization procedure  developed to analyze integrals of curvature invariants on manifolds with a squashed conical excess \cite{Fursaev:2013fta}, and its applications, e.g., quantum corrections to entanglement entropy. 





\noindent \noindent\section*{Acknowledgments}
We are grateful to Pablo Bueno, Luca Ciambelli, Elena C\'{a}ceres, Xi Dong, and Alejandro Vilar L\'{o}pez for discussions and useful correspondence. We especially thank Manus Visser for many fruitful discussions and initial collaboration. JK is supported by the Deutsche Forschungsgemeinschaft (DFG, German Research Foundation) under Germany’s Excellence Strategy through the W\"{u}rzburg-Dresden
Cluster of Excellence on Complexity and Topology in Quantum Matter - ct.qmat (EXC 2147, project-id 390858490), as well as through the German-Israeli Project Cooperation (DIP) grant ‘Holography and the Swampland’. JK thanks the Osk. Huttunen foundation and the Magnus Ehrnrooth foundation for support during earlier stages of this work. AS is supported by STFC consolidated grant ST/X000753/1 and was partially supported by the Simons Foundation via \emph{It from Qubit Collaboration} and by EPSRC when this work was initiated. AS thanks the Isaac Newton Institute for Mathematical Sciences, Cambridge, for support
and hospitality during the program \emph{Black holes: bridges
between number theory and holographic quantum information} (supported by EPSRC grant no EP/R014604/1) as this work was being completed.

\begin{appendix}

\section{Geometry of replicated manifolds and squashed cones} \label{app:repmanifolds}


Here we summarize the geometric features of replicated manifolds and squashed cones, including an explicit computation of their curvature components, generalizing previous results in the literature. 


\subsection{Replicated manifolds and squashed cones: a brief history}

Consider a $D$-dimensional Riemannian manifold endowed with Euclidean metric $g_{ab}$. Denote the coordinates on the manifold by $X^{a}$ with $a,b,\ldots=1,2,\ldots,D$. Assume the manifold has a codimension-2 surface, e.g., a conical defect or corner, $\mathcal{C}$, such that near $\mathcal{C}$, the line element locally takes the form (we describe the construction more carefully momentarily)
\beq ds^{2}=g_{ab}\,dX^{a}dX^{b}=\hat{g}_{pq}\,dy^{p}dy^{q}+h_{AB}\,d\hat{x}^{A}d\hat{x}^{B}+2V_{A}\,\rho^{2}\,d\tau d\hat{x}^{A}+\ldots\;.\label{eq:metgenapp}\eeq
Here the metric $\hat{g}_{pq}(y)=e^{2A(y)}g_{pq}(y)$ is the metric transverse to $\mathcal{C}$, described by Cartesian coordinates $y^{p}$, with indices $p,q=1,2$. Typically one works with polar coordinates $y^{p}=\{y^{1},y^{2}\}=\{\rho\cos\tau,\rho\sin\tau\}$, or complex coordinates $y^{p}=\{y^{1},y^{2}\}=\{z,\bar{z}\}=\{\rho e^{i\tau},\rho e^{-i\tau}\}$.\footnote{In terms of complex coordinates $(z,\bar{z})$, note that $dz=e^{i\tau}d\rho+izd\tau$ and $d\bar{z}=e^{-i\tau}d\rho-i\bar{z}d\tau$ such that $dzd\bar{z}=d\rho^{2}+\rho^{2}d\tau^{2}$, and $\bar{z}dz-zd\bar{z}=2i\rho^{2}d\tau$.} The coordinates of $\mathcal{C}$ are denoted by $\hat{x}^{A}$ for $A,B=1,\ldots,D-2$, where $h_{AB}$ and $V_{A}$ are at the moment generic tensors which depend on $(y^{p},\hat{x}^{A})$. The ellipsis refers to terms which ultimately can be ignored. 

Manifolds with a metric of the form (\ref{eq:metgenapp}) include both what we dub non-singular replicated manifolds and singular squashed cones. First consider non-singular replicated manifolds. To this end, one begins by constructing a special foliation of a base manifold $\mathcal{M}$ that encodes both the geometric structure of the background and $\mathcal{C}$. Such foliations can only be found perturbatively in the distance $y$ from $\mathcal{C}$.\footnote{At sufficient distances from $\mathcal{C}$ the coordinate system $(y^{p},\hat{x}^{A})$ breaks down due to caustics.} To leading order in distance from $\mathcal{C}$, the metric on $\mathcal{M}$ takes the form \cite{Fursaev:2013fta,Rosenhaus:2014woa},
\beq ds^{2}=d\rho^{2}+\rho^{2}d\tau^{2}+(\sigma_{AB}(\hat{x})+2y^{p}\,K_{(p)AB}(\hat{x})+\ldots)\,d\hat{x}^{A}d\hat{x}^{B}+2V_{A}\,\rho^{2}\,d\tau d\hat{x}^{A}+\ldots\;.\label{eq:nonsingrepmanapp}\eeq
Such a spacetime can be explicitly derived by employing a Gaussian normal coordinate expansion in $(\rho,\hat{x}^{A}$) with origin ($\rho=0$) at $\mathcal{C}$.\footnote{More precisely, select coordinates $\hat{x}^{A}$ along $\mathcal{C}$. Then, the two-dimensional space transverse to $\mathcal{C}$ is filled by geodesics radiating orthogonally from any point in $\mathcal{C}$. By construction, $g_{\rho A}=g_{\rho\tau}=0$.} Here $\sigma_{AB}$ is the induced metric of $\mathcal{C}$ embedded in $\mathcal{M}$, with extrinsic curvatures $K_{(1)AB}$ and $K_{(2)AB}$.\footnote{We use $K_{(p)AB}$ for notational compactness. For example, $K_{(1)AB}=L_{AB}$ and $K_{(2)AB}=Q_{AB}$.} 
The replicated manifold $(\mathcal{M}_{n},g)$ is obtained by periodically gluing integer $n\geq1$ copies of this space with cuts along a codimension-1 surface that end on $\mathcal{C}$.
 Notably, the replicated manifold $(\mathcal{M}_{n},g)$ has the same the foliation (\ref{eq:nonsingrepmanapp}), having no explicit dependence on $n$. The Euclidean time $\tau$, however, gains periodicity $\tau\sim \tau+2\pi n$ upon replication. Consequently, for integer $n>1$, the geometry (\ref{eq:nonsingrepmanapp}) acquires conical defects at $\rho=0$, with angular \emph{excess} $\Delta\tau=2\pi n-2\pi$. Historically, such a conical geometry is called a `squashed cone' \cite{Dowker:1994bj,Fursaev:2013fta} since it has a discrete group of transformations $\tau\to\tau+2\pi k$ for integer $k$, opposed to spacetimes with conical singularities such that $\mathcal{C}$ is a fixed point set of a global $U(1)$ isometry (as in the case of stationary black holes, e.g., \cite{Banados:1993qp,Carlip:1993sa,Fursaev:1995ef}).\footnote{A simple physical example distinguishing a squashed cone from a $U(1)$ symmetric cone is spherical versus planar Rindler space. In planar (static) Rindler space there is a Killing frame of reference such that the reduced density matrix of the Rindler wedge may be cast in terms of a local Rindler Hamiltonian generating Rindler time-translations. Consequently, non-integer powers of the reduced state can be geometrically described in terms of a standard Rindler metric with non-integer periodicity in Euclidean time. Alternatively, spherical Rindler space (a special case of (\ref{eq:nonsingrepmanapp})) is time dependent, however, globally, retains a Euclidean time-translation symmetry $\tau\to\tau+2\pi k$ for any integer $k$. Thus, only integer powers of the reduced density matrices of spherical Rindler observers have a geometric representation, i.e., spherical Rindler space whose Euclidean time coordinate is periodic for positive integer $n$.\label{ft:replica}} Since $2y^{p}K_{(p)AB}$ enter at linear order in $\rho$, the $U(1)$ isometry is broken at this order.

To remove the conical defect on $(\mathcal{M}_{n},g)$, the geometry (\ref{eq:nonsingrepmanapp}), with $\tau\sim\tau+2\pi n$, 
may be regularized by introducing a regularization function $f_{n}(\rho,b)=\frac{\rho^{2}+n^{2}b^{2}}{\rho^{2}+b^{2}}$ as the $g_{\rho\rho}$ metric coefficient with regularization parameter $b$  \cite{Fursaev:1995ef,Fursaev:2013fta}. Near $\rho\approx0$ the regularized metric is 
\beq ds^{2}=n^{2}d\rho^{2}+\rho^{2}d\tau^{2}+(\sigma_{AB}+2\rho\cos{\tau}\, K_{(1)AB}+2\rho\sin{\tau}\, K_{(2)AB})\,d\hat{x}^{A}d\hat{x}^{B}+\ldots\;.\label{eq:singrepmanapp}\eeq
This metric on $\mathcal{M}_{n}$ no longer has a conical excess at $\mathcal{C}$, however, the $n$-dependence acquired by the $g_{\rho\rho}$ metric component yields divergent curvature invariants and equations of motion at $\rho=0$, and is thus an example of a `singular squashed cone'. 
To counteract the curvature singularities, a strategy is to replace the $\rho$ in the $AB$ metric components with $\rho^{n}c^{n-1}$ for an irrelevant constant $c$. Doing so leads to a regular metric in that the resulting geometry has no conical or curvature singularities at $\rho=0$ when $n>1$, dubbed a `regularized squashed cone' \cite{Fursaev:2013fta}.\footnote{To clarify, singularities in integrals of quadratic curvature invariants disappear when $n>1$, however, the singularity in the metric will disappear when $n>2$.} The Hartle--Hawking solution described in the main text belongs to this family of regularized cones: gravity performs the regularization dynamically.  Note that both squashed cone metrics (\ref{eq:nonsingrepmanapp}) and (\ref{eq:singrepmanapp}) depend explicitly on $\cos\tau$ and $\sin\tau$. Hence, the geometry cannot be understood at arbitrary non-integer values because a jump will occur in the extrinsic curvatures on the $\tau=0$ and $\tau=2\pi n$ hypersurfaces. This suggests an ordering with the replica trick: first perform all computations at integer values of $n$, and then analytically continue $n$ at the end of the calculation (consistent with Footnote \ref{ft:replica}).


Another conical geometry often considered is (see, e.g., Appendix B of \cite{Dong:2013qoa})
\beq ds^{2}=d\rho^{2}+\rho^{2}d\tau^{2}+2V_{A}\,\rho^{2}\,d\tau d\hat{x}^{A}+T\,\rho^{4}\,d\tau^{2}+(\sigma_{AB}+H_{AB})\,d\hat{x}^{A}d\hat{x}^{B}+\ldots\;,\label{eq:repmanDongv1}\eeq
where $\sigma_{AB}(\hat{x})$ is the induced metric of $\mathcal{C}$, and the metric functions $T, V_{A}$ and $H_{AB}$ are all specific expansions in $\rho^{2}$ and $\rho^{n}e^{\pm in\tau}$. In the literature, various expansions of the metric functions have been given, however, determining the correct expansion is intimately related to the `splitting problem', as we describe in the main text and in detail in Appendix \ref{app:solvingsplittingprobs}. Therefore, here we refrain from introducing the explicit form of the metric function expansions, and report the metric at leading order,
\beq ds^{2}=d\rho^{2}+\rho^{2}d\tau^{2}+(\sigma_{AB}+2\mathcal{K}_{(1)AB}\,\rho^{n}e^{in\tau}+2\mathcal{K}_{(2)AB}\,\rho^{n}e^{-in\tau})\,d\hat{x}^{A}d\hat{x}^{B}+\ldots\;,\label{eq:squashedconedonglead}\eeq
where we have momentarily neglected the $V_{A}\,\rho^{2}$ and $T\,\rho^{4}$ terms.  

It is worth comparing conical geometries (\ref{eq:nonsingrepmanapp}) and (\ref{eq:squashedconedonglead}). Both are valid metrics to be placed on the replicated space $\mathcal{M}_{n}$, each having a conical defect with the same angular excess $\Delta\tau$.\footnote{Upon orbifolding, $\mathcal{M}_{n}/\mathbb{Z}_{n}$, where $\tau\sim \tau+2\pi$, either geometry is regular. This is manifestly true for (\ref{eq:nonsingrepmanapp}) and also true for (\ref{eq:squashedconedonglead}) with positive integers $n$.} 
An obvious difference is that unlike the squashed cone (\ref{eq:nonsingrepmanapp}), the geometry (\ref{eq:squashedconedonglead}) explicitly depends on $n$. Consequently, the metric functions $\mathcal{K}_{(p)AB}$ are \emph{not} to be identified with extrinsic curvatures of $\mathcal{C}$ (except when $n=1$). Further,  metric (\ref{eq:squashedconedonglead}) is a squashed cone, though it breaks $U(1)$ isometry at higher than linear order in $\rho$ (when $n>1$).
Moreover, the geometry (\ref{eq:squashedconedonglead}) has a curvature singularity  when $1< n < 2$ (for integer $n\geq 2$ there is no singularity) which propagates into the equations of motion. Maintaining regularity thus requires imposing an additional set of conditions on the metric, which follow from solving the most divergent contributions to the equations of motion. Lastly, a physical distinction between metrics (\ref{eq:nonsingrepmanapp}) and (\ref{eq:squashedconedonglead}) is that the former is used when evaluating the entropy of fixed-area states (\ref{fursaev}), while the latter, upon rescaling $\tau\to\tau/n$, is used when computing the entropy of Hartle--Hawking states (\ref{maldacena}). 

For completion, as we further describe in the main text, Eq. (\ref{eq:metreghigh}), another common form of a regularized squashed cone is to take metric (\ref{eq:repmanDongv1}) and perform the coordinate transformations $\rho\to n\rho^{1/n}$ and $\tau\to \tau/n$, resulting in, to leading order (see Appendix B of \cite{Dong:2013qoa})
\beq 
\begin{split}
ds^{2}=e^{2A}\,(d\rho^{2}+\rho^{2}d\tau^{2})+&e^{4A}\,T\,\rho^{4}d\tau^{2}+2e^{2A}\,V_{A}\,\rho^{2}\,d\tau d\hat{x}^{A}\\
&+(\sigma_{AB}+2y^{p}\,\mathcal{K}_{(p)AB})\,d\hat{x}^{A}d\hat{x}^{B}+\ldots\;,\label{eq:singrepmanv2app}
\end{split}
\eeq
with $A(\rho)=-\epsilon\log\rho$ and $\epsilon=1-1/n$. The orbifold $\mathcal{M}_{n}/\mathbb{Z}_{n}$, where $\tau\sim \tau+2\pi$ has a conical deficit $\Delta\tau=2\pi-2\pi/n=2\pi\epsilon$.
Regularizing the conical singularities of (\ref{eq:singrepmanv2app}) is subtle. Initially, regularization was dealt with by appropriately modifying the conformal factor $e^{2A}$, namely, $A\to-\epsilon\log(\rho^{2}+a^{2})/2$ for $a>0$. Incidentally, however, this regularization method of the squashed cone metric considered by Dong (Eq. (3.22) of \cite{Dong:2013qoa}) is not necessarily a solution to the equations of motion of the gravitational theory in question \cite{Camps:2014voa,Miao:2014nxa,Miao:2015iba,Camps:2016gfs}.

\subsection{Curvature of squashed cones}

\noindent \textbf{General metric.} Let us now work out various curvature relations relevant for this work and beyond. It proves beneficial to work in holomorphic coordinates $(z,\bar{z})$.
We consider a metric of the general form
\beq 
\begin{split}
ds^{2}&=fdzd\bar{z}+h\,(\bar{z}dz-zd\bar{z})^{2}+(\sigma_{AB}+H_{AB})\,d\hat{x}^{A}d\hat{x}^{B}+2iV_{A}\,(\bar{z}dz-zd\bar{z})\,d\hat{x}^{A}\;,    
\end{split}
\label{eq:genmetholo}\eeq
with components
\beq
\begin{split}
 &g_{zz}=h(z,\bar{z},\hat{x})\,\bar{z}^{2}\;,\quad g_{\bar{z}\bar{z}}=h(z,\bar{z},\hat{x})\,z^{2}\;,\quad g_{z\bar{z}}=\frac{1}{2}f(z,\bar{z})-z\bar{z}\,h(z,\bar{z},\hat{x})\\
 &g_{AB}=\sigma_{AB}(\hat{x})+H_{AB}(z,\bar{z},\hat{x})\;,\quad g_{zA}=i\bar{z}\,V_{A}(z,\bar{z},\hat{x})\;,\quad g_{\bar{z}A}=-iz\,V_{A}(z,\bar{z},\hat{x})\;.
\end{split}
\eeq
Here we treat $(z,\bar{z})$ as general complex coordinates, i.e., they serve as either the complex coordinates of the Hartle--Hawking metric (\ref{generalgncomplex}) (where $f\to1$, $h\to T^{(n)}$, and $V_{A}\to V^{(n)}_{A}$), or the metric in $(\zeta,\bar{\zeta})$-coordinates (\ref{eq:metreghigh}) (where $f\to e^{2A_{\epsilon}}$, $h\to e^{4A_{\epsilon}}T^{(n)}$, and $V_{A}\to e^{2A_{\epsilon}}V^{(n)}_{A}$), recovering the squashed metric Eq. (3.22) of \cite{Dong:2013qoa}.

We will be interested in computing the components of Riemann curvature near $\rho=|z|=0$. For our purposes it is sufficient then to work to order $\mathcal{O}(|z|)$ in the Riemann tensor, which means we can neglect terms of order $\mathcal{O}(|z|^{2})$ and beyond in the Christoffel symbols. To this order the components of the inverse metric are
\beq 
\begin{split}
 g^{zz}&=-\frac{4z^{2}}{f^{2}}(h+V_{A}V_{B}\sigma^{AB})\;,\quad g^{\bar{z}\bar{z}}=-\frac{4\bar{z}^{2}}{f^{2}}(h+V_{A}V_{B}\sigma^{AB})\;,\quad g^{AB}=\sigma^{AB}\\
 g^{z\bar{z}}&=2f^{-1}+4(z\bar{z})f^{-2}(h+V_{A}V_{B}\sigma^{AB})\\
 g^{zA}&=2izf^{-1}\sigma^{AB}V_{B}\left[1+4\rho^{2}(f^{-1}h+f^{-1}V_{C}V_{D}\sigma^{CD})\right]\\
 g^{\bar{z}A}&=-2i\bar{z}f^{-1}\sigma^{AB}V_{B}\left[1+4\rho^{2}(f^{-1}h+f^{-1}V_{C}V_{D}\sigma^{CD})\right]\;.
\end{split}
\eeq
Subsequently, a straightforward exercise yields the Christoffel symbols to order $\mathcal{O}(|z|)$:
\beq 
\begin{split} 
\Gamma^{z}_{\;zz}&= f^{-1}[(\partial_{z}f-2z\bar{z}\partial_{z}h-\bar{z}^{2}\partial_{\bar{z}}h-4\bar{z}h)+2z\bar{z}\{ f^{-1}(h+V^{2})\partial_{z}f-V^{A}\partial_{z}V_{A}\}]\;,\\
\Gamma^{\bar{z}}_{\;\bar{z}\bar{z}}&= f^{-1}[(\partial_{\bar{z}}f-2z\bar{z}\partial_{\bar{z}}h-z^{2}\partial_{z}h-4zh)+2z\bar{z}\{ f^{-1}(h+V^{2})\partial_{\bar{z}}f-V^{A}\partial_{\bar{z}}V_{A}\}]\;,\\
\Gamma^{z}_{\;z\bar{z}}&=f^{-1}z(2h+z\partial_{z}h)\;,\quad \Gamma^{\bar{z}}_{\;z\bar{z}}=f^{-1}\bar{z}(2h+\bar{z}\partial_{\bar{z}}h)\;,\\
\Gamma^{z}_{\;\bar{z}\bar{z}}&=-2z^{2}f^{-2}(h+V_{A}V^{A})\partial_{\bar{z}}f+z^{2}f^{-1}\partial_{\bar{z}}h+2z^{2}f^{-1}\sigma^{AB}V_{B}\partial_{\bar{z}}V_{A}\;,\\
\Gamma^{\bar{z}}_{\;zz}&=-2\bar{z}^{2}f^{-2}(h+V_{A}V^{A})\partial_{z}f+\bar{z}^{2}f^{-1}\partial_{z}h+2\bar{z}^{2}f^{-1}\sigma^{AB}V_{B}\partial_{z}V_{A}\;,\\
\Gamma^{z}_{\;AB}&=-f^{-1}\partial_{\bar{z}}H_{AB}-izf^{-1}[(\partial_{A}V_{B}+\partial_{B}V_{A})-2\gamma^{C}_{\;AB}V_{C}]\;,\\
\Gamma^{\bar{z}}_{\;AB}&=-f^{-1}\partial_{z}H_{AB}+i\bar{z}f^{-1}[(\partial_{A}V_{B}+\partial_{B}V_{A})-2\gamma^{C}_{\;AB}V_{C}]\;,\\
\Gamma^{A}_{\;zB}&=\frac{1}{2}\sigma^{AD}\partial_{z}H_{BD}+\frac{i\bar{z}}{2}\sigma^{AD}F_{BD}-\bar{z}f^{-1}\sigma^{AC}V_{C}[2V_{B}+\bar{z}\partial_{\bar{z}}V_{B}+z\partial_{z}V_{B}]\;,\\
\Gamma^{A}_{\;\bar{z}B}&=\frac{1}{2}\sigma^{AD}\partial_{\bar{z}}H_{BD}-\frac{iz}{2}\sigma^{AD}F_{BD}-zf^{-1}\sigma^{AC}V_{C}[2V_{B}+\bar{z}\partial_{\bar{z}}V_{B}+z\partial_{z}V_{B}]\;,\\
\Gamma^{z}_{\;zA}&=-if^{-1}(2V_{A}+z\partial_{z}V_{A}+\bar{z}\partial_{\bar{z}}V_{A})+izf^{-1}\sigma^{DC}V_{C}\partial_{z}H_{AD}\;,\\
\Gamma^{\bar{z}}_{\;\bar{z}A}&=if^{-1}(2V_{A}+z\partial_{z}V_{A}+\bar{z}\partial_{\bar{z}}V_{A})-i\bar{z}f^{-1}\sigma^{DC}V_{C}\partial_{\bar{z}}H_{AD}\;,\\
\Gamma^{z}_{\;\bar{z}A}&=izf^{-1}\sigma^{BD}V_{B}\partial_{\bar{z}}H_{AD}\;,\quad \Gamma^{\bar{z}}_{\;zA}=-i\bar{z}f^{-1}\sigma^{BD}V_{B}\partial_{z}H_{AD}\;,\\
\Gamma^{A}_{\;zz}&=-i\bar{z}f^{-1}\sigma^{AB}V_{B}\partial_{z}f+i\bar{z}\sigma^{AD}\partial_{z}V_{D}\;,\\
\Gamma^{A}_{\;\bar{z}\bar{z}}&=izf^{-1}\sigma^{AB}V_{B}\partial_{\bar{z}}f-iz\sigma^{AD}\partial_{\bar{z}}V_{D}\;,\quad \Gamma^{A}_{\;z\bar{z}}=\frac{i}{2}\sigma^{AD}(\bar{z}\partial_{\bar{z}}V_{D}-z\partial_{z}V_{D})\;,\\
\Gamma^{A}_{\;BC}&=\gamma^{A}_{\;BC}[\sigma]+i\bar{z}\sigma^{AD}V_{D}\partial_{\bar{z}}H_{BC}-iz\sigma^{AD}V_{D}\partial_{z}H_{BC}+\bar{\gamma}^{A}_{\;BC}[H]\;,
\end{split}
\eeq
where we used $\sigma^{AB}$ to raise and lower indices, e.g., $\sigma^{AB}V_{A}V_{B}=V^{A}V_{A}\equiv V^{2}$, $F_{BD}\equiv\partial_{B}V_{D}-\partial_{D}V_{B}$, $\gamma^{A}_{\;BC}$ is the Christoffel symbol for  metric $\sigma_{AB}(\hat{x})$, 
and 
\beq \bar{\gamma}^{A}_{\;BC}[H]\equiv\frac{1}{2}\sigma^{AD}(\partial_{B}H_{CD}+\partial_{C}H_{BD}-\partial_{D}H_{BC})\;.\eeq
To derive these expressions we have ignored terms of order $\mathcal{O}(|z|^{2})$ and higher.\footnote{Some contributions implicitly include higher powers. Take, for example, $\Gamma^{z}_{\;zz}$. For our purposes, at leading order $f\sim\mathcal{O}(|z|^{0})$, and likewise for $h$ and $V_{A}$, while $z$- or $\bar{z}$-derivatives of these functions can produce terms of order $\mathcal{O}(|z|^{-1})$. Consequently, $\Gamma^{z}_{\;zz}$ includes terms ranging from $\mathcal{O}(|z|^{-1})$ up to $\mathcal{O}(|z|^{2})$ (and higher, depending on the Taylor expansion of $h$).}

For our purposes, it is sufficient to specialize the metric (\ref{eq:genmetholo}) such that $f=e^{2A}$, $h=e^{4A}T$, and $V_{A}\to e^{2A}V_{A}$, where $T$ and $V_{A}$ are in principle functions of $(z,\bar{z},\hat{x}^{B})$ coordinates, and $A=-\epsilon\log(z\bar{z})/2$ for a real parameter $\epsilon$. With these metric functions the geometry is a generalization of the squashed cones (\ref{eq:squashedconedonglead}) and (\ref{eq:singrepmanv2app}). The Christoffel symbols simplify to
\beq
\begin{split}
&\Gamma^{z}_{\;zz}=-\frac{\epsilon}{z}-4\bar{z}Te^{2A}(1-\epsilon)\;, \quad \Gamma^{\bar{z}}_{\;\bar{z}\bar{z}}=-\frac{\epsilon}{\bar{z}}-4zTe^{2A}(1-\epsilon)\;,\\
 &\Gamma^{z}_{\;z\bar{z}}=2zTe^{2A}(1-\epsilon)\;,\quad \Gamma^{\bar{z}}_{\;\bar{z}z}=2\bar{z}Te^{2A}(1-\epsilon)\;,\quad \Gamma^{z}_{\;\bar{z}\bar{z}}=\Gamma^{\bar{z}}_{\;zz}=0\;,\\
 &\Gamma^{z}_{\;AB}=-e^{-2A}\partial_{\bar{z}}H_{AB}-iz[(\partial_{A}V_{B}+\partial_{B}V_{A})-2\gamma^{C}_{\;AB}V_{C}]\;,\\
&\Gamma^{\bar{z}}_{\;AB}=-e^{-2A}\partial_{z}H_{AB}+i\bar{z}[(\partial_{A}V_{B}+\partial_{B}V_{A})-2\gamma^{C}_{\;AB}V_{C}]\;,\\
&\Gamma^{A}_{\;zB}=\frac{1}{2}\sigma^{AD}\partial_{z}H_{BD}+\frac{i\bar{z}}{2}e^{2A}\sigma^{AD}F_{BD}-2\bar{z}\sigma^{AC}V_{B}V_{C}e^{2A}(1-\epsilon)\;,\\
&\Gamma^{A}_{\;\bar{z}B}=\frac{1}{2}\sigma^{AD}\partial_{\bar{z}}H_{BD}-\frac{iz}{2}e^{2A}\sigma^{AD}F_{BD}-2z\sigma^{AC}V_{B}V_{C}e^{2A}(1-\epsilon)\;,\\
&\Gamma^{z}_{\;zA}=-2iV_{A}(1-\epsilon)-i(z\partial_{z}V_{A}+\bar{z}\partial_{\bar{z}}V_{A})+iz\sigma^{DC}V_{C}\partial_{z}H_{AD}\;,\\
&\Gamma^{\bar{z}}_{\;\bar{z}A}=2iV_{A}(1-\epsilon)+i(z\partial_{z}V_{A}+\bar{z}\partial_{\bar{z}}V_{A})-i\bar{z}\sigma^{DC}V_{C}\partial_{\bar{z}}H_{AD}\;,\\
&\Gamma^{z}_{\;\bar{z}A}=izV^{D}\partial_{\bar{z}}H_{AD}\;,\quad \Gamma^{\bar{z}}_{\;zA}=-i\bar{z}V^{D}\partial_{z}H_{AD}\;,\quad \Gamma^{A}_{\;z\bar{z}}=\frac{i}{2}e^{2A}(\bar{z}\partial_{\bar{z}}V^{A}-z\partial_{z}V^{A})\;,\\
&\Gamma^{A}_{\;zz}=i\bar{z}e^{2A}\sigma^{AD}\partial_{z}V_{D}\;,\quad \Gamma^{A}_{\;\bar{z}\bar{z}}=-ize^{2A}\sigma^{AD}\partial_{\bar{z}}V_{D}\;,\\
&\Gamma^{A}_{\;BC}=\gamma^{A}_{\;BC}+iV^{D}(\bar{z}\partial_{\bar{z}}H_{BC}-z\partial_{z}H_{BC})+\bar{\gamma}^{A}_{\;BC}\;,
\end{split}
\eeq
where $F_{BD}\equiv\partial_{B}V_{D}-\partial_{D}V_{B}$.

\noindent \textbf{Riemann curvature.} We attain the following components of the Riemann tensor\footnote{Depending on the expansions of $T, U_{A}$, and $H_{AB}$, the Riemann tensor components (\ref{eq:Rabcdcompsv1}) may include terms of order $\mathcal{O}(\rho)$ and higher, however, it would be wrong to conclude these are the only such contributions at that order. This is because in deriving (\ref{eq:Rabcdcompsv1}) we have chosen to neglect terms which vanish as $\rho\to0$.}
\beq 
\begin{split}
&\hspace{-4mm}R_{pqrs}=12 e^{4A}T(1-\epsilon)^{2}\hat{\varepsilon}_{pq}\hat{\varepsilon}_{rs}\;,\\
&\hspace{-4mm} R_{pqrD}=e^{2A}\hat{\varepsilon}_{pq}(3-2\epsilon)\partial_{r}V_{D}\;,\\
&\hspace{-4mm} R_{pqCD}=2\hat{\varepsilon}_{pq}(1-\epsilon)e^{2A}F_{CD}+\frac{\sigma^{AF}}{4}[(\partial_{p}H_{DF})(\partial_{q}H_{AC})-(\partial_{p}H_{CF})(\partial_{q}H_{AD})]\;,\\
& \hspace{-4mm} R_{pBqD}=e^{2A}[\hat{\varepsilon}_{pq}(1-\epsilon)F_{BD}+4\hat{g}_{pq}(1-\epsilon)^{2}V_{B}V_{D}]+\frac{1}{4}\sigma^{FC}(\partial_{p}H_{DF})(\partial_{q}H_{BC})\\
&-\frac{1}{2}\partial_{p}\partial_{q}H_{BD}+\delta^{z}_{p}\delta^{z}_{q}(\partial_{z}A)(\partial_{z}H_{BD})+\delta^{\bar{z}}_{p}\delta^{\bar{z}}_{q}(\partial_{\bar{z}}A)(\partial_{\bar{z}}H_{BD})\;,\\
& \hspace{-4mm} R_{pBCD}=\frac{1}{2}\left[\nabla_{D}(\partial_{p}H_{BC})-\nabla_{C}(\partial_{p}H_{BD})\right]+(1-\epsilon)\hat{\varepsilon}_{pq}\hat{g}^{qr}[V_{D}\partial_{r}H_{BC}-V_{C}\partial_{r}H_{BD}]\;,\\
&\hspace{-4mm} R_{ABCD}=\widehat{R}_{ABCD}+\frac{e^{-2A}}{4}\hat{g}^{pq}\left[(\partial_{p}H_{AD})(\partial_{q}H_{BC})-(\partial_{p}H_{AC})(\partial_{q}H_{BD})\right]\;.
\end{split}
\label{eq:Rabcdcompsv1}\eeq
Here $\hat{\varepsilon}_{z\bar{z}}=\frac{i}{2}=-\hat{\varepsilon}_{\bar{z}z}$, $\hat{\varepsilon}_{zz}=\hat{\epsilon}_{\bar{z}\bar{z}}=0$, while $\hat{g}_{ab}=\hat{g}_{ba}$ with $\hat{g}_{z\bar{z}}=\frac{1}{2}$ and $\hat{g}_{zz}=\hat{g}_{\bar{z}\bar{z}}=0$, and inverse $\hat{g}^{ab}=(\hat{g}_{ab})^{-1}$. Moreover, $\nabla_{A}$ refers to the (Van der Waerden-Bortolotti) covariant derivative compatible with $\sigma_{AB}$, and $\widehat{R}_{ABCD}$ is the associated Riemann tensor. Substituting in $H_{AD}=2\mathcal{K}_{(p)AB}y^{p}+\mathcal{Q}_{(pq)AB}y^{p}y^{q}$, we recover Eq. (3.23) of \cite{Dong:2013qoa}, up to factors of $\epsilon$ (which we do not set to zero at this stage), and importantly, the term proportional to $\delta^{z}_{p}\delta^{z}_{q}$ in $R_{pBqD}$. In fact, it is this term which renders the Riemann tensor divergent, e.g., 
\beq R_{zBzD}=-\frac{\epsilon}{z}\mathcal{K}_{(z)BD}+\ldots\;.\label{eq:divergencezeta}\eeq
Alternatively, when $\epsilon=0$ (equivalently, $A=0$) and $H_{AB}=2\mathcal{K}_{(z)AB}z^{n}+2\mathcal{K}_{(\bar{z})AB}\bar{z}^{n}$, we still find a divergence in $R_{zBzD}$ coming from the $\partial^{2}_{z}H_{BD}$, yielding
\beq R_{zBzD}=-\frac{n(n-1)}{z}z^{n-1}\mathcal{K}_{(z)BD}+\ldots\;,\label{eq:divergencez}\eeq
consistent with Eq. (19) of \cite{Camps:2014voa}. Meanwhile, the metric (\ref{eq:nonsingrepmanapp}) (where $n=1$) has no such curvature singularities. Note \eqref{eq:divergencezeta} and \eqref{eq:divergencez} are related by a coordinate transformation $z \rightarrow n z^{1\slash n}$, $\bar{z} \rightarrow n \bar{z}^{1\slash n}$ along with $\mathcal{K}_{(z,\bar{z})AB}\rightarrow n^{-n}\, \mathcal{K}_{(z,\bar{z})AB}$. This is the coordinate transformation used to write the Hartle--Hawking metric \eqref{generalgncomplex} as \eqref{eq:metreghigh}.

To derive the components (\ref{eq:Rabcdcompsv1}) (and Ricci curvature expressions below), it is useful to know 
\beq
\begin{split}
& \hspace{-4mm} R^{z}_{\;\bar{z}z\bar{z}}=-12T^{2}z^{2}e^{4A}(1-\epsilon)^{2}\;,\quad R^{\bar{z}}_{\;\bar{z}z\bar{z}}=-6Te^{2A}(1-\epsilon)^{2}\;,\\
& \hspace{-4mm} R^{A}_{\;\bar{z}z\bar{z}}=-\frac{3i}{2}e^{2A}(1-\epsilon)\sigma^{AD}\partial_{\bar{z}}V_{D}\;,\quad R^{z}_{\;z\bar{z}A}=-i(3-2\epsilon)\partial_{\bar{z}}V_{A}\;,\quad R^{z}_{\;\bar{z}zA}=0\;,\\
 & \hspace{-4mm} R^{z}_{\;zzA}= -i(3-2\epsilon)\partial_{z}V_{A}-iz\partial^{2}_{z}V_{A}+i\sigma^{DC}V_{C}(z\partial^{2}_{z}H_{AD}+\epsilon\partial_{z}H_{AD})\;,\\
 & \hspace{-4mm} R^{z}_{\;\bar{z}\bar{z}A}=iz\sigma^{BD}V_{B}\partial^{2}_{\bar{z}}H_{AD}+\frac{i\epsilon z}{\bar{z}}\sigma^{BD}V_{B}\partial_{\bar{z}}H_{AD}\;,\\
 & \hspace{-4mm} R^{A}_{\;zzB}=\frac{\epsilon}{2z}\sigma^{AD}\partial_{z}H_{BD}+\frac{1}{2}\sigma^{AD}\partial_{z}^{2}H_{BD}+\frac{1}{4}\sigma^{AD}\sigma^{FC}\partial_{z}H_{FD}\partial_{z}H_{BC}\;,\\
 & \hspace{-4mm} R^{A}_{\;z\bar{z}B}=\frac{1}{2}\sigma^{AD}\partial_{z}\partial_{\bar{z}}H_{BD}+\frac{e^{2A}}{2}(1-\epsilon)[i\sigma^{AD}(\partial_{B}V_{D}-\partial_{D}V_{B})-4\sigma^{AC}V_{B}V_{C}(1-\epsilon)]\\
&+\frac{1}{4}\sigma^{AD}\sigma^{FC}\partial_{z}H_{BC}\partial_{\bar{z}}H_{FD}\;,\\
& \hspace{-4mm} R^{z}_{\;\bar{z}CD}= -\frac{1}{2}e^{-2A}\sigma^{AF}\left[(\partial_{\bar{z}}H_{CF})(\partial_{\bar{z}}H_{AD})-(\partial_{\bar{z}}H_{DF})(\partial_{\bar{z}}H_{AC})\right]\;,\\
& \hspace{-4mm} R^{\bar{z}}_{\;\bar{z}CD}= 2i(1-\epsilon)F_{CD}-\frac{1}{2}e^{-2A}\sigma^{AF}\left[(\partial_{z}H_{CF})(\partial_{\bar{z}}H_{AD})-(\partial_{z}H_{DF})(\partial_{\bar{z}}H_{AC})\right]\;,\\
& \hspace{-4mm} R^{A}_{\;zCD}=-\frac{1}{2}\left[\nabla_{D}(\partial_{z}H^{A}_{\;C})-\nabla_{C}(\partial_{z}H^{A}_{\;D})\right]-i(1-\epsilon)(V_{D}\partial_{z}H^{A}_{\;C}-V_{C}\partial_{z}H^{A}_{\;D})\;,\\
 & \hspace{-4mm} R^{z}_{\;BzD}= -e^{-2A}\partial_{z}\partial_{\bar{z}}H_{BD}+\frac{e^{-2A}}{2}(\partial_{\bar{z}}H^{C}_{\;D})(\partial_{z}H_{BC})+4(1-\epsilon)^{2}V_{B}V_{D}-i(1-\epsilon)F_{BD}\;, \\
 & \hspace{-4mm} R^{\bar{z}}_{\;BzD}=-e^{-2A}\partial^{2}_{z}H_{BD}-\frac{\epsilon}{z}e^{-2A}\partial_{z}H_{BD}+\frac{e^{-2A}}{2}\sigma^{FC}(\partial_{z}H_{DF})(\partial_{z}H_{BC})\;,\\
& \hspace{-4mm} R^{z}_{\;BCD}= e^{-2A}\nabla_{D}(\partial_{\bar{z}}H_{BC})+2i(1-\epsilon)e^{-2A}V_{C}\partial_{\bar{z}}H_{BD}-[C\leftrightarrow D]\\
& \hspace{-4mm} R^{A}_{\;BCD}= \widehat{R}^{A}_{\;BCD}+\frac{e^{-2A}}{2}\sigma^{AF}[(\partial_{z}H_{FD})(\partial_{\bar{z}}H_{BC})-(\partial_{z}H_{FC})(\partial_{\bar{z}}H_{BD})]\\
 &+\frac{e^{-2A}}{2}\sigma^{AF}[(\partial_{\bar{z}}H_{FD})(\partial_{z}H_{BC})-(\partial_{\bar{z}}H_{FC})(\partial_{z}H_{BD})]\;,
\end{split}\label{eq:Rabcdcomps}
\eeq
and similarly for their complex conjugates.

It is also useful to know the Riemann tensor components with raised indices, 
\beq
\begin{split}
&R^{pq}_{\;\;\;rs}=12Te^{-4A}(1-\epsilon)^{2}\hat{\varepsilon}^{pq}\hat{\varepsilon}_{rs}\;,\\
&R^{z\bar{z}}_{\;\;\;rD}=-2ie^{-2A}(3-2\epsilon)\partial_{r}V_{D}+4ie^{-2A}V^{B}z\left(\frac{1}{2}\partial_{\bar{z}}\partial_{r}H_{BD}-\delta^{\bar{z}}_{r}(\partial_{\bar{z}}A)(\partial_{\bar{z}}H_{BD})\right)\\
 &+4ie^{-2A}V^{B}\bar{z}\left(\frac{1}{2}\partial_{z}\partial_{r}H_{BD}-\delta^{z}_{r}(\partial_{z}A)(\partial_{z}H_{BD})\right)\;,\\
&R^{z\bar{z}}_{\;\;\;CD}=-4ie^{-2A}(1-\epsilon)F_{CD}+e^{-4A}\left[(\partial_{\bar{z}}H^{A}_{\;D})(\partial_{z}H_{AC})-(\partial_{\bar{z}}H^{A}_{\;C})(\partial_{z}H_{AD})\right]\;,\\
&R^{zB}_{\;\;\;zD}=i(\epsilon-1)F^{B}_{\;D}+4(1-\epsilon)^{2}V^{B}V_{D}+\frac{1}{2}e^{-2A}(\partial_{\bar{z}}H^{C}_{\;D})(\partial_{z}H^{B}_{\;C})-e^{-2A}\partial_{\bar{z}}\partial_{z}H^{B}_{\;D}\;,\\
&R^{zB}_{\;\;\;\bar{z}D}=\frac{1}{2}e^{-2A}(\partial_{\bar{z}}H^{C}_{\;D})(\partial_{\bar{z}}H^{B}_{\;C})-e^{-2A}\partial^{2}_{\bar{z}}H^{B}_{\;D}+2e^{-2A}(\partial_{\bar{z}}A)(\partial_{\bar{z}}H^{B}_{\;D})\;,\\
&R^{zB}_{\;\;\;CD}=e^{-2A}[\nabla_{D}(\partial_{\bar{z}}H^{B}_{\;C})-\nabla_{C}(\partial_{\bar{z}}H^{B}_{\;D})]-2i(1-\epsilon)e^{-2A}[V_{D}\partial_{\bar{z}}H^{B}_{\;C}-V_{C}\partial_{\bar{z}}H^{B}_{\;D}]\;,\\
&R^{AB}_{\;\;\;\;CD}=\widehat{R}^{AB}_{\;\;\;\;CD}+\frac{e^{-2A}}{4}\hat{g}^{pq}[(\partial_{p}H^{A}_{\;D})(\partial_{q}H^{B}_{\;C})-(\partial_{p}H^{A}_{\;C})(\partial_{q}H^{B}_{\;D})]\;.
\end{split}
\label{eq:riemanncomponentstwoup}
\eeq
 where $\hat{\varepsilon}^{pq}=g^{pu}g^{qt}\hat{\varepsilon}_{ut}$, such that $\hat{\varepsilon}^{z\bar{z}}\approx-2ie^{-4A}$, and $\hat{\varepsilon}^{\bar{z}z}=2ie^{-4A}$. Substituting in the appropriate metric recovers Eq. (40) of \cite{Camps:2014voa}.

\noindent \textbf{Ricci curvature.} The Ricci tensor components are:
\beq
\begin{split}
 &R_{pq}=4e^{2A}(1-\epsilon)^{2}\hat{g}_{pq}(3T+V^{2})+\frac{1}{4}(\partial_{p}H^{BC})(\partial_{q}H_{BC})-\frac{1}{2}\partial_{p}\partial_{q}H\\
 &+\delta_{p}^{z}\delta_{q}^{z}(\partial_{z}A)(\partial_{z}H)+\delta^{\bar{z}}_{p}\delta^{\bar{z}}_{q}(\partial_{\bar{z}}A)(\partial_{\bar{z}}H)\;,\\
&R_{zD}=-i(3-2\epsilon)\partial_{z}V_{D}-iz\partial^{2}_{z}V_{D}+iV^{B}(z\partial_{z}^{2}H_{BD}+\epsilon\partial_{z}H_{BD})\\
&+\frac{1}{2}[\nabla_{C}(\partial_{z}H^{C}_{\;D})-\nabla_{D}(\partial_{z}H)]-i(1-\epsilon)(V_{D}\partial_{z}H-V_{C}\partial_{z}H^{C}_{\;D})\;,\\
&R_{BD}=\widehat{R}_{BD}-2e^{-2A}\partial_{z}\partial_{\bar{z}}H_{BD}+\frac{e^{-2A}}{2}\biggr[2(\partial_{z}H^{C}_{\;D})(\partial_{\bar{z}}H_{BC})-(\partial_{z}H)(\partial_{\bar{z}}H_{BD})\\
&+2(\partial_{\bar{z}}H^{C}_{\;D})(\partial_{z}H_{BC})-(\partial_{\bar{z}}H)(\partial_{z}H_{BD})\biggr]+8(1-\epsilon)^{2}V_{B}V_{D}\;,  
\end{split}
\label{eq:Riccigencomps}\eeq
where $H\equiv \sigma^{AB}H_{AB}$, $\sigma^{AB}F_{AB}=0$, and $r_{BD}$ is the Ricci tensor with respect to metric $\sigma_{AB}$, i.e., $\widehat{R}_{BD}=\sigma^{AC}\widehat{R}_{ABCD}$. The Ricci scalar is then
\beq
\begin{split}
 \hspace{-6mm} R&=\widehat{R}+(1-\epsilon)^{2}[24T+16V^{2}]+e^{-2A}(3(\partial_{z}H^{BC})(\partial_{\bar{z}}H_{BC})-(\partial_{z}H)(\partial_{\bar{z}}H)-4\partial_{z}\partial_{\bar{z}}H)\,,
\end{split}
\eeq
with Ricci scalar $\widehat{R}$ with respect to  $\sigma_{AB}$, and we used $R\approx 2g^{z\bar{z}}R_{z\bar{z}}+\sigma^{BD}R_{BD}$ with $g^{z\bar{z}}\approx 2e^{-2A}$. If $H_{AB}=2\mathcal{K}_{(p)AB}y^{p}$, we recover the Ricci scalar in Eq. (B.4) of \cite{Bhattacharyya:2014yga}.

Notice the Ricci tensor components $R_{zz}$ and $R_{\bar{z}\bar{z}}$ have a $1/z$ divergence. The Ricci scalar, however, has no such divergence since components $R_{zz}$ and $R_{\bar{z}\bar{z}}$ do not enter into its computation at leading order in $\rho$. Correspondingly, the Einstein tensor components $G_{pq}=R_{pq}-\frac{1}{2}Rg_{pq}$ and  hence Einstein's field equations have the same $1/z$ divergence.


\section{Solving the splitting problem for Hartle--Hawking metrics} \label{app:solvingsplittingprobs}

Here we detail the splittings for Hartle--Hawking states in Einstein gravity by solving Einstein's equations perturbatively in distance around the codimension-2 surface $\mathcal{C}$ for a metric $g_{n}$ satisfying the Hartle--Hawking boundary condition \eqref{maldacena}.


\subsection{Replica symmetric expansion of the Hartle--Hawking metric}\label{subapp:replicasymmetricexp}

We consider metrics of the type (\ref{eq:repmanDongv1}), repeated here for convenience,
\begin{equation}
    ds^{2} = dzd\bar{z}+T^{(n)}\,(\bar{z}dz - z d\bar{z})^2+ 2iV_{A}^{(n)}\,(\bar{z}dz - z d\bar{z})\, d\hat{x}^A + h_{AB}^{(n)}\,d\hat{x}^Ad\hat{x}^B\;,
    \label{HHmetricapp}
\end{equation}
with $h^{(n)}_{AB}=\sigma_{AB}+H^{(n)}_{AB}$, and where the component functions have series expansions
\begin{align}
    T^{(n)} &= \mathcal{T}_0+\mathcal{T}_1\,(z\bar{z}) + \ldots\label{Texp}\\
    V_{A}^{(n)} &= \mathcal{U}_A + \mathcal{V}_{(z)A}\,z^n + \mathcal{V}_{(\bar{z})A}\,\bar{z}^n + \ldots \label{Vexp}\\
    H_{AB}^{(n)} &= 2\mathcal{K}_{(z)AB}\,z^n + 2\mathcal{K}_{(\bar{z})AB}\,\bar{z}^n+\mathcal{Q}_{(zz)AB}\, z^{2n} + \mathcal{Q}_{(\bar{z}\bar{z})AB}\,\bar{z}^{2n} + 2\mathcal{Q}_{(z\bar{z})AB}\,(z\bar{z}) + \ldots
    \label{Hexp}
\end{align}
 Each coefficient $\mathcal{T}_{k}$, $\mathcal{U}_A$, $\mathcal{V}_{(p)A}$, $\mathcal{K}_{(p)AB}$ and $\mathcal{Q}_{(pq)AB}$, while arbitrary functions of $\hat{x}^{A}$, have their own expansion in powers of $(z\bar{z})^{n-1}$; in particular, we denote the coefficient of $(z\bar{z})^{l(n-1)}$ by $\mathcal{T}_{k,l} $, $\mathcal{U}_{lA}$, $\mathcal{V}_{l(p)A}$, $\mathcal{K}_{l(p)AB}$ or $\mathcal{Q}_{l(pq)AB}$. For example,
\begin{align}
    \mathcal{Q}_{(z\bar{z})AB} &= \mathcal{Q}_{0(z\bar{z})AB} + \mathcal{Q}_{1(z\bar{z})AB}\,(z\bar{z})^{n-1}+ \mathcal{O}\bigl((z\bar{z})^{2(n-1)}\bigr)\ .
\end{align}
Hence, $\mathcal{Q}_{0(z\bar{z})AB}$ is the coefficient of $(z\bar{z})$ and $\mathcal{Q}_{1(z\bar{z})AB}$ the coefficient of $(z\bar{z})^n$ in $H^{(n)}_{AB}$. Writing $z = \rho e^{i\tau\slash n}$ and $ \bar{z} = \rho e^{-i\tau\slash n}$, the metric satisfies the Hartle--Hawking boundary condition \eqref{maldacena} when $\rho \rightarrow 0$. The only non-zero components of the Riemann tensor at $\mathcal{C}$ are given by
\begin{align}
R_{z\bar{z}z\bar{z}}'\lvert_{\mathcal{C}}& = -3\mathcal{T}_{0,0}\;,\nonumber\\
R_{z\bar{z}AB}'\lvert_{\mathcal{C}} &= i\,(\partial_A\mathcal{U}_{0B}-\partial_B\mathcal{U}_{0A})\;,\nonumber\\
R_{zA\bar{z}B}'\lvert_{\mathcal{C}}&= \frac{i}{2}\,(\partial_A\mathcal{U}_{0B}-\partial_B\mathcal{U}_{0A})-2\mathcal{Q}_{0(z\bar{z})AB}\;,\nonumber\\
R_{zAzB}'\lvert_{\mathcal{C}}&=2\,\mathcal{U}_{0A}\mathcal{U}_{0B} -2\mathcal{Q}_{0(zz)AB}\;,\nonumber\\
R_{ABCD}'\lvert_{\mathcal{C}}&= \widehat{R}_{ABCD}'\;.
\label{eq:gntensorsapp}
\end{align}
Here $\widehat{R}_{ABCD}'$ denotes the Riemann tensor of $\sigma_n$ which is different from the Riemann tensor $\widehat{R}_{ABCD}$ of $\sigma_1$.

The metric at $n=1$ is denoted by $g_1$ where
\begin{equation}
    ds^{2} = dzd\bar{z}+T\,(\bar{z}dz - z d\bar{z})^2+ 2iV_{A}\,(\bar{z}dz - z d\bar{z})\, d\hat{x}^A + h_{AB}\,d\hat{x}^Ad\hat{x}^B\;, 
    \label{HHmetricappunity}
\end{equation}
with $T \equiv T^{(1)} $, $V_{A}\equiv V_{A}^{(1)}$ and $h_{AB}\equiv h_{AB}^{(1)}$. We write the expansions of the metric functions at $n = 1$ as
\begin{align}
    T &= T_0+T_1\,(z\bar{z}) + \ldots\label{Texp1}\\
    V_{A} &= U_A + V_{(z)A}\,z + V_{(\bar{z})A}\,\bar{z} + \ldots \label{Vexp1}\\
    H_{AB} &= 2K_{(z)AB}\,z + 2K_{(\bar{z})AB}\,\bar{z}+Q_{(zz)AB}\, z^2 + Q_{(\bar{z}\bar{z})AB}\,\bar{z}^2 + 2Q_{(z\bar{z})AB}\,(z\bar{z}) + \ldots
    \label{Hexp1}
\end{align}
where all coefficients $T_{0},T_{1},U_{A}$ etc. are independent of $z$ and $\bar{z}$. These coefficients are related to the $n\rightarrow 1$ limit of the $n > 1$ coefficients via
\begin{gather}
    T_0 = \sum_{k=0}^{\infty}\mathcal{T}_{0,k}\lvert_{n=1},\quad U_{A} = \sum_{k=0}^{\infty}\mathcal{\mathcal{U}}_{kA}\lvert_{n=1},\quad V_{(p)A} = \sum_{k=0}^{\infty}\mathcal{\mathcal{V}}_{k(p)A}\lvert_{n=1}\;,\nonumber\\
    K_{(p)AB} = \sum_{k=0}^{\infty}\mathcal{\mathcal{K}}_{k(p)AB}\lvert_{n=1},\quad  Q_{(pq)AB} = \sum_{k=0}^{\infty}\mathcal{Q}_{k(pq)AB}\lvert_{n=1}\;.
    \label{eq:infiniteseriesapp}
\end{gather}
Hence, from the Riemann tensor components (\ref{eq:Rabcdcompsv1}), we find
\begin{align}
R_{z\bar{z}z\bar{z}}\lvert_{\mathcal{C}}& = -3T_{0}\;,\nonumber\\
R_{z\bar{z}pA}\lvert_{\mathcal{C}}&= \frac{3}{2}\,i\,\partial_{p}V_{(p)A}\;,\nonumber\\
R_{z\bar{z}AB}\lvert_{\mathcal{C}} &= i\,(\partial_AU_B-\partial_BU_A)+K_{(z)B}^{C}K_{(\bar{z})CA}-K_{(z)A}^{C}K_{(\bar{z})CB}\;,\nonumber\\
R_{zA\bar{z}B}\lvert_{\mathcal{C}}&= \frac{i}{2}\,(\partial_AU_B-\partial_BU_A)+K_{(z)A}^{C}K_{(\bar{z})CB}+K_{(\bar{z})A}^{C}K_{(z)CB}-2Q_{(z\bar{z})AB}\;,\nonumber\\
R_{zAzB}\lvert_{\mathcal{C}}&=2\,U_{A}U_{B}+K_{(z)A}^{C}K_{(\bar{z})CB}+K_{(\bar{z})A}^{C}K_{(z)CB}-2Q_{(zz)AB}\;,\nonumber\\
R_{zABC}\lvert_{\mathcal{C}}&= \widehat{\nabla}_{C}K_{(z)AB} - \widehat{\nabla}_{B}K_{(z)AC}+2i\,[U_C K_{(z)AB}-U_B K_{(z)AC}]\;,\nonumber\\
R_{ABCD}\lvert_{\mathcal{C}}&= \widehat{R}_{ABCD}+\bigl[2K_{(z)AD}K_{(\bar{z})BC} - 2K_{(z)AC}K_{(\bar{z})BD}+ (z\leftrightarrow \bar{z})\bigr]\;.
    \label{eq:g1riemannsapp}
\end{align}
By substituting the Riemann tensors of $g_n$ (\ref{eq:gntensorsapp}) at $n = 1$ into the Riemann tensors of $g_1$ (\ref{eq:g1riemannsapp}) and using (\ref{eq:infiniteseriesapp}) we arrive at the following off-shell splitting relations (we do not write $\lvert_{\mathcal{C}}$ for clarity)
\begin{align}
    R_{z\bar{z}z\bar{z}}'\big\lvert_{n = 1} &= R_{z\bar{z}z\bar{z}} + \sum_{k=1}^\infty  3\mathcal{T}_{0,k}\big\lvert_{n=1}\;,\nonumber\\
    R_{z\bar{z}pA}'\big\lvert_{n = 1}&=R_{z\bar{z}pA}- \frac{3}{2}\,i\sum_{k=1}^{\infty}\partial_{p}\mathcal{V}_{k(p)A}\big\lvert_{n=1}\;,\label{eq:splitrels}\\
    R_{ABCD}'\big\lvert_{n = 1}&= R_{ABCD}-\sum_{k=0}^{\infty}\sum_{l=0}^{\infty}\bigl[2\mathcal{K}_{k(z)AD}\mathcal{K}_{l(\bar{z})BC} - 2\mathcal{K}_{k(z)AC}\mathcal{K}_{l(\bar{z})BD}+ (z\leftrightarrow \bar{z})\bigr]\;\nonumber,
\end{align}
and so forth.

\paragraph{Gauge transformations of the frame bundle.} 

Consider a Hartle--Hawking metric $g_n$ of the form \eqref{generalgn}. Under a diffeomorphism $F$ it transforms as
\begin{equation}
\widetilde{g}_{n ab}(x) = \frac{\partial F^{c}}{\partial x^{a}}\frac{\partial F^{d}}{\partial x^{b}}\,g_{n cd}(F(x))\;.
\label{eq:HHdiffeo}
\end{equation}
We will focus on the diffeomorphism
\begin{equation}
F^{\tau}(x) = \tau - 2n\lambda(\hat{x}),\quad F^{\rho}(x) = \rho,\quad F^{A}(x) = \hat{x}^{A}
\label{eq:Udiffeomorphism}
\end{equation}
that acts non-trivially only on the $ \tau $-direction. Then the metric $ \widetilde{g}_n $ takes the same form as \eqref{generalgn}, but with transformed functions
\begin{align}
\widetilde{V}_{A}^{(n)}(x) &= V_{A}^{(n)}(F(x))+\partial_A \lambda\\
\widetilde{H}_{AB}^{(n)}(x) &= H_{AB}^{(n)}(F(x))+\bigl[4V_{A}^{(n)}(F(x))\,\partial_B\lambda + 2\rho^{2}\,(\partial_A \lambda)(\partial_B\lambda) + (A\leftrightarrow B)\bigr]
\end{align}
and with $\widetilde{T}^{(n)}= T^{(n)}$ being invariant since it does not involve $\tau$. By substituting the expansions \eqref{Texp} - \eqref{Hexp}, it follows that the first few coefficients transform as
\begin{gather}
\widetilde{\mathcal{U}}_{0A}=\mathcal{U}_{0A}+\partial_A \lambda,\quad \widetilde{\mathcal{Q}}_{0(z\bar{z})AB} = \mathcal{Q}_{0(z\bar{z})AB}+\bigl[2\,\mathcal{U}_{0A}\,(\partial_{B}\lambda) + (\partial_A \lambda)(\partial_B\lambda) + (A\leftrightarrow B)\bigr],\nonumber\\
\widetilde{\mathcal{V}}_{(z)A} = e^{-2in\lambda}\,\mathcal{V}_{(z)A},\quad \widetilde{\mathcal{K}}_{(z)AB} = e^{-2in\lambda}\,\mathcal{K}_{(z)AB},\quad \widetilde{\mathcal{Q}}_{(zz)AB} = e^{-4in\lambda}\,\mathcal{Q}_{(zz)AB}\;
\label{eq:redefinition}
\end{gather}
with the transformations of the rest of the coefficients being readily obtainable. Hence the diffeomorphism \eqref{eq:Udiffeomorphism} can be interpreted as a $U(1)$ gauge transformation \eqref{eq:redefinition} for which $\mathcal{U}_{0A}$ acts as the connection and for which other coefficients are charged: the charge of the coefficient of $z^{nk}\bar{z}^{nl}$ is $2n\,(l-k)$. The diffeomorphism \eqref{eq:Udiffeomorphism} with $n = 1$ can also be applied to the $n = 1$ metric \eqref{HHmetricappunity} in which case $U_A$ is the gauge connection.

\subsection{Einstein gravity}\label{subapp:Einsteinsplitting}

The task now is to solve the coefficients appearing in the splitting relations \eqref{eq:splitrels} by imposing the equations of motion of the bulk gravitational theory. Let us show this explicitly for Einstein gravity. 

At leading order, the Ricci tensor of the metric \eqref{HHmetricapp} is
\begin{align}
	R_{zz}' &= \frac{1}{4}(\partial_{z}H^{(n)BC})(\partial_{z}H_{BC}^{(n)})-\frac{1}{2}\,\partial_{z}^{2}H^{(n)}\label{Rzz}\\
	R_{z\bar{z}}' &= 6T^{(n)} + 2V_A^{(n)} V^{(n)A} +\frac{1}{4}(\partial_{z}H^{(n)BC})(\partial_{\bar{z}}H_{BC}^{(n)})-\frac{1}{2}\partial_{z}\partial_{\bar{z}}H^{(n)}\label{Rzzb}\\
	R_{AB}' &= \widehat{R}_{AB}' -2\partial_z\partial_{\bar{z}}H_{AB}^{(n)} + (\partial_z H^{(n)C}_{A})(\partial_{\bar{z}}H_{CB}^{(n)})+ (\partial_{\bar{z}} H^{(n)C}_{A})(\partial_{z}H_{CB}^{(n)})\nonumber\\
 &-\frac{1}{2}\,(\partial_z H^{(n)})(\partial_{\bar{z}}H_{AB}^{(n)})- \frac{1}{2}\,(\partial_{\bar{z}} H^{(n)})(\partial_{z}H_{AB}^{(n)}) + 8\,V_{A}^{(n)}V_{B}^{(n)}\;.\label{RAB}
\end{align}
These expressions are exact up to terms that vanish in the double limit $n\rightarrow 1$ and $\rho \rightarrow 0$: the coefficients of terms that are finite or divergent in $\rho \rightarrow 0$ after setting $n=1$ are exact. Using the expansions \eqref{Texp} - \eqref{Hexp}, we get for the first few exact terms
\begin{align}
R_{zz}' &= -n(n-1)\,\mathcal{K}_{0(z)}\,z^{n-2} + \Bigl[n^2\mathcal{K}_{0(z)}^{AB}\mathcal{K}_{0(z)AB}-n(2n-1)\,\mathcal{Q}_{0(zz)}\Bigr]\,z^{2n-2} + \ldots\\
R_{z\bar{z}}' &=6\mathcal{T}_{0,0} -\mathcal{Q}_{0(z\bar{z})}+2\,\mathcal{U}_{0A}\,\mathcal{U}_{0}^A+\Bigl[6\mathcal{T}_{0,1}+n^{2}\,\mathcal{K}_{0(z)}^{AB}\mathcal{K}_{0(\bar{z})AB}\Bigr.\label{eq:Rzzbexp}\\
&\hspace{5.5cm}\Bigl.-n^{2}\mathcal{Q}_{1(z\bar{z})}+4\,\mathcal{U}_{0A}\,\mathcal{U}_{1}^{A}\Bigr]\,(z\bar{z})^{n-1} + \ldots\nonumber\\
 R_{AB}' &=  \widehat{R}_{AB}'-4\mathcal{Q}_{0(z\bar{z})AB}+2\,\mathcal{U}_{0A}\,\mathcal{U}_{0B}+\Bigl[4n^{2}\,\mathcal{K}_{0(z)A}^{C}\mathcal{K}_{0(\bar{z})BC}-2n^{2}\,\mathcal{K}_{0(z)}\mathcal{K}_{0(\bar{z})AB}\Bigr.\label{eq:RABexp}\\
 &\hspace{5cm}\Bigl.-4n^{2}\,\mathcal{Q}_{1(z\bar{z})AB}+8\,\mathcal{U}_{0A}\,\mathcal{U}_{1B} + (z\leftrightarrow \bar{z})\Bigr]\,(z\bar{z})^{n-1}+ \ldots\nonumber
\end{align}
and their complex conjugates. All indices are understood to be contracted with $\sigma_{AB}$. Upon setting $n = 1$ we see that all terms shown are either finite or divergent when $\rho \rightarrow 0$, so their coefficients are exact. There are also an infinite number of higher-order terms of the form $(z\bar{z})^{l(n-1)}$ with $l> 1$ whose coefficients are exact and can be computed.


Let us then impose vacuum Einstein's equations $R_{ab}' = 0$. Without loss of generality, we will solve Einstein's equations in the gauge (coordinate system) where
\begin{equation}
\mathcal{U}_{0A} = 0.
\label{eq:Ugaugeappendix}
\end{equation}
which is also assumed in \cite{Miao:2014nxa,Miao:2015iba,Camps:2016gfs}. Since Einstein's equations are diffeomorphism covariant, the solution in a general gauge is obtained by a gauge transformation \eqref{eq:redefinition}. In the end, all diffeomorphism invariant quantities constructed from the metric (such as gravitational actions) are invariant under the gauge transformation so there is no loss of generality. 

\paragraph{Leading coefficients.} First, the leading terms in the $zz$ and $\bar{z}\bar{z}$ components of $R_{ab}' = 0$ imply that $\overbar{g}_n$ satisfies
\begin{equation}
    \sigma^{AB}\,\overbar{\mathcal{K}}_{0(z)AB} = \sigma^{AB}\,\overbar{\mathcal{K}}_{0(\bar{z})AB} = 0\;,
    \label{vanishingKtrace}
\end{equation}
while the subleading terms give
\begin{equation}
    \sigma^{AB}\,\overbar{\mathcal{Q}}_{0(zz)AB} = \frac{n}{2n-1}\,\overbar{\mathcal{K}}_{0(z)}^{AB}\overbar{\mathcal{K}}_{0(z)AB},\quad \sigma^{AB}\,\overbar{\mathcal{Q}}_{0(\bar{z}\bar{z})AB} = \frac{n}{2n-1}\,\overbar{\mathcal{K}}_{0(\bar{z})}^{AB}\overbar{\mathcal{K}}_{0(\bar{z})AB}\;.
\end{equation}
The remaining $z\bar{z}$ and $AB$ components of $R_{ab} = 0$ give four equations
\begin{gather}
    6\mathcal{T}_{0,0} -\mathcal{Q}_{0(z\bar{z})} = 0,\quad 6\mathcal{T}_{0,1}+n^{2}\,\mathcal{K}_{0(z)}^{AB}\mathcal{K}_{0(\bar{z})AB}-n^{2}\mathcal{Q}_{1(z\bar{z})} = 0,\label{splittingequations}\\
    \widehat{R}_{AB}-4\mathcal{Q}_{0(z\bar{z})AB} = 0,\quad 2\mathcal{K}_{0(z)A}^{C}\mathcal{K}_{0(\bar{z})BC}-\mathcal{K}_{0(z)}\mathcal{K}_{0(\bar{z})AB}-2\mathcal{Q}_{1(z\bar{z})AB}+ (z\leftrightarrow \bar{z}) = 0.\nonumber
\end{gather}
From the second line of \eqref{splittingequations} we get the on-shell relations
\begin{equation}
\overbar{\mathcal{Q}}_{0(z\bar{z})AB} = \frac{1}{4}\widehat{R}_{AB}', \quad \overbar{\mathcal{Q}}_{1(z\bar{z})AB} = \mathcal{K}_{0(z)A}^{C}\mathcal{K}_{0(\bar{z})BC}-\frac{1}{2}\,\mathcal{K}_{0(z)}\mathcal{K}_{0(\bar{z})AB}+ (z\leftrightarrow \bar{z}).
\label{eq:leadingsol1}
\end{equation}
Substituting to the first two equations gives
\begin{equation}
\overbar{\mathcal{T}}_{0,0} = \frac{1}{24}\widehat{R}',\quad \overbar{\mathcal{T}}_{0,1} = \frac{n^{2}}{6}\left(\mathcal{K}_{0(z)}^{AB}\mathcal{K}_{0(\bar{z})AB}-\mathcal{K}_{0(z)}\mathcal{K}_{0(\bar{z})}\right).
\label{eq:leadingsol2}
\end{equation}
Imposing also \eqref{vanishingKtrace}, we can write the first-order coefficients as
\begin{equation}
    \overbar{\mathcal{Q}}_{1(z\bar{z})AB} = \overbar{\mathcal{K}}_{0(z)A}^{C}\overbar{\mathcal{K}}_{0(\bar{z})BC}+\overbar{\mathcal{K}}_{0(\bar{z})A}^{C}\overbar{\mathcal{K}}_{0(z)BC},\quad \overbar{\mathcal{T}}_{0,1} = \frac{n^{2}}{6}\,\overbar{\mathcal{K}}_{0(z)}^{AB}\overbar{\mathcal{K}}_{0(\bar{z})AB}\ .
    \label{eq:campsgeneral}
\end{equation}

\paragraph{Vanishing of higher-order coefficients.} The complete solution of the splitting problem for a general replica symmetric metric \eqref{HHmetricapp} requires solving for an infinite number of higher-order coefficients of $(z\bar{z})^{l(n-1)}$. We have checked numerically for small values of $k$ that Einstein's equations require that
\begin{align}
    \overbar{\mathcal{U}}_{kA} = \overline{\mathcal{K}}_{k(p)AB}= \overline{\mathcal{Q}}_{k(pp)AB}= 0,\quad k\geq 1\ ,\quad \overbar{\mathcal{T}}_{0,k} = \overbar{\mathcal{Q}}_{k(z\bar{z})AB}= 0,\quad k\geq 2\,,
    \label{eq:vanishinghigherapp}
\end{align}
which has been proven generally in \cite{Camps:2014voa} except for the $\mathcal{Q}$ coefficients. Assuming \eqref{eq:vanishinghigherapp} holds, the on-shell in Einstein gravity (in the gauge \eqref{eq:Ugaugeappendix})
\begin{align}
    \overbar{T}^{(n)} &= \overbar{\mathcal{T}}_{0,0}+\overbar{\mathcal{T}}_{0,1}\,(z\bar{z})^{n-1}+\overbar{\mathcal{T}}_{1,0}\,(z\bar{z}) + \ldots\nonumber\\
    \overbar{V}_{A}^{(n)} &= \overbar{\mathcal{V}}_{0(z)A}\,z^n + \overbar{\mathcal{V}}_{0(\bar{z})A}\,\bar{z}^n + \ldots \nonumber\\
    \overbar{H}_{AB}^{(n)} &= 2\overbar{\mathcal{K}}_{0(z)AB}\,z^n + 2\overbar{\mathcal{K}}_{0(\bar{z})AB}\,\bar{z}^n+\overbar{\mathcal{Q}}_{0(zz)AB}\, z^{2n} + \overbar{\mathcal{Q}}_{0(\bar{z}\bar{z})AB}\,\bar{z}^{2n} + 2\overbar{\mathcal{Q}}_{0(z\bar{z})AB}\,(z\bar{z})\nonumber\\
    &\hspace{8cm} + 2\overbar{\mathcal{Q}}_{1(z\bar{z})AB}\,(z\bar{z})^n + \ldots
    \label{eq:einsteinonshell}
\end{align}
where all coefficients are independent of $z,\bar{z}$. Thus only the powers $(z\bar{z})$, $(z\bar{z})^{n-1}$ and $(z\bar{z})^n$ appear as has been assumed in \cite{Camps:2016gfs}.

\paragraph{The splitting relations.} By the vanishing of the higher-order coefficients the infinite series \eqref{eq:infiniteseriesapp} truncate on-shell as
\begin{gather}
    \overbar{T}_{0} = \overbar{\mathcal{T}}_{0,0}\big\lvert_{n=1}\, +\, \overbar{\mathcal{T}}_{0,1}\big\lvert_{n=1},\quad \overbar{U}_{A} = \overbar{\mathcal{U}}_{0A}\big\lvert_{n=1},\quad \overbar{K}_{(p)AB} = \overbar{\mathcal{K}}_{0(p)AB}\big\lvert_{n=1}\\
    \overbar{Q}_{(pq)AB} = \overbar{\mathcal{Q}}_{0(pq)AB}\big\lvert_{n=1}+\overbar{\mathcal{Q}}_{1(pq)AB}\big\lvert_{n=1}\ .
\end{gather}
Hence by \eqref{vanishingKtrace} the trace of the extrinsic curvature of $\mathcal{C}$ in $\overbar{g}_1$ vanishes
\begin{equation}
    \sigma^{AB}\,\overbar{K}_{(p)AB} = 0,\quad p = z,\bar{z},
\end{equation}
which is the area extremization prescription of Einstein gravity. Using \eqref{eq:campsgeneral} at $n = 1$ gives
\begin{align}
    \overbar{T}_{0} &= \overbar{\mathcal{T}}_{0,0}\lvert_{n=1}\,+\,\frac{1}{6}\,\overbar{K}_{(z)}^{AB}\overbar{K}_{(\bar{z})AB}\\
    \overbar{Q}_{(z\bar{z})AB} &= \overbar{\mathcal{Q}}_{0(z\bar{z})AB}\big\lvert_{n=1}+\overbar{K}_{(z)A}^{C}\,\overbar{K}_{(\bar{z})BC}+\overbar{K}_{(\bar{z})A}^{C}\,\overbar{K}_{(z)BC}\ ,
\end{align}
We can write these in terms of the Riemann tensor components of the metric $\lim_{n\rightarrow 1}\overbar{g}_n$ \eqref{eq:gntensorsapp} and $\overbar{g}_1$ \eqref{eq:g1riemannsapp} respectively as
\begin{align}
    \overbar{R}'_{z\bar{z}z\bar{z}}\big\lvert_{n=1}\, &= \overbar{R}_{z\bar{z}z\bar{z}}+ \frac{1}{2}\,\overbar{K}_{(z)}^{AB}\,\overbar{K}_{(\bar{z})AB}\\
    \overbar{R}_{zA\bar{z}B}'\big\lvert_{n=1}\, &= \overbar{R}_{zA\bar{z}B} +\overbar{K}_{(z)A}^{C}\,\overbar{K}_{(\bar{z})BC}+\overbar{K}_{(\bar{z})A}^{C}\,\overbar{K}_{(z)BC}
    \label{Einsteinsplittingapp}
\end{align}
The remaining splitting relations in Einstein gravity are fixed by \eqref{eq:vanishinghigherapp} to be
\begin{align}
 \overbar{R}_{z\bar{z}pA}'\big\lvert_{n = 1}&= \overbar{R}_{z\bar{z}pA}-\frac{3}{2}\,i\,\partial_{p}\overbar{V}_{(p)A},\nonumber\\
 \overbar{R}_{z\bar{z}AB}'\big\lvert_{n=1}&=\overbar{R}_{z\bar{z}AB} -\overbar{K}_{(z)B}^{C}\overbar{K}_{(\bar{z})CA}+\overbar{K}_{(z)A}^{C}\overbar{K}_{(\bar{z})CB}\;,\nonumber\\
\overbar{R}_{zAzB}'\big\lvert_{n=1}&=\overbar{R}_{zAzB}-\overbar{K}_{(z)A}^{C}\overbar{K}_{(\bar{z})CB}-\overbar{K}_{(\bar{z})A}^{C}\overbar{K}_{(z)CB}\;,\nonumber\\
\overbar{R}_{zABC}\big\lvert_{n=1}&=\overbar{R}_{zABC}-\widehat{\nabla}_{C}\overbar{K}_{(z)AB} + \widehat{\nabla}_{B}\overbar{K}_{(z)AC}-2i\,[U_C \overbar{K}_{(z)AB}-U_B \overbar{K}_{(z)AC}]\;,\nonumber\\
\overbar{R}_{ABCD}\big\lvert_{n=1}&=\overbar{R}_{ABCD}- \bigl[2\overbar{K}_{(z)AD}\overbar{K}_{(\bar{z})BC} - 2\overbar{K}_{(z)AC}\overbar{K}_{(\bar{z})BD}+ (z\leftrightarrow \bar{z})\bigr]\;.
\label{eq:trivialsplittings}
\end{align}

\paragraph{Fixing the induced metric.} Perturbatively solving Einstein's equations around $\rho = 0$ does not determine the induced metric $\sigma$. In particular, the $n$-dependence of $\sigma$ is not fixed, because $n$ appears as a multiplicative factor in front of $d\tau^2$ that can be removed by a coordinate transformation. However, we know that the full non-perturbative solution $\overbar{g}_n$ fixes $\sigma$ to $\overbar{\sigma}_n$, but this is not visible in the perturbative expansion. Let us illustrate this in a simple example where the non-perturbative solution is available.

Vacuum Einstein's equations in $D=5$ dimensions has the one-parameter family of solutions $\overbar{g}_n$ given by
\begin{equation}
ds^{2} = d\rho^{2}+ \overbar{T}^{(n)}(\rho)\,d\tau^{2} + \overbar{h}^{(n)}(\rho)\,ds^{2}_{S^{3}}\;,
\label{eq:sphericalBH}
\end{equation}
where the functions are given by
\begin{equation}
\overbar{T}^{(n)}(\rho) =\frac{\rho^{2}}{\rho^{2}+n^{2}},\quad \overbar{h}^{(n)}(\rho) = \rho^{2} + n^{2}\;.
\label{eq:nonpertspherical}
\end{equation}
In the limit $ \rho \rightarrow 0 $, $ \overbar{T}^{(n)} $ has the expansion
\begin{equation}
\overbar{T}^{(n)}(\rho) = \frac{\rho^{2}}{n^{2}}-\frac{\rho^{4}}{n^{4}} + \mathcal{O}(\rho^{6}).
\end{equation}
so that the metric \eqref{eq:sphericalBH} satisfies the Hartle--Hawking boundary condition \eqref{maldacena} at $\rho = 0$. From $\overbar{h}^{(n)}$ we can identify that $ \overbar{\sigma}_n $ is the round metric on a sphere of radius $n$:
\begin{equation}
\overbar{\sigma}_{nAB}\,d\hat{x}^{A}d\hat{x}^{B} = n^{2}ds^{2}_{S^{3}}\;.
\end{equation}
The metric \eqref{eq:sphericalBH} is simply the $ D=5 $ Schwarzschild black hole in the radial coordinate $\rho$. The standard radial coordinate is given by
\begin{equation}
r^{2} = \rho^{2} + n^{2}
\end{equation}
in which the metric takes the familiar form
\begin{equation}
ds^{2} =\biggl(1-\frac{n^{2}}{r^{2}}\biggr)\,d\tau^{2}+ \biggl(1-\frac{n^{2}}{r^{2}}\biggr)^{-1}dr^{2}+ r^{2}\,ds^{2}_{S^{3}}\ .
\end{equation}

Let us now solve Einstein's equations perturbatively and compare to the non-perturbative solution. To this end, consider the following series expansions of the metric functions
\begin{equation}
T(\rho) = \frac{\rho^{2}}{n^{2}}-\frac{4\mathcal{T}_{0}}{n^{2}}\,\rho^{4} + \mathcal{O}(\rho^{6}),\quad h(\rho) = L^{2}+2\mathcal{Q}_0\,\rho^{2}+\mathcal{O}(\rho^{4}).
\label{eq:Thpertexpansions}
\end{equation}
where $\mathcal{T}_0,\mathcal{Q}_0$ are unknown coefficients and the radius $ L $ of $ \sigma $ is a free parameter. In this case, we have
\begin{equation}
\widehat{R}_{AB} = \frac{2}{L^{2}}\,\sigma_{AB},
\end{equation}
so that the Ricci tensors \eqref{eq:Rzzbexp} and \eqref{eq:RABexp} are
\begin{equation}
R_{z\bar{z}} = 6T_0 - \frac{3}{L^{2}}\,\mathcal{Q}_0 + \mathcal{O}(\rho^{2}) ,\quad R_{AB} = \frac{2}{L^{2}}\,(2\mathcal{Q}_0 - 1)\,\sigma_{AB} + \mathcal{O}(\rho^{2}).
\end{equation}
The Einstein equation is thus solved by (these also follow directly from \eqref{eq:leadingsol1} and \eqref{eq:leadingsol2})
\begin{equation}
\overbar{\mathcal{T}}_0 = \frac{1}{4L^{2}},\quad \overbar{\mathcal{Q}}_0  = \frac{1}{2},
\end{equation}
giving the expansions
\begin{equation}
\overbar{T}^{(n)}(\rho) = \frac{\rho^{2}}{n^{2}}-\frac{\rho^{4}}{n^{2}L^{2}} + \mathcal{O}(\rho^{6}),\quad \overbar{h}^{(n)}(\rho) = L^{2}+\rho^{2}+\mathcal{O}(\rho^{4}).
\label{eq:pertsolution}
\end{equation}
The expansions \eqref{eq:Thpertexpansions} can be extended to higher-orders and it follows that Einstein's equations fix all coefficients in terms of $ n $ and $ L $. However, $ L $ is not fixed in terms of $ n $ meaning that $\sigma$ is not fixed to $\overbar{\sigma}_n$ by the perturbative expansion. We can see that if we do this manually by setting $ L=n $, the perturbative solution \eqref{eq:pertsolution} reduces to the Schwarzschild solution \eqref{eq:nonpertspherical}.

\subsection{Pure Lovelock gravity}\label{app:lovelockextremization}

For pure Lovelock gravity solving subleading parts of the equations of motion $E_{(m)b}^a = 0$ to solve the splitting problem is difficult and has not been done in the literature so far. In this appendix instead of solving the full splitting problem, we will only solve the leading term in the $E^{\bar{z}}_{(m)z}$ component which gives the equation for the embedding of $\mathcal{C}$ (the generalization of the area extremization prescription of Einstein gravity).

The equation of motion tensor of pure Lovelock gravity is given by \eqref{lovelockEOM} from which we get
\begin{align}
E^{\bar{z}}_{(m)z} &= -\frac{1}{2^{m+1}}\delta^{\bar{z}a_1b_1\ldots a_mb_m}_{z c_1d_1\ldots c_md_m}R^{c_1d_1}_{a_1b_1}\cdots R^{c_md_m}_{a_mb_m}\\
&= -\frac{4m}{2^{m+1}}\delta^{B_1A_2B_2\ldots A_mB_m}_{D_1C_2D_2\ldots C_mD_m}R^{\bar{z}D_1}_{zB_1}R^{C_2D_2}_{A_2B_2}\cdots R^{C_mD_m}_{A_mB_m} + \ldots\ .
\label{Ezzbarapp}
\end{align}
Substituting the expansion \eqref{Hexp} to \eqref{eq:riemanncomponentstwoup}, we get
\begin{equation}
R'^{\bar{z}A}_{zB} = -n(n-1)\,\mathcal{K}_{0(z)B}^{A}\,z^{n-2} + \ldots\;.
\end{equation}
Substituting this to \eqref{Ezzbarapp} gives
\begin{equation}
E'^{\bar{z}}_{(m)z} = -2m\,n(n-1)\,\mathfrak{E}'^{A}_{(m-1)B} \mathcal{K}_{0(z)A}^{B}\,z^{n-2} + \ldots
\end{equation}
where we have identified
\begin{equation}
\mathfrak{E}^{A}_{(m)B} = -\frac{1}{2^{m+1}}\delta^{AA_1B_1\ldots A_mB_m}_{BC_1D_1\ldots C_mD_m}R^{C_2D_2}_{A_2B_2}\cdots R^{C_mD_m}_{A_mB_m}
\end{equation}
which is the equation of motion tensor \eqref{lovelockEOM} with all Riemann tensors replaced by their corner projected counterparts \eqref{eq:cornerprojectedriemann}. Including the condition obtained from the complex conjugate equation $E^{z}_{(m)\bar{z}} = 0$, we get that $\overbar{g}_n$ satisfies the conditions
\begin{equation}
    \overbar{\mathfrak{E}}_{(m-1)ab}'\,\overbar{\mathcal{K}}_{0(p)}^{ab} = 0\,\quad p = z,\bar{z}.
\end{equation}

\section{Calculation of corner angles} \label{app:Thetam}

Here we compute the opening angles of the wedges obtained by cutting open the Hartle--Hawking and fixed-area manifolds at the codimension-2 surface where the Euclidean time shrinks to zero size.

Let $m,k$ be positive integers and consider the manifold $(\mathcal{M}_m,g_k)$ such that near the codimension-2 surface $\mathcal{C}$ we have
\begin{equation}
    ds^{2}=d\rho^{2}+\frac{\rho^{2}}{k^{2}}\,d\tau^{2}+\sigma_{AB}\,d\hat{x}^{A}d\hat{x}^{B}+\ldots,\quad \tau \sim \tau + 2\pi m.
    \label{eq:metrickm}
\end{equation}
The subscript $m$ on $\mathcal{M}_m$ indicates the periodic identification of $\tau$ while the subscript $k$ of $g_k$ indicates the coefficient of $d\tau^2$. This metric has a conical singularity at with angular deficit or excess given by
\begin{equation}
    \Delta\tau = 2\pi\,(1-mk^{-1}).
\end{equation}
The two cases of interest in the main text are $\mathbb{Z}_n$-quotients of Hartle--Hawking states ($k = n$,  $m = 1$) and fixed-area states ($k = 1$,  $m = n$). In the former case, there is always a conical deficit, while in the latter case, it is always a conical excess, assuming $n > 1$.

Cutting the manifold $(\mathcal{M}_m,g_k)$ open along the surface $\mathcal{B} = \{\rho>0,\tau=0\}$ produces a wedge shaped manifold $(\mathcal{W}_m,g_k)$ with two boundaries $\mathcal{B}_1,\mathcal{B}_2$ corresponding to two sides of the cut:
\begin{gather}
\mathcal{B}_\alpha = \{\rho>0,\tau=\tau_\alpha\}, \quad\alpha=1,2,\\
    \tau_1 = \epsilon,\quad \tau_2 = 2\pi m-\epsilon,
\end{gather}
where $\epsilon \rightarrow 0^+$ is a small regulator. The angle between the normal vectors of the two wedges is given by
\begin{equation}
    \cos{\Theta_{m,k}} = g_{kab}\,n^a_1n^b_2\lvert_{\rho = 0}.
    \label{eq:Thetamkapp}
\end{equation}
To compute these, we need to move to a coordinate system that covers the origin $\rho = 0$. We introduce Cartesian coordinates
\begin{equation}
    y^{1} = \rho \cos{(\tau/k)},\quad y^{2} = \rho\sin{(\tau/k)}.
\end{equation}
in which the metric \eqref{eq:metrickm} takes the form
\begin{equation}
    ds^{2}=(dy^1)^2 +(dy^2)^2+\sigma_{AB}\,d\hat{x}^{A}d\hat{x}^{B}+\ldots\ .
\end{equation}
With the aforementioned Cartesian coordinates, the outward-pointing unit normal vectors of the two boundaries are
\begin{equation}
n_{\alpha}\lvert_{\mathcal{B}_\alpha}\,= \frac{\rho}{k}\,\partial_{\tau}\Bigr\lvert_{\tau=\tau_\alpha} = \sin{\frac{\tau_{\alpha}}{k}}\,\partial_{y^1} - \cos{\frac{\tau_{\alpha}}{k}}\,\partial_{y^2}, \quad \alpha = 1,2.  
\end{equation}
Substituting to \eqref{eq:Thetamkapp} gives
\begin{equation}
\Theta_{m,k} = \pi - (\tau_2-\tau_1)\,k^{-1} = \pi\,(1-2mk^{-1}).
\label{anglenonregular}
\end{equation}
Thus for Hartle--Hawking states and fixed-area states we get
\begin{equation}
    \Theta_{1,n} = \pi\,(1 - 2n^{-1}),\quad \Theta_{n,1} = \pi\,(1 - 2n)
\end{equation}
respectively. Here the important observation is that we are not restricting to the principal branch of arc cosine (similar to \cite{Takayanagi:2019tvn}). In particular, $ \mathcal{B}_2 $ lives on the sheet $ m $. From this it follows the angle between the two tangent vectors $r_{1,2}^a$ of $\mathcal{B}_{1,2}$ is given by $ \pi-\Theta_{m,k}  = 2\pi mk^{-1} $ so that the angular deficit or excess is $ \Delta \tau = (\pi-\Theta_{m,k})-2\pi = \Theta_{m,k} + \pi $.

\section{Variational identities}\label{app:variations}

Here we set our notation characterizing hypersurfaces and derive expressions for variations of various geometric quantities under an infinitesimal variation of the inverse metric $\delta g^{ab} $ on a manifold with two boundaries that intersect at a corner. We make frequent use of these variational identities.


\subsection{Notation for hypersurfaces}\label{app:hypersurfaceconventions}

Consider a $D$-dimensional Euclidean manifold $(\mathcal{M},g)$ with two codimension-1 surfaces $ \mathcal{B}_\alpha \subset \mathcal{M} $ that intersect at a codimension-2 surface $\mathcal{C} = \mathcal{B}_1\cap \mathcal{B}_2$. The coordinates on $ \mathcal{M} $ are denoted by $ X^{a} $ with $ a=1,\ldots,D $ and the worldvolume coordinates of $ \mathcal{B}_{\alpha} $ are denoted by $ \tilde{x}^{i}_\alpha $ with $ i=1,\ldots,D-1 $. Similarly, coordinates of $\mathcal{C}$ are denoted by $ \hat{x}^{A} $ where $ A = 1,\ldots,D-2 $. We will denote the embedding of the surface $ \mathcal{B}_\alpha\subset \mathcal{M} $ by $ X^a = \tilde{E}^a_{\alpha}(\tilde{x}_\alpha) $ and the embedding of $\mathcal{C}\subset \mathcal{M}$ by $X^a = E^a(\hat{x}^A) $. Tangent vectors are defined as
\begin{equation}
    e^{a}_A = \frac{\partial E^a}{\partial \hat{x}^A},\quad \tilde{e}^{\,a}_{\alpha i} = \frac{\partial \tilde{E}^a_\alpha}{\partial \tilde{x}^i_\alpha}.
\end{equation}
In this article, a rank-2 tensor field on $\mathcal{M}$ is denoted by $T_{ab}$ and its pull-backs to the different surfaces are denoted as
\begin{equation}
    T_{AB} \equiv e^a_A\,e^b_B\,T_{ab}\lvert_{\mathcal{C}},\quad T_{\alpha ij} \equiv \tilde{e}^{\,a}_{\alpha i}\,\tilde{e}^{\,b}_{\alpha j}\,T_{ab}\lvert_{\mathcal{B}_{\alpha}}
\end{equation}
with the restriction to the surface included in the definition. These definitions are generalized to a tensor of any rank.

Let $ S_\alpha(X) $ be a function that vanishes at the surface $\mathcal{B}_\alpha$ as $S_\alpha(\tilde{E}_\alpha(\tilde{x}_\alpha)) = 0$. The outward-pointing unit normal vector field $ n^{a}_{\alpha} = g^{ab}\,n_{\alpha b} $ of $ \mathcal{B}_{\alpha} $ is given by
\begin{equation}
n_{\alpha a} = \Omega_{\alpha}\,\partial_a S_{\alpha}, \quad \Omega^{-2}_{\alpha} = g^{ab}\,\partial_a S_{\alpha}\,\partial_b S_{\alpha},
\label{nOmega}
\end{equation}
and the projector onto $ \mathcal{B}_{\alpha} $ is
\begin{equation}
h_{\alpha ab} = g_{\alpha ab} - n_{\alpha a}n_{\alpha b}.
\label{Bprojector}
\end{equation}
We will denote the unit normalized orthogonal projection of $ n_{2} $ onto $ \mathcal{B}_1 $ by $ r_1 $ with a similar definition of $ r_2 $, in other words,
\begin{equation}
r_{1}^{a} = \csc{\Theta}\,h^{a}_{1 b}\,n_{2}^{b}, \quad r_{2}^{a} = \csc{\Theta}\,h^{a}_{2 b}\,n_{1}^{b},
\end{equation}
where the intersection angle is defined via
\begin{equation}
\cos{\Theta} = g_{ab}\,n^{a}_{1}n^{b}_{2}.
\label{intersectionangleapp}
\end{equation}
Using \eqref{Bprojector} we get explicitly
\begin{equation}
r_{1}^{a} = -\cot{\Theta}\,n_{1}^{a}+\csc{\Theta}\,n_{2}^{a}, \quad r_{2}^{a} = \csc{\Theta}\,n_{1}^{a}-\cot{\Theta}\,n_{2}^{a},
\label{r1r2}
\end{equation}
which also allows us to express $ \{n_1^{a},r_1^{a}\} $ as linear combinations of $ \{n_2^{a},r_2^{a}\} $ as
\begin{align}
n_{2}^{a} &= n_{1}^{a}\cos{\Theta}+r_{1}^{a}\sin{\Theta}\nonumber\\
r_{2}^{a} &= n_{1}^{a}\sin{\Theta}-r_{1}^{a}\cos{\Theta}.
\label{vectorrotation}
\end{align}
We also define the projector onto $\mathcal{C}$ as
\begin{equation}
\sigma_{ab} = g_{ab} - n_{\alpha a}n_{\alpha b} - r_{\alpha a}r_{\alpha b}.
\label{sigma}
\end{equation}
The extrinsic curvature of $\mathcal{B}_{\alpha}$ is defined as
\begin{equation}
K_{\alpha ab} = h^{c}_{\alpha a}\,h^{d}_{\alpha b}\,\nabla_c n_{\alpha d}
\end{equation}
and the two  extrinsic curvatures of $\mathcal{C}$ are given by 
\begin{equation}
L_{\alpha ab} = \sigma^{c}_{a}\,\sigma^{d}_{b}\,\nabla_c n_{\alpha d}, \quad Q_{\alpha ab} = \sigma^{c}_{a}\,\sigma^{d}_{b}\,\nabla_c r_{\alpha d}.
\label{twoextrinsics}
\end{equation}
Due to \eqref{vectorrotation}, they are related as
\begin{align}
L_{2}^{ab} &= L_{1}^{ab}\cos{\Theta}+Q_{1}^{ab}\sin{\Theta}\nonumber\\
Q_{2}^{ab} &= L_{1}^{ab}\sin{\Theta}-Q_{1}^{ab}\cos{\Theta}.
\label{rotextrLQ}
\end{align}

\subsection{Variations with respect to the inverse metric}

We consider variations $ \delta g^{ab} $ of the inverse bulk metric and assume that the embeddings of $ \mathcal{B}_{\alpha}$ are kept fixed under the variation $\delta E_{\alpha} = 0$. Thus also the embedding of $ \mathcal{C} $ is fixed.
From \eqref{nOmega} it follows that \cite{Jiang:2018sqj}
\begin{equation}
\delta n_{\alpha a} = \delta \omega_{\alpha}\,n_{\alpha a}, \quad \delta n^{a}_{\alpha} = -\delta \omega_{\alpha}\,n^{a}_{\alpha} -\deltabar A^{a}_{\alpha}
\label{deltan}
\end{equation}
where we have defined
\begin{equation}
\delta \omega_{\alpha} \equiv  -\frac{1}{2}\,n_{\alpha a}n_{\alpha b}\,\delta g^{ab}, \quad \deltabar A^{a}_{\alpha} \equiv -h^{a}_{\alpha b}\,n_{\alpha c}\,\delta g^{bc},
\label{deltaomega}
\end{equation}
and $ \deltabar A^{a}_{\alpha}  $ is tangent to $ \mathcal{B}_{\alpha}  $ since $ n_{\alpha a}\,\deltabar A^{a}_{\alpha}  = 0 $. The notation $ \deltabar $ indicates $ \deltabar A^a_\alpha $ is not a total variation of any vector field. Using \eqref{deltan}, the variation of the metric can be decomposed as \cite{Jiang:2018sqj}
\begin{equation}
\delta g^{ab} = \delta h^{ab}_{\alpha}- 2\,\delta \omega_{\alpha}\,n^{a}_{\alpha} n^{b}_{\alpha} -2\,n^{(a}_{\alpha} \deltabar A^{b)}_{\alpha}.
\label{deltag1}
\end{equation}
For the $ \mathcal{B}_{\alpha} $ projected Christoffel symbols, one can prove the formula (see also \cite{Cano:2018ckq})
\begin{equation}
h^{d}_{\alpha a}\,h^{e}_{\alpha b}\,h^c_{\alpha f}\, \delta \Gamma^{f}_{de} = \delta \widetilde{\Gamma}^{c}_{\alpha ab}+K_{\alpha ab}\,\deltabar A^{c}_{\alpha},
\label{deltaGamma}
\end{equation}
where $\widetilde{\Gamma}^{c}_{\alpha ab}$ denote the induced Christoffel symbols built from $ h_{\alpha ab} $. 

Using \eqref{deltan}, the variation of the intersection angle \eqref{intersectionangleapp} can be written in two ways as
\begin{equation}
    \delta\Theta =(\delta\omega_1 - \delta \omega_2)\,\cot{\Theta} + r_{1a}\,\deltabar A^{a}_{1} = -(\delta\omega_1 - \delta \omega_2)\,\cot{\Theta} + r_{2a}\,\deltabar A^{a}_{2}.
    \label{deltaThetamid}
\end{equation}
Adding the two expressions together gives
\begin{equation}
	\delta\Theta = \frac{1}{2}\,(r_{1a}\,\deltabar A^{a}_{1}+r_{2a}\,\deltabar A^{a}_{2}) = -\frac{1}{2}\,(n_{1a}r_{1b}+n_{2a} r_{2b})\,\delta g^{ab}.
	\label{deltaThetanoBCapp}
\end{equation}

\subsection{Variations under periodic boundary conditions}\label{app:dirichletvariations}
 
We will now restrict to variations $ \delta g^{ab} $ that satisfy periodic boundary conditions at the boundaries $\mathcal{B}_\alpha$ as
\begin{equation}
    \delta h_1^{ab}\lvert_{\mathcal{B}_1} = \delta h_2^{ab}\lvert_{\mathcal{B}_2}.
    \label{periodicapp}
\end{equation}
First, equation \eqref{deltag1} combined with the periodic boundary condition implies the algebraic equation
\begin{equation}
\delta \omega_2\,n^{a}_2n^{b}_2+\deltabar A^{(a}_2n^{b)}_2=\delta \omega_1\,n^{a}_1n^{b}_1+\deltabar A^{(a}_1n^{b)}_1, \quad \text{at }\mathcal{C}.
\label{maineq}
\end{equation}
As shown in \cite{Jiang:2018sqj}, this equation has the solution\footnote{Note that in \cite{Jiang:2018sqj} this equation arises for Dirichlet boundary conditions $\delta h_1^{ab}\lvert_{\mathcal{B}_1} = \delta h_2^{ab}\lvert_{\mathcal{B}_2} = 0$, but it also applies for our periodic boundary conditions.}
\begin{equation}
    \delta \omega_{1} = \delta \omega_{2} \equiv \delta \omega, \quad r_{\alpha a}\,\deltabar A^a_\alpha = \tan{\Theta}\,\delta\omega\quad \deltabar \hat{A}^{a}_1 = \deltabar \hat{A}^{a}_2 = 0,
    \label{eq:solution}
\end{equation}
where we have defined the projection $\deltabar\hat{A}^{a}_\alpha = \sigma^a_b\,\deltabar A^b_\alpha$ and all quantities are evaluated at $\mathcal{C}$. Combining with \eqref{deltaThetamid} gives
\begin{equation}
\delta \omega_{\alpha} = \cot{\Theta}\,\delta \Theta , \quad \deltabar A^{a}_{\alpha} = r^{a}_{\alpha}\,\delta\Theta, \quad \text{at }\mathcal{C}.
\label{deltaomegaAwedge}
\end{equation}
Using \eqref{vectorrotation} and \eqref{deltan}, we get for the variations of the normal vectors
\begin{gather}
\delta n_{\alpha a}=\cot{\Theta}\,\delta\Theta\,n_{\alpha a}\\
\delta n_{1}^{a}\, = - \csc{\Theta}\,\delta\Theta\,n^{a}_{2}, \quad \delta n_{2}^{a} = - \csc{\Theta}\,\delta\Theta\,n^{a}_{1},\quad \text{at }\mathcal{C}.
\end{gather}
Substituting this into the variation of \eqref{r1r2} gives
\begin{equation}
\delta r_{\alpha a} = n_{\alpha a}\,\delta \Theta, \quad \delta r_{\alpha}^{a} =0,\quad \text{at }\mathcal{C}.
\label{deltarcorner}
\end{equation}
One can see that these are consistent with $\delta(n_{\alpha}^an_{\alpha a}) = \delta(r_{\alpha}^ar_{\alpha a}) = 0$.

We will now impose a Dirichlet boundary condition at the corner,
\begin{equation}
    \delta \sigma^{ab}\lvert_{\mathcal{C}}\, = 0,
    \label{dirichletapp}
\end{equation}
in addition to the periodic boundary condition \eqref{periodicapp} at the edges. Since $h_{\alpha}^{ab} = \sigma_{\alpha}^{ab}  - r_{\alpha}^ar_{\alpha}^b$, the formula \eqref{deltarcorner} implies
\begin{equation}
    \delta h_{\alpha}^{ab}\lvert_{\mathcal{C}}\, = 0.
    \label{hdirichletapp}
\end{equation}
Considering variations of extrinsic curvatures \eqref{twoextrinsics}, we have
\begin{equation}
\delta Q_{\alpha AB} = e^{a}_{A}\,e^{b}_{B}\,\nabla_{a}\delta r_{\alpha b}\lvert_{\mathcal{C}}\, -\, e^{a}_{A}\,e^{b}_{B}\,r_{\alpha c}\,\delta\Gamma_{ab}^{c}
\label{deltaQfirst}
\end{equation}
where we used that the embedding of $ \mathcal{C} $ is kept fixed. Employing \eqref{deltarcorner}, we find for the first term
\begin{equation}
e^{a}_{A}\,e^{b}_{B}\,\nabla_{a}\delta r_{\alpha b}\lvert_{\mathcal{C}}\, = L_{\alpha AB}\,\delta \Theta.
\end{equation}
Using \eqref{deltaGamma}, we get for the second term
\begin{equation}
e^{a}_{A}\,e^{b}_{B}\,r_{\alpha c}\,\delta\Gamma_{ab}^{c}\lvert_{\mathcal{C}}\, = L_{\alpha AB}\,r_{\alpha a}\,\deltabar A_{\alpha}^{a}\lvert_{\mathcal{C}}\, =  L_{\alpha AB}\,\delta \Theta,
\end{equation}
where we used that $ \delta \widetilde{\Gamma}^{c}_{\alpha ab}\lvert_{\mathcal{C}}\, = 0 $ due to \eqref{hdirichletapp}. Substituting these two equations into \eqref{deltaQfirst} yields
\begin{equation}
\delta Q_{\alpha AB} = 0.
\label{deltaQdir}
\end{equation}
Substituting to the variation of the rotation relation \eqref{rotextrLQ} gives
\begin{equation}
\delta L_{1 AB} = -\csc{\Theta}\,\delta \Theta\,L_{2 AB}, \quad \delta L_{2 AB} = -\csc{\Theta}\,\delta \Theta\,L_{1 AB}.
\label{deltaLdir}
\end{equation}
We are working with variations that satisfy a Dirichlet boundary condition \eqref{dirichletapp} at the corner so that \eqref{deltaQdir} - \eqref{deltaLdir} apply to $ \delta L^{A}_{\alpha B} $ and $ \delta Q^{A}_{\alpha B} $ as well.


\section{Variations of gravitational actions in the presence of corners}\label{app:actionvariation}

In this appendix we compute variations of gravitational actions in the presence of corners. For Einstein and Lovelock gravity, we consider actions with and without boundary and corner terms for various boundary conditions.

\subsection{$F(\text{Riemann})$ gravity}

Consider general diffeomorphism invariant $F(\text{Riemann})$ theory of gravity with Euclidean action
\begin{equation}
I_{\mathcal{M}}[g] = -\int_{\mathcal{M}} d^D x \sqrt{g}\,F(g^{ab},R_{abcd}).
\label{Friemannapp}
\end{equation}
The variation of the Lagrangian under variations of the bulk inverse metric is given by
\begin{equation}
\delta(\sqrt{g}\, F) = \sqrt{g}\,(E_{ab}\,\delta g^{ab}+\nabla_a\,\deltabar v^{a})\;,
\end{equation}
where the equation of motion tensor $E_{ab}$ and $\deltabar v^{a}$ are (see, e.g.,  \cite{Padmanabhan:2013xyr})
\begin{align}
	E_{ab} &= P_{acde}R_{b}^{\;\;\,cde}-\frac{1}{2}\,g_{ab}F +2\nabla^{c}\nabla^{d}P_{acbd}\,,\\ \deltabar v^{c} &= 2P_{a}^{\;\;bcd}\delta \Gamma^{a}_{bd}+2\delta g_{bd}\nabla_aP^{abcd}.
	\label{variationquantities}
\end{align}
Here we introduced the tensor
\begin{equation}
	P_{abcd}  \equiv \frac{\partial F}{\partial R^{abcd}},
	\label{Ptensor}
\end{equation}
which manifestly inherits all of the algebraic symmetries of the Riemann tensor.  Clearly, the equations of motion will in general involve higher than second-order derivatives of the metric. The total variation of the action \eqref{Friemann} then is
\begin{equation}
\delta I_{\mathcal{M}}[g] = -\int_{\mathcal{M}} d^D x \sqrt{g}\,E_{ab}\,\delta g^{ab}  - \int_{\partial \mathcal{M}}d^{D-1} x \sqrt{h}\,n_{a}\,\deltabar v^{a}\;,
\label{actionvariation}
\end{equation}
where $ h_{ab} $ is the induced metric of the boundary $ \partial \mathcal{M} $ and $ n^{a} $ is its outward-pointing unit normal vector. In Euclidean signature, the boundary variation can be decomposed as
\begin{equation}
n_{a}\,\deltabar v^{a} = n_{a}\,\deltabar W^{a} + D_{a}\,\deltabar Y^{a}
\label{eq:nvvariation}
\end{equation}
where the two terms are given by \cite{Jiang:2018sqj}
\begin{align}
	n_a\,\deltabar W^{a} &\equiv -4\Psi_{ab}\,\delta K^{ab}+(2n^{a}\nabla^{d}P_{acbd}+6\Psi_{ab}\,K^{a}_{c})\,\delta h^{bc} - 2n^{d}P_{acbd} D^{a}\delta h^{bc}\label{deltabarU}\\
	\deltabar Y^{a} &\equiv -2\Psi_{b}^a\,\deltabar A^{b}.\label{deltabarY}
\end{align}
Here $ K_{ab} = h_{a}^{c}h^{d}_{b}\,\nabla_{c}n_d $ is the extrinsic curvature of $ \partial \mathcal{M} $, $ D_{a} $ is the covariant derivative compatible with the induced metric $ h_{ab} $, $ \deltabar A^{a} = -h^{a}_{b}\,n_c\,\delta g^{bc} $, and we have defined the symmetric tensor
\begin{equation}
	\Psi_{ab} \equiv P_{acbd}\,n^{c}n^{d}.
	\label{Psiab}
\end{equation}

We will now consider a wedge $(\mathcal{W},g)$ with two boundary components $ \partial \mathcal{W} = \mathcal{B}_{1}\cup \mathcal{B}_2 $ that intersect at a corner $\mathcal{C} = \mathcal{B}_1\cap \mathcal{B}_2$. Since $\partial \mathcal{B}_{1,2}= \mathcal{C}$, we get by using \eqref{eq:nvvariation} the variation (without any boundary terms added)
\begin{align}
	\delta I_{\mathcal{W}}[g]&= -\int_{\mathcal{W}} \hspace{-1mm} d^{D} x \sqrt{g}\,E_{ab}\,\delta g^{ab}\label{totalvariation}\\
 &-\sum_{\alpha =1}^{2}\int_{\mathcal{B}_{\alpha}} \hspace{-2mm} d^{D-1} x \sqrt{h_\alpha}\,n_{\alpha a}\,\deltabar W^{a}_{\alpha} - \sum_{\alpha =1}^{2}\int_{\partial\mathcal{B}_{\alpha}} d^{D-2} x \sqrt{\sigma}\, r_{\alpha a}\,\deltabar Y^{a}_{\alpha}.\nonumber	
\end{align}
The corner variation can be written explicitly as
\begin{equation}
    -\sum_{\alpha =1}^{2}\int_{\partial\mathcal{B}_{\alpha}} d^{D-2} x \sqrt{\sigma}\, r_{\alpha a}\,\deltabar Y^{a}_{\alpha} = 2\sum_{\alpha =1}^{2}\int_{\mathcal{C}} d^{D-2} x \sqrt{\sigma}\, r_{\alpha a}\Psi^a_{\alpha b}\,\deltabar A^b_{\alpha}\;,
    \label{cornervarmid}
\end{equation}
where we substituted \eqref{deltabarY} and used that all curvature invariants of the metric are finite at $\mathcal{C}$ to push the integrals from $\partial \mathcal{B}_\alpha$ to $\mathcal{C}$. By inserting a Kronecker delta $\delta^a_b = \sigma^a_b + n^a_{\alpha}n_{\alpha b}+r^a_{\alpha}r_{\alpha b}$ in each term, e.g., $r_{1a}\Psi^{a}_{1b}\,\deltabar A^{b}_{1}=r_{1a}\Psi^{a}_{1b}\,\delta^{b}_{c}\,\deltabar A^{c}_{1}$, this can be decomposed as
\begin{align}
	-\sum_{\alpha =1}^{2}\int_{\mathcal{C}} d^{D-2} x \sqrt{\sigma}\, r_{\alpha a}\,\deltabar Y^{a}_{\alpha} &= \int_{\mathcal{C}}d^{D-2} x \sqrt{\sigma}\, \Psi\,\delta \Theta \label{cornervariationdecomposition}\\
 &+2\int_{\mathcal{C}}d^{D-2} x \sqrt{\sigma}\,(r_{1a}\Psi^a_{1 b}\,\deltabar \hat{A}_{1}^{b}+r_{2a}\Psi^a_{2 b}\,\deltabar \hat{A}_{2}^{b}),\nonumber
\end{align}
where we introduced the Wald entropy density \cite{Wald:1993nt,Iyer:1994ys}
\begin{equation}
    \Psi \equiv 4\Psi_{\alpha ab}\,r^a_{\alpha}r^b_{\alpha} = 4P_{abcd}\,n_{\alpha}^{[a} r^{b]}_\alpha n_{\alpha}^{[c} r^{d]}_\alpha,
\label{eq:Waldentdensityapp}\end{equation} 
defined the corner projection $	\deltabar\hat{A}^{a}_{\alpha} \equiv \sigma^{a}_{b}\,\deltabar A^{b}_{\alpha}$, and used the formula \eqref{deltaThetanoBCapp} for the variation $\delta\Theta$. Also note, via \eqref{Psiab}, $\Psi_{\alpha ab}\,n_{\alpha}^{b}=0$ due to the algebraic symmetries of $P_{abcd}$. Hence the variation (\ref{totalvariation}) is given by
\begin{align}
    \delta I_{\mathcal{W}}[g] &=  -\int_{\mathcal{W}} \hspace{-1mm} d^{D} x \sqrt{g}\,E_{ab}\,\delta g^{ab}-\sum_{\alpha =1}^{2}\int_{\mathcal{B}_{\alpha}} \hspace{-2mm} d^{D-1} x \sqrt{h_\alpha}\,n_{\alpha a}\,\deltabar W^{a}_{\alpha}\nonumber\\
    &+\int_{\mathcal{C}}d^{D-2} x\sqrt{\sigma}\, \Psi\,\delta\Theta+2\int_{\mathcal{C}}d^{D-2} x \sqrt{\sigma}\,(r_{1a}\Psi^a_{1 b}\,\deltabar \hat{A}_{1}^{b}+r_{2a}\Psi^a_{2 b}\,\deltabar \hat{A}_{2}^{b})\;.
    \label{eq:actionvarnoBCs}
\end{align}

\paragraph{Periodic boundary conditions at the edges.} As in Appendix \ref{app:dirichletvariations}, we will impose periodic boundary conditions at the edges $\mathcal{B}_{1,2}$ and a Dirichlet boundary condition at the corner $\mathcal{C}$, i.e.,
\begin{equation}
    h_1^{ab}\lvert_{\mathcal{B}_1}\, = h_2^{ab}\lvert_{\mathcal{B}_2},\quad \sigma^{ab}\lvert_{\mathcal{C}}\, = \text{fixed}.
    \label{eq:periodicdirichlet}
\end{equation}
As proven in Appendix \ref{app:dirichletvariations}, this amounts to a Dirichlet condition at the corner
\begin{equation}
    \delta \sigma^{ab}|_{\mathcal{C}} = \delta h_1^{ab}|_{\mathcal{C}} = \delta h_2^{ab}|_{\mathcal{C}}  = 0\;,
    \label{Dirichletperiodic}
\end{equation}
which also implies $\deltabar\hat{A}^{a}_{1} = \deltabar\hat{A}^{a}_{2} = 0$ as written in \eqref{eq:solution}. Hence the last term of second line vanishes leaving us with
\begin{equation}
    \delta I_{\mathcal{W}}[g] =  -\int_{\mathcal{W}} \hspace{-1mm} d^{D} x \sqrt{g}\,E_{ab}\,\delta g^{ab}-\sum_{\alpha =1}^{2}\int_{\mathcal{B}_{\alpha}} \hspace{-2mm} d^{D-1} x \sqrt{h_\alpha}\,n_{\alpha a}\,\deltabar W^{a}_{\alpha}+\int_{\mathcal{C}}d^{D-2} x\sqrt{\sigma}\, \Psi\,\delta\Theta
    \label{cornervarPBC}
\end{equation}
under periodic boundary conditions \eqref{eq:periodicdirichlet} for the metric.

\paragraph{Dirichlet boundary conditions at the edges.} For completeness, we also give the formula when we impose Dirichlet boundary conditions at the edges $\mathcal{B}_{1,2}$, i.e.,
\begin{equation}
    h_1^{ab}\lvert_{\mathcal{B}_1}\, = \text{fixed},\quad  h_2^{ab}\lvert_{\mathcal{B}_2}\, = \text{fixed},\quad \sigma^{ab}\lvert_{\mathcal{C}}\, = \text{fixed}\;,
\end{equation}
which amounts to
\begin{equation}
     \delta h_1^{ab}|_{\mathcal{B}_1} = \delta h_2^{ab}|_{\mathcal{B}_2}  = 0\;.
     \label{eq:dirichletedges}
\end{equation}
In this case only the first term of \eqref{deltabarU} survives and we get
\begin{equation}
    \delta I_{\mathcal{W}}[g] =  -\int_{\mathcal{W}} \hspace{-1mm} d^{D} x \sqrt{g}\,E_{ab}\,\delta g^{ab}+\sum_{\alpha =1}^{2}\int_{\mathcal{B}_{\alpha}} \hspace{-2mm} d^{D-1} x \sqrt{h_\alpha}\,4\Psi_{\alpha ab}\,\delta K^{ab}_{\alpha}+\int_{\mathcal{C}}d^{D-2} x\sqrt{\sigma}\, \Psi\,\delta\Theta\,.
    \label{cornervarDBC}
\end{equation}




\subsection{Einstein gravity}\label{subapp:Einsteinsplittings}


In Einstein gravity, we have $P_{abcd}=\frac{1}{2}(g_{ac}g_{bd}-g_{ad}g_{bc})$ which gives
\begin{equation}
    \Psi_{ab} = \frac{1}{2}\,h_{ab},\quad \Psi = 2,\quad n_a\,\deltabar W^{a} = -2h_{ab}\,\delta K^{ab}+3K_{ab}\,\delta h^{ab},\quad \deltabar Y^{a} =-\deltabar A^{a},
\end{equation}
where we used $\nabla^aP_{abcd} = 0$ and $n_a\,D^{a}\delta h^{bc} = n_b\,D^{a}\delta h^{bc} = 0$. In the case of Einstein gravity, the second term in \eqref{cornervariationdecomposition} vanishes without imposing any boundary conditions because $r_{\alpha a}\,\deltabar \hat{A}^a_\alpha = 0$. Hence varying the Einstein--Hilbert term alone yields
\begin{align}
    -\delta\int_{\mathcal{W}} d^D x \sqrt{g}\,R&=  -\int_{\mathcal{W}} \hspace{-1mm} d^{D} x \sqrt{g}\,G_{ab}\,\delta g^{ab}\label{cornervar}\\
    &+\sum_{\alpha =1}^{2}\int_{\mathcal{B}_{\alpha}} \hspace{-2mm} d^{D-1} x \sqrt{h_\alpha}\,(2h_{ab}\,\delta K^{ab}-3K_{ab}\,\delta h^{ab})+\int_{\mathcal{C}}d^{D-2} x\sqrt{\sigma}\, 2\delta\Theta\nonumber
\end{align}
valid for all metric variations. Now supplement the Einstein action a Gibbons--Hawking--York boundary term
\begin{equation}
    I_{\mathcal{W}}[g] = -\int_{\mathcal{W}} d^D x \sqrt{g}\,R -\sum_{\alpha=1}^{2}\int_{\mathcal{B}_{\alpha}}d^{D-1}x\sqrt{h_{\alpha}}\,2K_{\alpha}\;.
\label{eq:wedgeactGHYnocorn}\end{equation}
The variation of a Gibbons--Hawking--York term without imposing any boundary conditions is given by
\begin{equation}
    \delta(\sqrt{h}\,2K) = \sqrt{h}\,[2h_{ab}\,\delta K^{ab}-(2K_{ab} +K\,h_{ab})\,\delta h^{ab}]\;.
\end{equation}
where we used $\delta h_{ab} = -\,g_{ac}g_{bd}\,\delta h^{cd}$. Hence combining with \eqref{cornervar} gives
\begin{align}
    \delta I_{\mathcal{W}}[g]&= -\int_{\mathcal{W}}d^{D}x\sqrt{g}\,G_{ab}\,\delta g^{ab}\label{eq:einsteinvariationnocornertermapp}\\
    &-\sum_{\alpha=1}^{2}\int_{\mathcal{B}_{\alpha}}d^{D-1}x\sqrt{h_{\alpha }}\,(K_{\alpha ab} -K_\alpha\,h_{\alpha ab})\,\delta h^{ab}_{\alpha}+2\int_{\mathcal{C}}d^{D-2}x\sqrt{\sigma}\,\delta \Theta,\nonumber
\end{align}
valid for all metric variations. Let us further supplement the action (\ref{eq:wedgeactGHYnocorn}) with the Hayward corner term,
\begin{equation}
    I_{\mathcal{W}}[g]+I_{\mathcal{C}}[g] = -\int_{\mathcal{W}} d^D x \sqrt{g}\,R -\sum_{\alpha=1}^{2}\int_{\mathcal{B}_{\alpha}}d^{D-1}x\sqrt{h_{\alpha}}\,2K_{\alpha} - \int d^{D-2}x\sqrt{\sigma}\,2\Theta\;.
\end{equation}
The variation of the Hayward term is
\begin{equation}
    \delta(\sqrt{\sigma}\,2\Theta) =\sqrt{\sigma}\,(2\,\delta\Theta - \Theta\,\sigma_{ab}\,\delta\sigma^{ab})
\end{equation}
so that combining with \eqref{eq:einsteinvariationnocornertermapp} gives
\beq
\begin{split}
    &\delta I_{\mathcal{W}}[g]+\delta I_{\mathcal{C}}[g]\\
    &= -\int_{\mathcal{W}}d^{D}x\sqrt{g}\,G_{ab}\,\delta g^{ab}-\sum_{\alpha=1}^{2}\int_{\mathcal{B}_{\alpha}}d^{D-1}x\sqrt{h_{\alpha }}\,(K_{\alpha ab} -K_\alpha h_{\alpha ab})\,\delta h^{ab}_{\alpha}\\
    &-\int_{\mathcal{C}}d^{D-2}x\sqrt{\sigma}\,(-\Theta\,\sigma_{ab})\,\delta\sigma^{ab}\;,
    \label{eq:einsteinvariationapp}
\end{split}
\eeq
valid for all metric variations. From this expression one can identify the boundary and corner stress tensors of Einstein gravity.


\subsection{Pure Lovelock gravity}\label{subapp:lovelockactionvars}

In pure Lovelock gravity of order $m$, $\nabla^aP_{abcd} = 0$ as in Einstein gravity so that
\begin{equation}
	n_a\,\deltabar W^{a} = -4\Psi_{ab}\,\delta K^{ab}+6\Psi_{ab}\,K^{a}_{c}\,\delta h^{bc} - 2n^{d}P_{acbd}\, D^{a}\delta h^{bc}\;.
\label{eq:naWLove}\end{equation}
Consider the action of pure Lovelock gravity on a wedge with boundary terms included at the edges
\begin{equation}
    I_{\mathcal{W}}[g] = -\int_{\mathcal{W}} d^D x \sqrt{g}\,\mathcal{R}_{(m)} -\sum_{\alpha=1}^{2}\int_{\mathcal{B}_{\alpha}}d^{D-1}x\sqrt{h_{\alpha}}\,B_{\alpha}\;.
\label{eq:LoveactbdrynoC}\end{equation}
It is known from previous work on manifolds without corners that \cite{Myers:1987yn,Teitelboim:1987zz,Miskovic:2007mg}
\begin{equation}
    \delta(\sqrt{h}\,B) + \sqrt{h}\,n_a\,\deltabar W^{a} = \sqrt{h}\, (\widetilde{T}_{ab}\,\delta h^{ab}+D_a\,\deltabar\widetilde{Y}^a)
\label{eq:deltaBLove}\end{equation}
where the boundary stress tensor $\widetilde{T}_{ab}$ is given by \eqref{lovelockboundarystress}. The total derivative term $D_a\,\deltabar\widetilde{Y}^a$ arises from the variation of $B$ under $\delta h^{ab}$ after integration by parts but the expression for $\deltabar\widetilde{Y}^a$ has not been written down in the literature. Combining identities (\ref{eq:naWLove}) and \eqref{eq:deltaBLove} together with the general variation (\ref{eq:actionvarnoBCs}), varying the Lovelock action (\ref{eq:LoveactbdrynoC}) returns
\begin{align}
    \delta I_{\mathcal{W}}[g] &=  -\int_{\mathcal{W}} \hspace{-1mm} d^{D} x \sqrt{g}\,E_{ab}\,\delta g^{ab}-\sum_{\alpha =1}^{2}\int_{\mathcal{B}_{\alpha}} \hspace{-2mm} d^{D-1} x \sqrt{h_\alpha}\,\widetilde{T}_{ab}\,\delta h^{ab}+\int_{\mathcal{C}}d^{D-2} x\sqrt{\sigma}\, \Psi\,\delta\Theta\nonumber\\
    &+\int_{\mathcal{C}}d^{D-2} x \sqrt{\sigma}\,[r_{1a}(2\Psi^a_{1 b}\,\deltabar \hat{A}^{b}_1-\deltabar\widetilde{Y}^a_1)+r_{2a}(2\Psi^a_{2 b}\,\deltabar \hat{A}^{b}_2-\deltabar\widetilde{Y}^a_2)]\;,
    \label{eq:actionvarnoBCsLove}
\end{align}
valid for all metric variations. We expect the second line to vanish identically (see below). Let us now include a corner term to the action as
\begin{equation}
    I_{\mathcal{W}}[g]+I_{\mathcal{C}}[g] = -\int_{\mathcal{W}} d^D x \sqrt{g}\,\mathcal{R}_{(m)} -\sum_{\alpha=1}^{2}\int_{\mathcal{B}_{\alpha}}d^{D-1}x\sqrt{h_{\alpha}}\,B_{\alpha}-\int_{\mathcal{C}}d^{D-2} x \sqrt{\sigma}\,C\;.
\end{equation}
By the definition of the corner term \eqref{boundarycornertermrequirement}, its variation is given by
\begin{equation}
    \delta(\sqrt{\sigma}\,C) = \sqrt{\sigma}\,\biggl[\Psi\,\delta\Theta + \biggl(\Phi_{ab}-\frac{1}{2}\,C\,\sigma_{ab}\biggr)\,\delta\sigma^{ab}\biggr]\;,
\end{equation}
where $\Phi_{ab}$ comes from the variation of $C$ under $\delta \sigma^{ab}$ after integration by parts and we have neglected a total derivative term which vanishes inside integrals over the corner. Then the action (\ref{eq:LoveactbdrynoC}) supplemented with a corner term has the variation
\begin{align}
    \delta I_{\mathcal{W}}[g]+\delta I_{\mathcal{C}}[g] &=  -\int_{\mathcal{W}} \hspace{-1mm} d^{D} x \sqrt{g}\,E_{ab}\,\delta g^{ab}-\sum_{\alpha =1}^{2}\int_{\mathcal{B}_{\alpha}} \hspace{-2mm} d^{D-1} x \sqrt{h_\alpha}\,\widetilde{T}_{ab}\,\delta h^{ab}\nonumber\\
    &+\int_{\mathcal{C}}d^{D-2} x\sqrt{\sigma}\, \biggl(\Phi_{ab}-\frac{1}{2}\,C\,\sigma_{ab}\biggr)\,\delta\sigma^{ab}\nonumber\\
    &+\int_{\mathcal{C}}d^{D-2} x \sqrt{\sigma}\,[r_{1a}(2\Psi^a_{1 b}\,\deltabar \hat{A}^{b}_1-\deltabar\widetilde{Y}^a_1)+r_{2a}(2\Psi^a_{2 b}\,\deltabar \hat{A}^{b}_2-\deltabar\widetilde{Y}^a_2)]
    \label{eq:actionvarcornernoBCsC}
\end{align}
We have also calculated this variation using the smoothing method presented in Section \eqref{subsec:cornerderivations} (with details given in Appendix \ref{app:cornerstress}) and the result is given by \eqref{generaltheoryvariation},
\begin{align}
&\delta I_{\mathcal{W}}[g]+\delta I_{\mathcal{C}}[g]\\
&= -\int_{\mathcal{W}}d^{D}x\sqrt{g}\,E_{ab}\,\delta g^{ab}-\sum_{\alpha=1}^{2}\int_{\mathcal{B}_{\alpha}}d^{D-1}x\sqrt{h_{\alpha }}\,\widetilde{T}_{\alpha ab}\,\delta h^{ab}_{\alpha}-\int_{\mathcal{C}}d^{D-2}x\sqrt{\sigma}\,\widehat{T}_{ab}\,\delta \sigma^{ab}\;,\nonumber
\end{align}
with the boundary and corner stress tensors given by \eqref{lovelockboundarystress} and \eqref{lovelockcornerstress} respectively. Comparing to \eqref{eq:actionvarcornernoBCsC} implies
\begin{equation}
   \Phi_{ab}-\frac{1}{2}\,C\,\sigma_{ab} =  \widehat{T}_{ab} ,\quad \deltabar\widetilde{Y}^a = 2\Psi^a_{b}\,\deltabar \hat{A}^{b}.
   \label{eq:implication}
\end{equation}
Hence the second line of \eqref{eq:actionvarnoBCsLove} and the third line of \eqref{eq:actionvarcornernoBCsC} vanish identically without imposing any boundary conditions. This justifies the use of equation \eqref{lovelockvariation} in the main text in the derivation of the Hartle--Hawking entropy. We have not managed to prove \eqref{eq:implication} directly by computing variations, but the validity of the smoothing method requires it. It would be preferable to have a direct proof of the result.

\section{Derivation of Lovelock corner term and stress tensor}\label{app:cornerstress}

In this appendix we derive the corner term and stress tensor of Einstein, pure Gauss--Bonnet and general Lovelock gravities using the smoothing method presented in Section \ref{subsec:cornerderivations}. We recall the corresponding formulae \eqref{eq:CTtildelimits} here for convenience
\begin{align}
    C_{(m)} &= \lim_{\epsilon\rightarrow 0}\int_0^{\Theta}d\theta\,M\,B_{(m)}\\
   \widehat{T}_{(m)ab}\,\delta \sigma^{ab} &= \lim_{\epsilon\rightarrow 0}\int_0^{\Theta}d\theta\,M\,\widetilde{T}_{(m)ab}\,\delta Z^{ab}\;,
   \label{eq:CTtildelimitsapp}
\end{align}
where $B_{(m)}$ and $\widetilde{T}_{(m)ab}$ are given by \eqref{eq:LovelockGH} and \eqref{lovelockboundarystress} respectively.


The metric near the corner can always be written in the form \cite{Hayward:1993my, Cano:2018ckq}
\begin{equation}
    ds^2 = N^2du^2 + Z_{ij}\,dx^idx^j = N^2du^2 +M^2d\theta^2 + h_{AB}\,d\hat{x}^Ad\hat{x}^B
\end{equation}
and the cap $\mathcal{C}_\epsilon$ is given by the circular arc $0 < \theta< \Theta$ at $ u = \epsilon$. When $\epsilon\rightarrow 0$, the origin $u = 0$ is near the corner so that $h_{AB}$ has the expansion
\begin{equation}
    h_{AB} = \sigma_{AB} + L_{1AB}\,u\cos{\theta} + Q_{1AB}\,u\sin{\theta} +\mathcal{O}(u^2)  
\end{equation}
where $L_{1AB},Q_{1AB}$ are the two extrinsic curvatures of $\mathcal{C}$. We will be focusing on metrics which satisfy
\begin{equation}
    M\lvert_{u = 0}\, = 0,\quad \lim_{\epsilon\rightarrow 0}\frac{\partial_u M}{N}\bigg\lvert_{u = \epsilon}=1\;.
    \label{eq:MNlimit}
\end{equation}
The extrinsic curvature of the arc $\mathcal{C}_{\epsilon}$ is given by 
\begin{equation}
    K_{ij} = \frac{1}{2N}\,\partial_u Z_{ij}\lvert_{u = \epsilon}\;, 
\end{equation}
whose non-zero components in these coordinates being
\begin{equation}
    K_{\theta}^\theta = \frac{\partial_u M}{MN}\bigg\lvert_{u = \epsilon},\quad K^A_B = \frac{1}{2N}h^{AB}\partial_uh_{AB}\lvert_{u = \epsilon}\;.
    \label{eq:Kcomponents}
\end{equation}
In the near corner limit $\epsilon \rightarrow 0$, it follows from \eqref{eq:MNlimit} that
\begin{align}
    K_{\theta}^\theta &= \frac{1}{M} + \ldots\label{eq:divergentKthetatheta}\\
    K^A_B &= L^A_{1B}\cos{\theta} + Q_{1B}^A\sin{\theta} + \ldots\label{eq:Krotation}
\end{align}
where ellipsis denote subleading terms in $\epsilon$. Hence $K^{\theta}_{\theta}$ diverges in the limit $\epsilon\rightarrow 0$.

\subsection{Einstein gravity}

Let us first consider Einstein gravity. The corner term is given by
\begin{equation}
C_{(1)}=\lim_{\epsilon\rightarrow 0}\int_{0}^{\Theta} d\theta\,M\,2K= \lim_{\epsilon\rightarrow 0}\int_{0}^{\Theta} d\theta\,M\,2(K^{\theta}_{\theta} + K^{A}_{A}) = 2\Theta\;,
\end{equation}
which is the Hayward term. For the corner stress tensor, we get
\begin{align}
\widehat{T}_{(1)ab}\,\delta \sigma^{ab}&=\lim_{\epsilon\rightarrow 0}\int_{0}^{\Theta} d\theta\,M\,(K^{i}_{j} - K \delta^{i}_{j})\,\delta h^{j}_{i}\\
&= \lim_{\epsilon\rightarrow 0}\int_{0}^{\Theta} d\theta\,M\,\left[- K^{A}_{A}\,\delta h^{\theta}_{\theta}+\left(K^{A}_{B} - K^{\theta}_{\theta}\delta^{A}_{B} - K^{C}_{C}\delta^{A}_{B}\right)\,\delta \sigma^{B}_{A}\right]\;,
\end{align}
where we used $ K^{\theta}_{A} = 0 $ so that the $ \delta h^{\theta}_{A} $ term does not appear. The only term that gives a finite contribution in the limit $ \epsilon\rightarrow 0 $ is the term linear in $ K^{\theta}_{\theta} $ which diverges as $M^{-1}$ compensating against the overall factor of $M$. Hence we get after performing the remaining $ \theta $-integral
\begin{equation}
\widehat{T}_{(1)ab}\,\delta \sigma^{ab}= -\Theta\,\sigma_{AB}\,\delta \sigma^{AB},
\end{equation}
so that the corner stress tensor of Einstein gravity is given by
\begin{equation}
\widehat{T}_{(1)ab} = -\Theta\,\sigma_{ab}.
\end{equation}
This matches the corner stress tensor appearing in the variation \eqref{fulleinsteinvariation} of the Einstein--Hilbert action supplemented by boundary and corner terms.

\subsection{Gauss--Bonnet gravity}\label{app:GBcorner}

The boundary term of pure Gauss--Bonnet gravity is given by
\begin{equation}
B_{(2)}  = 2\delta^{ii_1j_1}_{jk_1l_1}K^{j}_{i}\left(R^{k_1l_1}_{i_1j_1} + \frac{4}{3}K_{i_1}^{k_1}K_{j_1}^{l_1}\right)\;. 
\end{equation}
Hence, the corner term is given by
\begin{align}
C_{(2)}&=\lim_{\epsilon\rightarrow 0}\int_{0}^{\Theta} d\theta\,M\,2\delta^{ii_1j_1}_{jk_1l_1}K^{j}_{i}\left(R^{k_1l_1}_{i_1j_1} + \frac{4}{3}K_{i_1}^{k_1}K_{j_1}^{l_1}\right)\\
&= \lim_{\epsilon\rightarrow 0}\int_{0}^{\Theta} d\theta\,2M\,\biggl[\delta^{\theta A_1B_1}_{\theta  C_1D_1}K^{\theta}_{\theta}\left(R^{C_1D_1}_{A_1B_1} + \frac{4}{3}K_{A_1}^{C_1}K_{B_1}^{D_1}\right)\biggr.\nonumber\\
&\quad \biggl.+2\delta^{A\theta B_1}_{B\theta D_1}K^{B}_{A}\left(R^{\theta D_1}_{\theta B_1} + \frac{4}{3}K_{\theta}^{\theta}K_{B_1}^{D_1}\right)+ \delta^{A \theta B_1}_{\theta C_1 D_1}K^{A}_{\theta}\left(R^{C_1D_1}_{\theta B_1} + \frac{4}{3}K_{\theta}^{C_1}K_{B_1}^{D_1}\right)\biggr]\;,
\label{Btermregularlimit}
\end{align}
and note the factor of two in the first term on the second line. Only terms linear in $ K^{\theta}_{\theta} $ survive the limit (the last term also vanishes by $ K_{\theta}^{A} = 0 $) with the result
\begin{equation}
C_{(2)} = \int_{0}^{\Theta} d\theta\, 2\delta^{A_1B_1}_{C_1D_1}\left(R^{C_1D_1}_{A_1B_1}+4K^{C_1}_{A_1}K^{D_1}_{B_1}\right)\;.
\label{Cintegral}
\end{equation}
Integrating over the first term gives simply a factor of $ \Theta $. For the second term, we can use \eqref{eq:Krotation} to perform the integral as
\begin{equation}
\int_{0}^{\Theta} d\theta\, K^{C}_{A}K^{D}_{B} = \frac{1}{2}\,\Theta\,(L_{1A}^{C}L_{1B}^{D}+Q_{1A}^{C}Q_{1B}^{D})+\frac{1}{2}\,(L_{1A}^{C}Q_{1B}^{D} + L_{2A}^{C}Q_{2B}^{D})\;,
\end{equation}
where we used that the rotation relation \eqref{rotextrLQ} implies
\begin{equation}
L_{1A}^{C}Q_{1B}^{D} + L_{2A}^{C}Q_{2B}^{D} = (L_{1A}^{C}L_{1B}^{D}-Q_{1A}^{C}Q_{1B}^{D})\sin{\Theta}\cos{\Theta}+2L_{1A}^{C}Q_{1B}^{D} \sin^{2}{\Theta}\;.
\end{equation}
Substituting this back into \eqref{Cintegral} gives
\begin{equation}
C_{(2)} = \Theta\,2\delta^{AB}_{CD}\left(R^{CD}_{AB} +2L_{1A}^{C}L_{1B}^{D}+2Q_{1A}^{C}Q_{1B}^{D}\right)+4\delta^{AB}_{CD}(L_{1A}^{C}Q_{1B}^{D} + L_{2A}^{C}Q_{2B}^{D})\;.
\end{equation}
Using the codimension-2 Gauss--Codazzi equation thus gives finally
\begin{equation}
C_{(2)} = 4\,\left(\Theta\,\widehat{R} + L_1Q_1+ L_2Q_2 -L_{1AB}Q_{1}^{AB}  -L_{2AB}Q_{2}^{AB}\right),
\end{equation}
where $ \widehat{R} $ is the Ricci scalar of $ \sigma_{AB} $. This matches with \cite{Cano:2018ckq}.

To obtain the corner stress tensor, we have to compute
\begin{align}
\widehat{T}_{(2)ab}\,\delta \sigma^{ab}&=\lim_{\epsilon\rightarrow 0}\int_{0}^{\Theta} d\theta\,M\,(-1)\,\delta h^{j}_{i}\, \delta^{iki_1j_1}_{jlk_1l_1}K^{l}_{k}\left(R^{k_1l_1}_{i_1j_1} + \frac{4}{3}K_{i_1}^{k_1}K_{j_1}^{l_1}\right)\\
&= \lim_{\epsilon\rightarrow 0}\int_{0}^{\Theta} d\theta\,M\,(-1)\biggl[\delta h^{B}_{A}\, \delta^{A ki_1j_1}_{B lk_1l_1}K^{l}_{k}\left(R^{k_1l_1}_{i_1j_1} + \frac{4}{3}K_{i_1}^{k_1}K_{j_1}^{l_1}\right)+\biggr.\\
&\biggl.+\delta h^{\theta}_{\theta}\, \delta^{\theta ki_1j_1}_{\theta lk_1l_1}K^{l}_{k}\left(R^{k_1l_1}_{i_1j_1} + \frac{4}{3}K_{i_1}^{k_1}K_{j_1}^{l_1}\right)+\delta h^{A}_{\theta}\, \delta^{\theta ki_1j_1}_{A lk_1l_1}K^{l}_{k}\left(R^{k_1l_1}_{i_1j_1} + \frac{4}{3}K_{i_1}^{k_1}K_{j_1}^{l_1}\right)\biggr]\;.\nonumber
\end{align}
In the second term and third terms, no $ K^{\theta}_{\theta} $ can appear since at least one $ \theta $ index is taken up by $ \delta h^{i}_{j} $. Hence only the first term gives a finite contribution in the limit. Then notice that the first term is exactly like the $ B $ term above, but with an extra pair of indices contracted with $ \delta h^{A}_{B} = \delta \sigma^{A}_{B} $. The limit of the first term is thus obtained in exactly the same way as \eqref{Btermregularlimit}:
\begin{align}
&\lim_{\epsilon\rightarrow 0}\int_{\mathcal{C}}d^{D-2}x\sqrt{\sigma}\int_{0}^{\Theta} d\theta\,M\,(-1)\,\delta h^{j}_{i}\, \delta^{iki_1j_1}_{jlk_1l_1}K^{l}_{k}\left(R^{k_1l_1}_{i_1j_1} + \frac{4}{3}K_{i_1}^{k_1}K_{j_1}^{l_1}\right)\\
&= \int_{\mathcal{C}}d^{D-2}x\sqrt{\sigma}\int_{0}^{\Theta} d\theta\,\delta h^{B}_{A}\, (-1)\,\delta^{A A_1B_1}_{B C_1D_1}\left(R^{C_1D_1}_{A_1B_1} + 4K_{A_1}^{C_1}K_{B_1}^{D_1}\right).
\end{align}
The integration over $ \theta $ works out the same way as above and we get the corner stress tensor of Gauss--Bonnet gravity
\begin{equation}
\widehat{T}^A_{(2)B} = 4\,\Theta\,\widehat{G}_{B}^{A} -2\delta^{AA_1B_1}_{BC_1D_1}(L_{1A_1}^{C_1}Q_{1B_1}^{D_1} + L_{2A_1}^{C_1}Q_{2B_1}^{D_1}),
\end{equation}
where $ \widehat{G}_{AB} $ is the Einstein tensor of the induced metric $ \sigma_{AB} $:
\begin{equation}
\widehat{G}_{B}^{A} = - \frac{1}{4}\delta^{A A_1B_1}_{B C_1D_1}R^{C_1D_1}_{A_1B_1} = \widehat{R}^{A}_{B} - \frac{1}{2}\widehat{R}\delta^{A}_{B}.
\end{equation}

\subsection{Pure Lovelock gravity}\label{app:lovelockcorner}

The previous calculations generalize straightforwardly to pure Lovelock gravity of order $m$. To this end, we expand the boundary term \eqref{eq:LovelockGH} as
\begin{equation}
B_{(m)} =\sum_{p=1}^{m}\frac{c_p}{2p-1} \delta^{i_1\ldots i_{2m-1}}_{j_1\ldots j_{2m-1}}K^{j_1}_{i_1} \cdots K^{j_{2p-1}}_{i_{2p-1}} R_{i_{2p}i_{2p+1}}^{j_{2p}j_{2p+1}} \cdots R_{i_{2m-2}i_{2m-1}}^{j_{2m-2}j_{2m-1}}
\label{eq:Bmexpanded}
\end{equation}
where the coefficients are
\begin{equation}
c_p = \frac{2m(2p-1)}{2^{m-p}}\binom{m-1}{p-1}\int_{0}^{1} ds\,(1-s^{2})^{p-1} = \frac{1}{2^{m-3p+2}}\frac{m!(p-2)!}{(2p-1)!(m-p)!}\;.
\label{eq:cpcoeffs}
\end{equation}
The corner term calculation limit works in the same way as in Gauss--Bonnet gravity with only terms linear in $K^\theta_\theta$ surviving the limit. The result is \cite{Cano:2018ckq}
\begin{equation}
C_{(m)} = \sum_{p=1}^{m}\int_{0}^{\Theta} d\theta\,c_p\,K^{2p-2}R^{m-p}\;,
\end{equation}
where the $ 2p-1 $ in the numerator of \eqref{eq:Bmexpanded} has cancelled and all indices are understood to be contracted with the generalized Kronecker delta symbol $ \delta^{A_1\ldots A_{2m-2}}_{B_1\ldots B_{2m-2}} $. After substituting \eqref{eq:Krotation}, the $\theta$-integration was done in \cite{Cano:2018ckq} with the result
\begin{align}
C_{(m)}&=\Theta\,\frac{2m}{2^{m-1}}\widehat{R}^{m-1}\label{lovelockcornermidstep}\\
&\qquad + \sum_{\alpha = 1}^{2}\sum_{l=2}^{m}\frac{m!(l-1)!}{2^{m-l-1}(m-l)!}R^{m-l}\,\text{Im}\,\biggl[\,\sum_{j=0}^{l-2}\frac{(L_{\alpha}-iQ_{\alpha})^{j}(L_{\alpha}+iQ_{\alpha})^{2l-2-j}}{j!\,(2l-2-j)!\,(l-j-1)}\,\biggr].\nonumber
\end{align}
Since all indices are contracted with a Kronecker delta this is explicitly \cite{Cano:2018ckq}
\begin{equation}
C_{(m)} = 2m\,\Theta\,\widehat{\mathcal{R}}_{(m-1)} + \mathcal{F}_{(m)}(L_1,Q_1) + \mathcal{F}_{(m)}(L_2,Q_2),
\end{equation}
where $ \mathcal{F}_{(m)}(L,Q) $ is defined in \eqref{canoF}.

To compute the corner stress tensor, expand the boundary stress tensor \eqref{lovelockboundarystress} as
\begin{equation}
\widetilde{T}_{(m)j}^i =\sum_{p=1}^{m}\frac{-c_p}{2\,(2p-1)} \delta^{ii_1\ldots i_{2m-1}}_{jj_1\ldots j_{2m-1}}K^{j_1}_{i_1} \cdots K^{j_{2p-1}}_{i_{2p-1}} R_{i_{2p}i_{2p+1}}^{j_{2p}j_{2p+1}} \cdots R_{i_{2m-2}i_{2m-1}}^{j_{2m-2}j_{2m-1}}\;,
\label{eq:Tmexpanded}
\end{equation}
where the coefficients are given by \eqref{eq:cpcoeffs}. The corner stress tensor becomes
\begin{align}
&\widehat{T}_{(m)B}^{A}\,\delta \sigma^{B}_{A}\nonumber\\
&=\lim_{\epsilon\rightarrow 0}\sum_{p=1}^{m}\int_{0}^{\Theta} d\theta\,M\,\frac{-c_p}{2\,(2p-1)}\,\delta h^{j}_{i}\,\delta^{ii_1\ldots i_{2m-1}}_{jj_1\ldots j_{2m-1}}K^{j_1}_{i_1} \cdots K^{j_{2p-1}}_{i_{2p-1}}R_{i_{2p}i_{2p+1}}^{j_{2p}j_{2p+1}} \cdots R_{i_{2m-2}i_{2m-1}}^{j_{2m-2}j_{2m-1}}\nonumber \\
&= \sum_{p=1}^{m}\int_{0}^{\Theta} d\theta\,\frac{-c_p}{2}\,\delta \sigma^{B}_{A}\,\delta^{A A_2\ldots A_{2m-1}}_{B B_2\ldots B_{2m-1}}K^{B_2}_{A_2} \cdots K^{B_{2p-1}}_{A_{2p-1}}R^{B_{2p}B_{2p+1}}_{A_{2p}A_{2p+1}} \cdots R^{B_{2m-2}B_{2m-1}}_{A_{2m-2}A_{2m-1}}.
\end{align}
After substituting \eqref{eq:Krotation}, the $ \theta $-integration works exactly in the same way as for the corner term above. Hence the result is given by \eqref{lovelockcornermidstep} multiplied by an overall factor of $-1\slash 2$ and with all indices contracted against $ \delta\sigma^B_A\,\delta^{A A_2\ldots A_{2m-1}}_{B B_2\ldots B_{2m-1}} $. Hence we get
\begin{equation}
\widehat{T}_{(m)ab} = 2m\,\Theta\,\widehat{E}_{(m-1)ab}  +\mathcal{F}_{(m)ab}(L_1,Q_1)+\mathcal{F}_{(m)ab}(L_2,Q_2),
\end{equation}
where the intrinsic equation of motion tensor
\begin{equation}
\widehat{E}_{(m)B}^{A}  =-\frac{1}{2}\frac{1}{2^{m}} \delta^{A A_1B_1\ldots A_mB_m}_{B C_1D_1\ldots C_mD_m}\widehat{R}^{C_1D_1}_{A_1B_1}\cdots \widehat{R}^{C_mD_m}_{A_mB_m}.
\end{equation}
and $\mathcal{F}_{(m)A}^B(L,Q)$ is given by \eqref{canoF} with indices contracted against $-\frac{1}{2}\,\delta^{A A_1\ldots A_{2m-2}}_{B B_1\ldots B_{2m-2}}$ which has two free indices.

\end{appendix}

\addcontentsline{toc}{section}{References}
\bibliography{higher_curvature.bib}
\bibliographystyle{JHEP}
	
\end{document}